\documentclass[%
twocolumn,
superscriptaddress,
 amsmath,amssymb,
 prx,
showkeys,
showpacs,
]{revtex4-2}
\usepackage{float}
\makeatletter
\let\newfloat\newfloat@ltx
\makeatother
\usepackage{algorithm}
\usepackage{algpseudocode}
\usepackage{graphicx}
\usepackage{dcolumn}
\usepackage{bm}
\usepackage{siunitx}
\usepackage{braket}
\usepackage{hyperref}
\usepackage{xcolor}
\usepackage{multirow}
\usepackage{array}
\usepackage{float}
\usepackage[title]{appendix}
\usepackage{etoolbox, chngcntr}
\usepackage[capitalize]{cleveref}
\usepackage{xr}
\usepackage{lipsum}
\usepackage{comment} 
\usepackage{yhmath} 
\usepackage{dsfont}

\makeatletter
\def\maketitle{
\@author@finish
\title@column\titleblock@produce
\suppressfloats[t]}
\makeatother

\newcommand{\ii}{\mathrm{i}}
\newcommand{\dt}[1]{\frac{\text{d}#1}{\text{d}t}}
\newcommand{\Tr}[1]{\textrm{Tr}\left\{ #1\right\}}

\floatstyle{plaintop}
\restylefloat{table}

\begin{document}
\begin{abstract}
    
\end{abstract}
\title{Mitigating crosstalk errors for simultaneous single-qubit gates on a superconducting quantum processor}
\author{Jaap J. Wesdorp}
\thanks{These two authors contributed equally. \\ 
jaap.wesdorp@meetiqm.com, eric@meetiqm.com
} 
\affiliation{ IQM Quantum Computers, Keilaranta 19, 02150 Espoo, Finland}

\author{Eric Hyyppä}
\thanks{These two authors contributed equally. \\ 
jaap.wesdorp@meetiqm.com, eric@meetiqm.com
} 
\affiliation{ IQM Quantum Computers, Keilaranta 19, 02150 Espoo, Finland}

\author{Joona Andersson}
\affiliation{ IQM Quantum Computers, Keilaranta 19, 02150 Espoo, Finland}

\author{Janos Adam}
\affiliation{ IQM Quantum Computers, Keilaranta 19, 02150 Espoo, Finland}

\author{Rohit Beriwal}
\affiliation{ IQM Quantum Computers, Keilaranta 19, 02150 Espoo, Finland}

\author{Ville Bergholm}
\affiliation{ IQM Quantum Computers, Keilaranta 19, 02150 Espoo, Finland}

\author{Saga Dahl}
\affiliation{ IQM Quantum Computers, Keilaranta 19, 02150 Espoo, Finland}

\author{Simone Diego Fasciati}
\affiliation{ IQM Quantum Computers, Keilaranta 19, 02150 Espoo, Finland}

\author{Alejandro Gomez Friero}
\affiliation{ IQM Quantum Computers, Keilaranta 19, 02150 Espoo, Finland}

\author{Zheming Gao}
\affiliation{ IQM Quantum Computers, Keilaranta 19, 02150 Espoo, Finland}

\author{Daria Gusenkova}
\affiliation{%
 IQM Quantum Computers, Georg-Brauchle-Ring 23-25, 80992 Munich, Germany
}

\author{Andrew Guthrie}
\affiliation{ IQM Quantum Computers, Keilaranta 19, 02150 Espoo, Finland}

\author{Johannes Heinsoo}
\affiliation{ IQM Quantum Computers, Keilaranta 19, 02150 Espoo, Finland}

\author{Tuukka Hiltunen}
\affiliation{ IQM Quantum Computers, Keilaranta 19, 02150 Espoo, Finland}

\author{Keiran Holland}
\affiliation{ IQM Quantum Computers, Keilaranta 19, 02150 Espoo, Finland}

\author{Amin Hosseinkhani}
\affiliation{%
 IQM Quantum Computers, Georg-Brauchle-Ring 23-25, 80992 Munich, Germany
}

\author{Sinan Inel}
\affiliation{ IQM Quantum Computers, Keilaranta 19, 02150 Espoo, Finland}

\author{Joni Ikonen}
\affiliation{ IQM Quantum Computers, Keilaranta 19, 02150 Espoo, Finland}

\author{Shan W. Jolin}
\affiliation{ IQM Quantum Computers, Keilaranta 19, 02150 Espoo, Finland}

\author{Kristinn Juliusson}
\affiliation{ IQM Quantum Computers, Keilaranta 19, 02150 Espoo, Finland}

\author{Seung-Goo Kim}
\affiliation{ IQM Quantum Computers, Keilaranta 19, 02150 Espoo, Finland}

\author{Anton Komlev}
\affiliation{ IQM Quantum Computers, Keilaranta 19, 02150 Espoo, Finland}

\author{Roope Kokkoniemi}
\affiliation{ IQM Quantum Computers, Keilaranta 19, 02150 Espoo, Finland}

\author{Otto Koskinen}
\affiliation{ IQM Quantum Computers, Keilaranta 19, 02150 Espoo, Finland}

\author{Joonas Kylmälä}
\affiliation{ IQM Quantum Computers, Keilaranta 19, 02150 Espoo, Finland}

\author{Alessandro Landra}
\affiliation{ IQM Quantum Computers, Keilaranta 19, 02150 Espoo, Finland}

\author{Julia Lamprich}
\affiliation{%
 IQM Quantum Computers, Georg-Brauchle-Ring 23-25, 80992 Munich, Germany
}
\author{Magdalena Lehmuskoski}
\affiliation{ IQM Quantum Computers, Keilaranta 19, 02150 Espoo, Finland}

\author{Nizar Lethif}
\affiliation{%
 IQM Quantum Computers, Georg-Brauchle-Ring 23-25, 80992 Munich, Germany
}

\author{Per Liebermann}
\affiliation{ IQM Quantum Computers, Keilaranta 19, 02150 Espoo, Finland}

\author{Tianyi Li}
\affiliation{ IQM Quantum Computers, Keilaranta 19, 02150 Espoo, Finland}

\author{Aleksi Lintunen}
\affiliation{ IQM Quantum Computers, Keilaranta 19, 02150 Espoo, Finland}

\author{Fabian Marxer}
\affiliation{ IQM Quantum Computers, Keilaranta 19, 02150 Espoo, Finland}

\author{Kunal Mitra}
\affiliation{ IQM Quantum Computers, Keilaranta 19, 02150 Espoo, Finland}

\author{Jakub Mro\.{z}ek}
\affiliation{ IQM Quantum Computers, Keilaranta 19, 02150 Espoo, Finland}

\author{Lucas Ortega}
\affiliation{ IQM Quantum Computers, Keilaranta 19, 02150 Espoo, Finland}

\author{Miha Papi\v{c}}
\affiliation{%
 IQM Quantum Computers, Georg-Brauchle-Ring 23-25, 80992 Munich, Germany
}

\author{Matti Partanen}
\affiliation{ IQM Quantum Computers, Keilaranta 19, 02150 Espoo, Finland}

\author{Alexander Plyushch}
\affiliation{ IQM Quantum Computers, Keilaranta 19, 02150 Espoo, Finland}

\author{Stefan Pogorzalek}
\affiliation{%
 IQM Quantum Computers, Georg-Brauchle-Ring 23-25, 80992 Munich, Germany
}

\author{Michael Renger}
\affiliation{%
 IQM Quantum Computers, Georg-Brauchle-Ring 23-25, 80992 Munich, Germany
}

\author{Jussi Ritvas}
\affiliation{ IQM Quantum Computers, Keilaranta 19, 02150 Espoo, Finland}

\author{Sampo Saarinen}
\affiliation{ IQM Quantum Computers, Keilaranta 19, 02150 Espoo, Finland}

\author{Indrajeet Sagar}
\affiliation{ IQM Quantum Computers, Keilaranta 19, 02150 Espoo, Finland}

\author{Matthew Sarsby}
\affiliation{ IQM Quantum Computers, Keilaranta 19, 02150 Espoo, Finland}

\author{Mykhailo Savytskyi}
\affiliation{ IQM Quantum Computers, Keilaranta 19, 02150 Espoo, Finland}

\author{Ville Selinmaa}
\affiliation{ IQM Quantum Computers, Keilaranta 19, 02150 Espoo, Finland}

\author{Ivan Takmakov}
\affiliation{ IQM Quantum Computers, Keilaranta 19, 02150 Espoo, Finland}

\author{Brian Tarasinski}
\affiliation{ IQM Quantum Computers, Keilaranta 19, 02150 Espoo, Finland}

\author{Francesca Tosto}
\affiliation{ IQM Quantum Computers, Keilaranta 19, 02150 Espoo, Finland}

\author{David Vasey}
\affiliation{ IQM Quantum Computers, Keilaranta 19, 02150 Espoo, Finland}

\author{Panu Vesanen}
\affiliation{ IQM Quantum Computers, Keilaranta 19, 02150 Espoo, Finland}

\author{Jeroen Verjauw}
\affiliation{%
 IQM Quantum Computers, Georg-Brauchle-Ring 23-25, 80992 Munich, Germany
}
\author{Alpo Välimaa}
\affiliation{ IQM Quantum Computers, Keilaranta 19, 02150 Espoo, Finland}

\author{Nicola Wurz}
\affiliation{%
 IQM Quantum Computers, Georg-Brauchle-Ring 23-25, 80992 Munich, Germany
}
\author{Hsiang-Sheng Ku}
\affiliation{%
 IQM Quantum Computers, Georg-Brauchle-Ring 23-25, 80992 Munich, Germany
}
\author{Frank Deppe}
\affiliation{%
 IQM Quantum Computers, Georg-Brauchle-Ring 23-25, 80992 Munich, Germany
}
\author{Juha Hassel}
\affiliation{ IQM Quantum Computers, Keilaranta 19, 02150 Espoo, Finland}

\author{Caspar Ockeloen-Korppi}
\affiliation{ IQM Quantum Computers, Keilaranta 19, 02150 Espoo, Finland}

\author{Wei Liu}
\affiliation{ IQM Quantum Computers, Keilaranta 19, 02150 Espoo, Finland}

\author{Jani Tuorila}
\affiliation{ IQM Quantum Computers, Keilaranta 19, 02150 Espoo, Finland}

\author{Chun Fai Chan}
\affiliation{ IQM Quantum Computers, Keilaranta 19, 02150 Espoo, Finland}

\author{Attila Geresdi}
\affiliation{%
 IQM Quantum Computers, Georg-Brauchle-Ring 23-25, 80992 Munich, Germany
}

\author{Antti Vepsäläinen}
\email{avepsalainen@meetiqm.com}
\affiliation{ IQM Quantum Computers, Keilaranta 19, 02150 Espoo, Finland}

\begin{abstract}

Single-qubit gates on superconducting quantum processors are typically implemented using microwave pulses applied through dedicated control lines. However, these microwave pulses may also drive other qubits due to crosstalk arising from capacitive coupling and wavefunction overlap in systems with closely spaced transition frequencies. Crosstalk and frequency crowding increase errors during simultaneous single-qubit operations relative to isolated gates, thus forming a major bottleneck for scaling superconducting quantum processors. In this work, we combine model-based qubit frequency optimization with pulse shaping to demonstrate crosstalk error mitigation in single-qubit gates on a 49-qubit superconducting quantum processor.  We introduce and experimentally verify an analytical model of simultaneous single-qubit gate error caused by microwave crosstalk that depends on a given pulse shape. By employing a model-based optimization strategy of qubit frequencies, we minimize the crosstalk-induced error across the processor and achieve a mean simultaneous single-qubit gate fidelity of 99.96\% for a 16-ns gate duration, approaching the mean individual gate fidelity. To further reduce the simultaneous error and required qubit frequency bandwidth on high-crosstalk qubit pairs, we introduce a crosstalk transition suppression (CTS) pulse shaping technique that minimizes the spectral energy around transitions inducing leakage and crosstalk errors. Finally, we combine CTS with model-based frequency optimization across the device and experimentally show a systematic reduction in the required qubit frequency bandwidth for high-fidelity simultaneous gates, supported by simulations of systems with up to 1000 qubits. By alleviating constraints on qubit frequency bandwidth for parallel single-qubit operations, this work represents an important step for scaling towards larger quantum processors.

\end{abstract}
\date{March 11, 2026}
\maketitle

\section{Introduction}
\label{sec: intro}

Superconducting circuits are one of the leading platforms for building scalable quantum processing units (QPUs) with quantum error correction (QEC) \cite{litinski2019game,krinner2022realizing, google2023suppressing, google2025quantum, lacroix2025scaling}. 
Current promising error-correcting codes rely on repeated error syndrome measurements and transversal operations \cite{gottesman1997stabilizer, versluis2017scalable, krinner2022realizing, lacroix2025scaling}, which require  simultaneous physical single-qubit operations across the QPU.
Additionally, near-term algorithms rely on parallelization to reduce the circuit depth
and to speed up computation~\cite{preskill2018quantum}.
In transmon-based QPUs, classical microwave \cite{nuerbolati2022canceling, kosen2024signal} and quantum hybridization crosstalk \cite{zhao2022quantum, heunisch2023tunable, lange2025cross} often degrade the fidelity of simultaneous single-qubit gates due to
errors in the qubit subspace and leakage to higher excited states \cite{theis2016simultaneous, google2023suppressing}.  
Such leakage errors may cause long-timescale correlated errors in QEC without additional leakage reduction units~\cite{mcewen2021removing, marques2023all, lacroix2025fast}, as error correcting codes do not correct for leakage~\cite{fowler2013coping}. 
Therefore, improved methods for crosstalk error mitigation are needed to allow further scaling of processor size towards fault-tolerant quantum computation.

\begin{figure}[h!]
    \centering
    \includegraphics[width=0.485\textwidth]{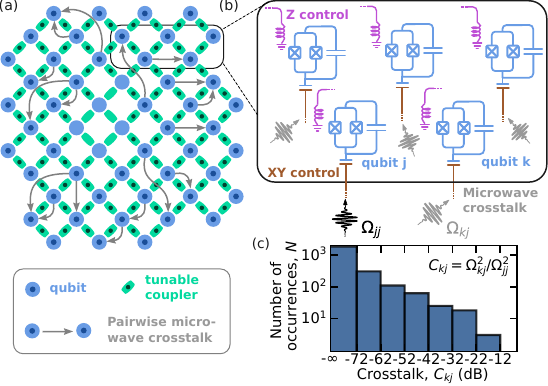}
    \caption{\textbf{Device overview and crosstalk statistics.} \textbf{(a)} Schematic of the IQM crystal device. Transmon qubits (blue) are connected through tunable couplers (cyan). The gray arrows indicate qubit control lines (beginning of the arrow) which have more than \SI{-30}{\decibel} crosstalk to another qubit (end of the arrows, respectively). The components without a dot in the center are not functional in this device, see \cref{sec: MW crosstalk and 1qb gate error}. \textbf{(b)} Partial circuit diagram of a group of qubits indicated in (a) to illustrate the effect of microwave drive crosstalk.  A qubit $j$ can be driven through its own control line causing Rabi oscillations with rate $\Omega_{jj}$ (black pulse), while a nearby qubit $k$ oscillates due to crosstalk with a smaller rate $\Omega_{kj}$ (grey pulses). Each qubit is frequency tunable using individual flux lines (Z control, pink) and controllable using drive lines (XY control, brown) - tunable couplers are not shown here. \textbf{(c)} Histogram of the crosstalk elements $\{C_{kj}\}$ of the 49 qubits used in this work. 
    } 
    \label{fig_main: device and crosstalk}
\end{figure}

Pulse shaping  provides a promising approach for mitigating leakage and crosstalk-induced errors. Conventionally, the derivative removal by adiabatic gate (DRAG) technique \cite{motzoi2009simple, chow2010optimized, lucero2010reduced} is applied to shape single-qubit drive pulses for weakly anharmonic qubits, such as transmons \cite{koch2007charge}. The DRAG technique suppresses the spectrum of the pulse at a single frequency, which enables the mitigation of the $|1\rangle-|2\rangle$ 
transition and hence single-qubit gates with a low leakage error of $\sim 10^{-5}$ in the absence of crosstalk~\cite{chen2016measuring, mckay2017efficient, hyyppa2024reducing, chiaro2025active}. 
Higher-derivative extensions of DRAG and other pulse shaping techniques have been theoretically proposed \cite{motzoi2013improving, schutjens2013single} and recently experimentally implemented~\cite{vesterinen2014mitigating, hyyppa2024reducing, li2024experimental, wang2025suppressing, matsuda2025selective, gao2025ultra, singh2026fast} to suppress the frequency spectrum at multiple transitions. This has been shown to reduce the excitation rate of undesired transitions and the impact of crosstalk in microwave-driven single- and two-qubit gates, while the pulses remain relatively simple to calibrate compared to closed-loop~\cite{kelly2014optimal,werninghaus2021leakage} and open-loop~\cite{caneva2011chopped, machnes2018tunable, berger2024dimensionality, genois2025quantum} optimal control techniques. Furthermore, these pulse shaping techniques can be seen as a complementary approach to active crosstalk cancellation~\cite{sung2021realization, nuerbolati2022canceling, google2023suppressing}.
Importantly, prior experimental studies on pulse shaping techniques for crosstalk mitigation have been limited to small subsystems comprising a few qubits~\cite{vesterinen2014mitigating, 
berger2024dimensionality, wang2025suppressing, matsuda2025selective} without demonstrating their scalability to large-scale QPUs.

Besides control techniques, signal crosstalk may be reduced in hardware using ground plane shielding via air-bridges~\cite{chen2014fabrication}, through silicon vias~\cite{vahidpour2017superconducting, alfaro-barrantes2020superconducting, yost2020solidstate, mallek2021fabrication, grigoras2022qubit}, flip-chip technology~\cite{yost2020solidstate, kosen2024signal}, or tiling modular 3D enclosures each containing a small number of qubits~\cite{ihssen2025low, spring2022high}.  
In the design stage, crosstalk can be accounted for by choosing the frequencies of the qubits to mitigate frequency crowding \cite{krinner2022realizing, osman2023mitigation, zhang2022highperformance, hertzberg2021laserannealing, zhang2025qplacer}.
However, crosstalk becomes increasingly difficult to predict at scale and the qubit frequencies must also satisfy other constraints related to, e.g., readout \cite{bengtsson2024modelbased} or two-qubit gates \cite{marxer2025999}. A complementary approach is \textit{in-situ} optimization of the qubit frequencies during calibration using, for example, flux-tunable qubits~\cite{ding2020systematic, klimov2024optimizing, lu2025neural, jiang2025generation}.
In order to calibrate large-scale devices in practice, these calibration methods should be fast, robust and scalable. This requires simple model-based approaches~\cite{bengtsson2024modelbased, klimov2020snake, klimov2024optimizing} that combine models of crosstalk error with other error mechanisms. 
Pulse shaping techniques can then be combined in the model-based optimization, while involving no additional hardware requirements.

In this work, we combine  model-based \textit{in-situ} frequency optimization with pulse shaping to mitigate crosstalk errors of simultaneous single-qubit gates on a 49-qubit superconducting processor.  
First, we study the statistics of microwave crosstalk of our device in Sec.~\ref{sec: MW crosstalk and 1qb gate error}. 
To understand and mitigate crosstalk, we discuss the experimental scaling of simultaneous gate errors with pulse amplitude, qubit-qubit detuning and distance. This is aided by an analytical model for the crosstalk-induced single-qubit gate error, which elucidates an important error mechanism arising from Rabi splitting during simultaneous drive pulses. In Sec.~\ref{sec: CT config results}, we experimentally mitigate the crosstalk-induced  single-qubit gate error by optimizing the qubit frequency configuration using error landscapes based on our analytical model. We demonstrate a mean simultaneous single-qubit gate fidelity of 99.96\%  approaching isolated gate fidelities for 16-ns $X_{\pi/2}$ gates at the cost of an increased qubit frequency bandwidth, quantified as the total span of qubit frequencies.

To reduce the required qubit frequency bandwidth, we introduce and demonstrate a crosstalk transition suppression (CTS) technique that utilizes pulse shaping to suppress the spectral energy of drive pulses around transition frequencies inducing crosstalk and leakage errors. In practice, we implement CTS using higher-derivative DRAG pulse shaping and off-resonant driving on qubit pairs with high microwave crosstalk, which is discussed in Sec.~\ref{sec: pulse shape engineering}. 
On an example high-crosstalk pair, we demonstrate up to a factor of $28 \pm 4$ reduction in the crosstalk-induced error during simultaneous $X_{\pi/2}$ gates using CTS compared to conventional cosine pulses.
Finally, in Sec.~\ref{sec: demonstration of reduced bandwidth on full QPU using pulse shaping techniques}, we combine CTS with qubit frequency optimization to 
mitigate simultaneous gate errors on 46 qubits and reduce the qubit frequency bandwidth  to below \SI{200}{\mega\hertz}, while retaining a mean crosstalk-induced $X_{\pi/2}$ error of $(4.4 \pm 1.0) \times 10^{-5}$. 
To further demonstrate the scalability of our approach, we simulate the frequency optimization assuming similar crosstalk statistics and design parameters as on the measured device. We verify that CTS systematically reduces the required average qubit frequency bandwidth 
 for larger devices with up to $1000$ qubits by on average $119\pm\SI{4}{\mega\hertz}$. Thus, the combination of CTS with model-based qubit frequency optimization enables high-fidelity parallel single-qubit operations while leaving a larger share of the qubit frequency budget for other operations, such as two-qubit gates or readout. 

\begin{figure*}
    \centering
    \includegraphics{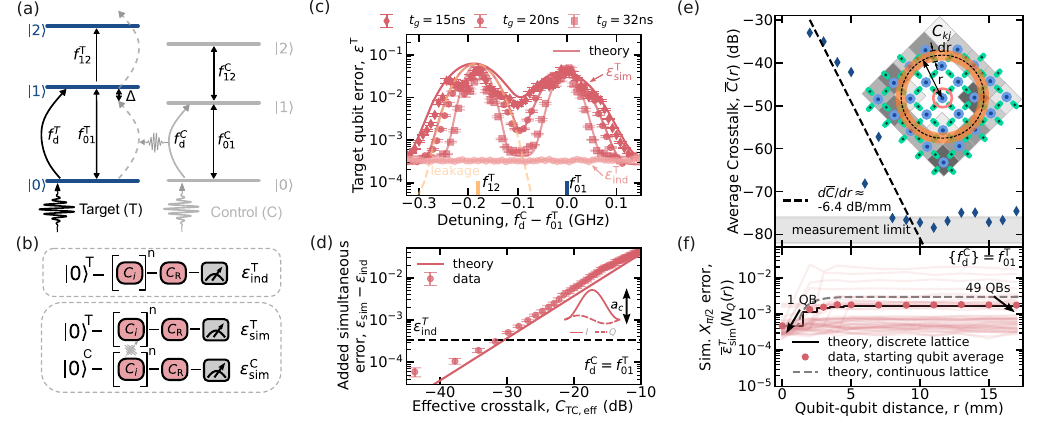}
    \caption{\textbf{The effect of microwave drive crosstalk on single-qubit gate error.} 
       \textbf{(a)} Schematic drawing of the relevant energy levels for a pair of transmon qubits, named target qubit T and control qubit C, shown in blue and grey respectively. Transition frequencies to the first ($f_{01}^\mathrm{T}$, $f_{01}^\mathrm{C}$) and from the first to the second excited state ($f_{12}^\mathrm{T}$, $f_{12}^\mathrm{C}$) are indicated for the target and control qubit, respectively. During operation, the target qubit is driven by both its own drive at frequency $f_\mathrm{d}^\mathrm{T}= f_{01}^\mathrm{T}$ and by the control qubit drive at detuned frequency $f_\mathrm{d}^\mathrm{C}=f_{01}^\mathrm{T}-\Delta/(2\pi)$ with an amplitude scaled by the crosstalk. \textbf{(b)} Clifford sequences used for measuring the individual and simultaneous $X_{\pi/2}$ error on the target and control, denoted by $\varepsilon_\mathrm{ind}^\mathrm{T}$, $\varepsilon_\mathrm{sim}^\mathrm{T}$, $\varepsilon_\mathrm{sim}^\mathrm{C}$  respectively, for the data in (c, d). \textbf{(c)} The crosstalk-induced gate error on the target qubit exhibits a double-peaked dependence on the control-target detuning $f_\mathrm{d}^\mathrm{C}-f_{01}^\mathrm{T}$, due to the resonance of $f_\mathrm{d}^\mathrm{C}$ with $f^\mathrm{T}_{01}$ (blue marker) or $f^\mathrm{T}_{12}$ (yellow marker) indicated in (a) for a pair of qubits with $C_\mathrm{TC}=-\SI{10}{\decibel}$. The peak centered at $f^\mathrm{T}_{12}$ is dominated by leakage error (yellow dashed line). Solid lines show the theoretical prediction with no fit parameters. \textbf{(d)} The crosstalk-induced error on the target qubit is proportional to the crosstalk. The experimental data at zero detuning ($\Delta=0)$ and analytical model are displayed as red dots and solid lines, respectively.  The effective crosstalk $C_\mathrm{TC,eff}$ is swept experimentally by scaling down the control pulse amplitude $a_\mathrm{C}$ to emulate a lower crosstalk (see inset and the main text). \textbf{(e)} The measured average crosstalk $\overline{C}(r)$ (blue diamonds, see main text) as a function of the qubit-qubit physical distance $r$. Each data point indicates the average crosstalk across target qubits 
       from control qubits in an annulus of radius $r$ and width $dr$ (see inset).  The dashed line indicates 
       an exponential fit to the first six data points above the measurement limit (grey shaded region). \textbf{(f)} The worst-case (fully resonant) simultaneous error saturates with distance due to the scaling of crosstalk with the on-chip lateral distance. The measured simultaneous $X_{\pi/2}$ error is shown for each target qubit (transparent red lines), when sweeping the number of included control qubits driven resonantly ($\{f_\mathrm{d}^\mathrm{C}\} = f_{01}^\mathrm{T}$) in simultaneous RB at an increasing distance $r$. Red dots indicate the error averaged over all 49 target qubits $\overline{\varepsilon}_\mathrm{sim}^\mathrm{T}$. Overlayed is the prediction of the resonant theory for the measured $\overline{C}(r)$ from (e), using the same discrete lattice (black solid steps) or an analytical integral assuming a continuous density (grey dashed line).
    }
    \label{fig_main: error from ct}
\end{figure*}

\section{Single-qubit gate error due to microwave crosstalk}
\label{sec: MW crosstalk and 1qb gate error}

The experiments in this work are performed on an IQM Crystal device with 54 superconducting transmon qubits and a similar square topology as on previous smaller IQM Crystal devices~\cite{ronkko2024onpremises, abdurakhimov2024technology}. Each qubit is connected to its neighbours via tunable couplers~\cite{yan2018tunable,sung2021realization}  as shown in \cref{fig_main: device and crosstalk}(a). This allows us to minimize the always-on entangling ZZ-interaction to below \SI{10}{\kilo\hertz} when the qubits are idled~\cite{yan2018tunable}, see~\cref{ap: qpu info and statistics}. Single-qubit operations are performed using capacitively coupled drive lines denoted as XY-control in~\cref{fig_main: device and crosstalk}(b). Additionally, the qubit transition frequencies are individually tunable using inductively coupled on-chip flux control lines denoted as Z-control in~\cref{fig_main: device and crosstalk}(b). 
A group of five qubits had broken Josephson junctions and were not usable, indicated without a central dot in~\cref{fig_main: device and crosstalk}(a). These qubits are not considered in the rest of the work. 

The crosstalk arising from unintended capacitive coupling between control lines and qubits is illustrated in schematics of a partial five qubit unit cell in~\cref{fig_main: device and crosstalk}(b).
A qubit $j$ can be driven through its own XY-control line to realize Rabi oscillations with rate $\Omega_{jj}$ using a drive frequency that is  resonant with its $\ket{0}\leftrightarrow\ket{1}$ transition frequency $f_{01}^j$. A nearby qubit
$k$ tuned on resonance with qubit $j$ then oscillates with a smaller rate $\Omega_{kj}$ due to crosstalk. We define the magnitude of the crosstalk as the drive line selectivity~\cite{spring2022high}
\begin{equation}
    C_{kj} = \Omega_{kj}^2/\Omega_{jj}^2.
    \label{eq: crosstalk definition}
\end{equation}

We first characterize the pair-wise crosstalk $C_{kj}$ on the device by measuring the Rabi rates $\Omega_{jj}, \Omega_{kj} \ \forall\ k, j\in \{1, \dots, 49\}$. The time and amplitude of a drive pulse for each qubit--control line pair is varied and we extract the resulting oscillation rate at maximum pulse amplitude~\cite{spring2022high}. If a qubit does not display at least a $\pi/2$ rotation at the maximum pulse length and amplitude, we consider the crosstalk to be below the measurement limit and calculate an upper bound on the crosstalk for that element, see Appendix~\ref{ap: crosstalk matric measurement} for more details. A histogram of all $49^2-49$ measured crosstalk matrix elements is shown in~\cref{fig_main: device and crosstalk}(c). The full matrix is shown in \cref{ap: crosstalk matric measurement}. 
The most detrimental elements that require crosstalk error mitigation using the methods described in this work, are those that are above $\SI{-30}{\decibel}$, shown in ~\cref{fig_main: device and crosstalk}(a). This level of crosstalk can cause simultaneous single-qubit $X_{\pi/2}$ errors up to $\sim3\times10^{-4}$ on the affected qubit, which is on the order of the single-qubit gate error obtained during isolated operation on our device, as shown in \cref{sec: experimental relation between crosstalk and 1qb gate error,sec:experimental_verification of ct error}. 
A key property of classical microwave crosstalk is that it typically displays a high degree of asymmetry, i.e. $C_{kj}\neq C_{jk}$. This is because the cross-coupling between the drive line of qubit $j$ and qubit $k$ depends on the routing of drive line $j$, which can strongly differ from the coupling of qubit $j$ due to the differently routed drive line of qubit $k$. 
 Thus, the most detrimental crosstalk elements on our device are one-directional, as shown in~\cref{fig_main: device and crosstalk}(a).

\subsection{Analytical model of single-qubit gate error due to crosstalk}
\label{sec: experimental relation between crosstalk and 1qb gate error}

To illustrate the detrimental effect of microwave crosstalk during simultaneous gates, we first focus on a single pair of qubits, where the source of the microwave crosstalk is referred to as control (C) and the impacted qubit is referred to as the target (T). 
The target qubit has two relevant transition frequencies, $f_{01}^\mathrm{T}:\ket{0}^\mathrm{T}\leftrightarrow\ket{1}^\mathrm{T}$ between the ground state and first excited state, and $f_{12}^\mathrm{T} : \ket{1}^\mathrm{T}\leftrightarrow\ket{2}^\mathrm{T}$ between the first and second excited state.
In a simultaneous single-qubit operation~[\cref{fig_main: error from ct}(a)], the target qubit is subjected to a microwave drive through its own drive line by a short pulse of length $t_{\rm g}$ at carrier frequency $f_\mathrm{d}^\mathrm{T}$.
At the same time, the control qubit, with transition frequencies $f_{01}^\mathrm{C}$ and $f_{12}^\mathrm{C}$ is driven by a second short pulse of length $t_{\rm g}$ at frequency  $f_\mathrm{d}^\mathrm{C}$ detuned  from $f_{01}^\mathrm{T}$ by $\Delta/(2\pi)= f_{01}^\mathrm{T} - f_\mathrm{d}^\mathrm{C}$, which affects the target qubit due to crosstalk. 

The Hamiltonian $H_\mathrm{T}$ of the target qubit, driven by an electromagnetic field from its own drive line and the additional perturbation  $H_{\rm TC}$, due to the control qubit drive are given by
\begin{align}\label{target_hamiltonian}
    &H_{\rm T}/\hbar = \omega^{\rm T}_{01}a^\dagger a + \frac{ \alpha^{\rm T}}{2}a^\dagger a^\dagger a a+ \ii  v^{\rm T}(t)(a^\dagger - a),\\
    &H_{\rm TC}/\hbar = \ii \lambda  v^{\rm C}(t)(a^\dagger - a), \label{eq: CT perturbation hamiltonian}
\end{align}
where $a^\dagger, a$ are the creation and annihilation operators in the target qubit subspace, $v^{j}(t)= [ s_I^j(t)  \cos(\omega_{\rm d}^{j}t + \phi^{j})+ s_Q^j(t) \sin(\omega^{j}_{\rm d}t + \phi^{j})]$ represent the applied driving fields modulated by the complex drive envelope $s^j(t)=s_I^j(t) - \ii s_Q^j(t)$. Here, $s_I^j(t)$ is the in-phase pulse envelope, $s_Q^j(t)$ is the quadrature envelope and index $j\in\{\mathrm{T},\mathrm{C}\}$. The drive phase is denoted by $\phi^j$ and $\omega_i^j$ denote angular frequency equivalents of the corresponding frequencies $f_i^j$ with the same sub- and superscripts ($\omega_i^j= 2 \pi f_i^j$). The target qubit anharmonicity  $\alpha^{\rm T}=\omega^{\rm T}_{12}-\omega^{\rm T}_{01}$ and $\lambda$ is related to the magnitude of drive crosstalk $C_\mathrm{TC}$ from the control drive line to the target qubit as $\lambda^2 = C_{\rm TC}$.

 We now evaluate the gate error of the target qubit during simultaneous driving of the control qubit. Throughout this work, we utilize and benchmark $X_{\pi/2}$ gates as the primitive single-qubit operation, which together with virtual $Z$ rotations enable an efficient implementation of universal single-qubit gates~\cite{mckay2017efficient}. During the gate, the time evolution of the perturbed density operator $\rho$ of the target qubit is given by the von Neumann equation $\dt{\rho} = -\frac{\ii}{\hbar}[H_{\rm T}+H_{\rm TC}, \rho]$ while the ideal evolution is given by  $\rho_{\rm ideal}(t_\mathrm{g})=U_{\rm ideal}(t_\mathrm{g}) \rho(0) U_{\rm ideal}^\dagger(t_\mathrm{g})$. Here, $\hat{U}_{\rm ideal} = \cos\frac{\theta(t)}{2}\sigma_x + e^{-\ii \phi}\sin\frac{\theta(t)}{2}\sigma_y$ is the ideal time-evolution of the target qubit under $H_{\rm T}$, where the rotation angle $\theta(t)$ of the target qubit is given by an integral of its envelope function, i.e, $\theta(t) = \int_0^t \text{d}t's_I^\mathrm{T}(t')$ at any time $0 \leq t \leq t_\mathrm{g}$ and $\sigma_i$ are the Pauli matrices acting on the lowest two levels, while performing identity on higher levels, see~\cref{ap: analytical error model} for more details.
We compute the fidelity under crosstalk as $\mathcal{F} = \Tr{[\rho_{\rm ideal}(t_\mathrm{g})\rho(t_\mathrm{g})]}$ and the error as $\mathcal{E} = 1 - \mathcal{F}$. By expanding up to second order in $\lambda$, an approximate expression for the average gate error $\mathcal{E}_{\rm av}$ over initial states can be expressed in terms of the energy spectral densities of the drive pulse applied to the control qubit
\begin{equation} 
    \begin{aligned}
    \mathcal{E}_{\rm av}
    &= \frac{\lambda^2}{12}\Bigl\{
    S[\Delta] + S_{\rm c}[\Delta] + S_{\rm s}[\Delta]\\
    &+ 3 S_{{\rm c}\frac{1}{2}}[\Delta + \alpha^{\rm T}]
    + 3 S_{{\rm s}\frac{1}{2}}[\Delta + \alpha^{\rm T}] 
    \Bigr\}+\mathcal{E}_\phi.
    \end{aligned}
\label{eq:mt_eps_av random phase}
\end{equation}
Here, $\Delta=2\pi (f_{01}^{\mathrm{T}} - f_{\mathrm{d}}^{\rm C})$, and the spectral densities are defined as $S_g[\omega] = |C_g[\omega]|^2$. Furthermore,  $C_g[\omega]$ denotes the Fourier transform of the drive pulse $s^{\rm C}(t)$ multiplied by a kernel function $g(t)$ that depends on the target qubit rotation angle $\theta(t)$ as $C_g[\omega] = \int_{-\infty}^\infty \text{d}te^{-\ii \omega t}s^{\rm C}(t)g(t)$, where $g(t)=1, g(t)=\sin\theta(t), g(t)=\cos\theta(t),g(t)=\sin\frac{\theta(t)}{2}, g(t)=\cos\frac{\theta(t)}{2}$ for $C[\omega], C_{\rm s}[\omega], C_{{\rm c}}[\omega], C_{{\rm s}\frac{1}{2}}[\omega], C_{{\rm c}\frac{1}{2}}[\omega]$, respectively.
The final term $\mathcal{E}_\phi$ indicates additional error terms that depend on the phase difference between the drive pulse on the target qubit and the drive pulse from the control qubit line arriving at the target qubit. See~\cref{ap: analytical error model} for the full expression and derivation of the error terms. 

The error terms $ S[\Delta], S_{\rm c}[\Delta]$ and $S_{\rm s}[\Delta]$ can be interpreted as off-resonant cross driving errors within the computational subspace of the target qubit arising from spectral power of the control qubit pulse near the target qubit transition frequency $f^{\rm T }_{01}$. Conversely, the error terms $S_{{\rm c}\frac{1}{2}}[\Delta + \alpha^{\rm T}]$ and  $S_{{\rm s}\frac{1}{2}}[\Delta + \alpha^{\rm T}]$ arise from spectral power of the control qubit pulse near the target qubit leakage transition $f^{\rm T}_{12}$, which causes leakage errors from off-resonant cross driving.
The terms including a $\theta(t)$ dependent kernel function arise because the target qubit and control qubit are simultaneously driven, leading to a Rabi splitting of the target qubit energy levels during the gate. These terms are important to account for when developing pulse shaping strategies to mitigate crosstalk errors for simultaneous gates, as shown in~\cref{sec: pulse shape engineering}.

\subsection{Experimental verification of crosstalk-induced single-qubit gate error }
\label{sec:experimental_verification of ct error}
To measure the effect of crosstalk on single-qubit gate errors experimentally, we perform simultaneous randomized benchmarking (RB)~\cite{magesan2011robust} sequences on the target-control pair to obtain the simultaneous $X_{\pi/2}$ error $\varepsilon_\mathrm{sim}^\mathrm{T}$. This is compared to the error $\varepsilon_\mathrm{ind}^\mathrm{T}$ obtained from an individual RB experiment on the target qubit to isolate the error coming from crosstalk. The circuit is shown in~\cref{fig_main: error from ct}(b). Each gate $C_i$ from the single-qubit Clifford group is decomposed into $\left\{I, X_{\pi/2},X_{-\pi/2}, Y_{\pi/2}, Y_{-\pi/2}\right\}$ operations for an average number of $N_\mathrm{g}=2.167$ single-qubit $\pi/2$-pulses per Clifford~\cite{epstein2014investigating}.  We calculate the average $X_{\pi/2}$ error on the target qubit from the obtained Clifford error $\varepsilon_{\mathrm{sim}, \mathrm{Cl}}^T$ using $\varepsilon^\mathrm{T}_\mathrm{sim}=\varepsilon^\mathrm{T}_{\mathrm{sim}, \mathrm{Cl}}/N_\mathrm{g}$ and similarly for $\varepsilon_\mathrm{ind}^\mathrm{T}$.  In the simultaneous RB experiments, we generate different random Clifford sequences for each qubit, which effectively randomizes the phase difference $\phi^\mathrm{C} - \phi^\mathrm{T}$ between the pulses of the target qubit and the control qubit within each gate sequence. As a result, the phase dependent terms average out and $\mathcal{E}_\phi\approx0$ in Eq.~\eqref{eq:mt_eps_av random phase}.
In the experiments of this section, we use raised cosine envelopes for the $I$-quadrature, $s^{\rm T}_I(t) = A_I^\mathrm{T}\left[1-\cos\left(2\pi t/t_\mathrm{g}\right)\right]/2$, where $A_I^\mathrm{T}=a_\mathrm{T}\Omega_\mathrm{TT}$, and DRAG-P~\cite{motzoi2009simple, chow2010optimized, lucero2010reduced} calibration for the $Q$-quadrature, see inset of~\cref{fig_main: error from ct}(d). We label this pulse shape as cosine DRAG throughout the rest of the manuscript and use it as a baseline for our experiments.

The dependence of the simultaneous gate error  on $\Delta$ and $t_\mathrm{g}$ is shown in~\cref{fig_main: error from ct}(c) for a qubit pair with strong crosstalk from control to target ($C_{\rm TC}=\SI{-10}{\decibel}$).
The simultaneous error on the target qubit $\varepsilon^\mathrm{T}_\mathrm{sim}$ has a clear double-peaked structure. The first peak in error occurs when the drive frequency of the control qubit $f_\mathrm{d}^\mathrm{C}$ is resonant with the $\ket{0}^\mathrm{T}\leftrightarrow\ket{1}^\mathrm{T}$ transition $f_{01}^\mathrm{T}$, corresponding to the terms $ S[\Delta], S_{\rm c}[\Delta]$ and $S_{\rm s}[\Delta]$ in~\cref{eq:mt_eps_av random phase}. The second peak arises  when $f_\mathrm{d}^\mathrm{C}$ is resonant with the $\ket{1}^\mathrm{T}\leftrightarrow\ket{2}^\mathrm{T}$ transition  $f_{12}^\mathrm{T}$, corresponding to the error terms $S_{{\rm c}\frac{1}{2}}[\Delta + \alpha^{\rm T}]$ and  $S_{{\rm s}\frac{1}{2}}[\Delta + \alpha^{\rm T}]$ in~\cref{eq:mt_eps_av random phase}.
Importantly, the error near $f_{12}^\mathrm{T}$ is dominated by leakage to $\ket{2}^\mathrm{T}$, which is an especially detrimental error channel for error correcting codes~\cite{mcewen2021removing, he2025experimental}.
The width of each error peaks scales as $1/t_{\mathrm{g}}$, because the spectral width of a pulse scales approximately inversely with its length. Thus, looking ahead at crosstalk mitigation protocols, the added simultaneous error at $\Delta\neq0,\Delta\neq\alpha^{\rm T}$ can be reduced by increasing $t_\mathrm{g}$, or, by increasing the detuning of the control qubit to the nearest of the transition frequencies $f_{01}^{\rm T}$ or $f_{12}^{\rm T}$.
As a control experiment, we  additionally measure the individual gate error $\varepsilon^\mathrm{T}_\mathrm{ind}$ on the target qubit (control qubit pulse off) before measuring $\varepsilon^\mathrm{T}_\mathrm{sim}$ for each  detuning. We see that $\varepsilon^\mathrm{T}_\mathrm{ind}$ remains approximately constant at $3 \times 10^{-4}$ during the experiment, indicating that indeed the error on the target qubit is due to crosstalk from the control qubit pulse. 

We compare the experimental $\varepsilon^\mathrm{T}_\mathrm{sim}$ to the predicted error of~\cref{eq:mt_eps_av random phase} (solid lines in~\cref{fig_main: error from ct}(c)), where we use only independently measured Hamiltonian parameters and the control and target pulse shapes as input. Without any free fit parameters, we find a good agreement between theory and experiment. In the comparison, we offset the theoretical $\mathcal{E}_{\rm av}$ by $\varepsilon^\mathrm{T}_\mathrm{ind}$ since~\cref{eq:mt_eps_av random phase} gives the simultaneous gate error neglecting decoherence and other error sources beyond crosstalk.
For cosine DRAG, the energy spectral density $S[\omega]$ can be approximated with a Gaussian function $S(\Delta, t_\mathrm{g})\propto e^{-(t_\mathrm{g}\Delta)^2/(2\sigma^2)}$ with $\sigma\approx2\pi \times 0.62$.
Furthermore the terms in~\cref{eq:mt_eps_av random phase} involving the target-pulse-dependent kernel functions cause an effective window averaging of the spectral density of the control envelope (see~\cref{ap: interpretation of analytical error model} for  more details).
Thus, for slowly varying spectral densities, such as that of cosine DRAG for small detunings, we have $S(\omega)\approx S_\mathrm{s}(\omega) + S_\mathrm{c}(\omega)$ 
and the error will approximately follow $\mathcal{E}_{\rm av}\propto \left(
    2S[\Delta] + 3 S[\Delta + \alpha^{\rm T}] 
    \right)$. 
This results in a Gaussian error peak centered at $\Delta=0$ and $\Delta=-\alpha^{\rm T}$, explaining the similarity between the error profile of~\cref{fig_main: error from ct}(c) and a double-Gaussian shape.

To complete the picture of crosstalk-induced error, we illustrate the scaling of the simultaneous gate error with the crosstalk strength at $\Delta=0$ in~\cref{fig_main: error from ct}(d). 
We perform the same experiment as before, but vary the pulse amplitude $a_\mathrm{C}$ on the control qubit during the simultaneous RB sequence, while keeping its length fixed.  This emulates an effective crosstalk strength $C_\mathrm{TC,eff} = C_\mathrm{TC}\left(\frac{a_\mathrm{C}}{a_0}\right)^2$, where $a_0$ is the calibrated true $X_{\pi/2}$ amplitude on the control qubit. 
The difference between the measured simultaneous error and individual error scales quadratically with the control pulse amplitude, and thus linearly with the crosstalk strength. 
The solid line indicates the predicted scaling following~\cref{eq:mt_eps_av random phase} at $\Delta=0$, yielding $\mathcal{E}_{\rm av}(\Delta=0)\approx c_0 \cdot C_{\rm TC,eff}$ for $c_0\approx0.37$.
The intuition about crosstalk errors obtained in this section serves as the basis for crosstalk error mitigation in Secs.~\ref{sec: CT config results}-\ref{sec: demonstration of reduced bandwidth on full QPU using pulse shaping techniques}.  

\subsection{Scaling of single-qubit gate error for larger devices}
\label{sec: scaling of crosstalk and error for larger devices}

An important requirement for scaling towards larger qubit numbers is that the error from crosstalk is bounded as the chip size increases.
We quantify the dependence of crosstalk on the physical qubit-qubit distance in~\cref{fig_main: error from ct}(e). The ensemble average crosstalk for an annulus at distance $r$ with width $\mathrm{d}r$ is defined as
\begin{equation}\label{eq:mt_ct_distance}
\overline{C}(r)=\frac{1}{N_\mathrm{Q}}\sum_{j\in N_\mathrm{Q}}\frac{1}{N_{q_j(r,r+\mathrm{d}r)}}\sum_{k\in q_j(r, r+\mathrm{d}r)}C_{jk}
\end{equation}
where $q_j(r,r+\mathrm{d}r)$ are the qubits present in the annulus centered at qubit $j$ within a distance $[r, r+\mathrm{d}r)$, and the double sum comes from averaging over annuli centered at all of the qubits, see inset of~\cref{fig_main: error from ct}(e). 
To provide an estimate for the scaling with distance, we fit a linear slope to the data points expressed in decibel above the \SI{-76}{\decibel} sensitivity limit of our measurement and find $\mathrm{d}\overline{C}/\mathrm{d}r\approx-\SI{6.4}{\decibel}/\SI{}{mm}$, which compares favorably to recently reported values in a flip-chip device~\cite{kosen2024signal}. Although there are many factors that can contribute to crosstalk, such as control line routing, we suspect the faster fall-off of crosstalk with distance is caused by the inclusion of through-silicon-vias (TSV's) in the routing layer that provide additional shielding (see \cref{ap: qpu info and statistics}).

To give an estimate for the worst-case classical crosstalk error, we consider a situation where all control drives are on resonance with a specific target qubit. For each target qubit, we measure the scaling of simultaneous gate error under resonant pulses from all other qubits within a disk of radius $r$. The average of this error over the 49 target qubits is displayed in~\cref{fig_main: error from ct}(f) with red circles. 
For $r=0$, the measured error is thus the individual single-qubit gate error and for $r\rightarrow\infty$ it is the fully simultaneous error caused by the control pulses of all of the other 48 qubits on the chip. 
The error increases initially with $r$ when more nearby qubits are included, but then remains approximately constant for $r>\SI{4}{\milli\meter}$ because the crosstalk drops off exponentially with distance in this device. 
For individual  qubits, the error shown in~\cref{fig_main: error from ct}(f) typically  exhibits a discrete jump when a control qubit with strong crosstalk to that specific target is included.

 Using the scaling of error with crosstalk strength found in~\cref{sec:experimental_verification of ct error}, we first compute the expected mean error
 $\overline{\epsilon}_{\rm sim}(r)\approx \frac{c_0}{N_\mathrm{Q}}\sum_{j\in N_\mathrm{Q}}\sum_{k\in q_j(0, r)}C_{jk}$ 
 for the discrete qubit lattice used in the experiment~[solid black line in~\cref{fig_main: error from ct}(f)], thus taking into account the edge qubits similarly as in the experiment. The discrete theory provides a good match with the experiment.
 
To provide insight into how this error would scale for larger devices, we additionally compute the expected average crosstalk-induced error for a continuous qubit density $\sigma_\mathrm{Q}$. Here, we use the exponential fit to the measured distance scaling of crosstalk in~\cref{fig_main: error from ct}(e) and we compute the expected average crosstalk-induced error $\overline{\epsilon}_{\rm sim}(r)\approx 2\pi c_0 \sigma_\mathrm{Q}\int_0^r  \overline{C}(r')r'dr'$, shown as a dashed line in~\cref{fig_main: error from ct}(f).
The continuous approximation predicts a slightly higher average error than the measured values due to edge effects on a finite-sized QPU.
For strong enough scaling with distance as is the case here, the integral converges and thus the total error remains bounded as the chip size grows.
This is a key requirement for scaling up to larger devices as it implies one only needs local mitigation for cancellation of crosstalk-induced errors.

\begin{figure*}[ht]
    \centering
    \includegraphics[width=\textwidth]{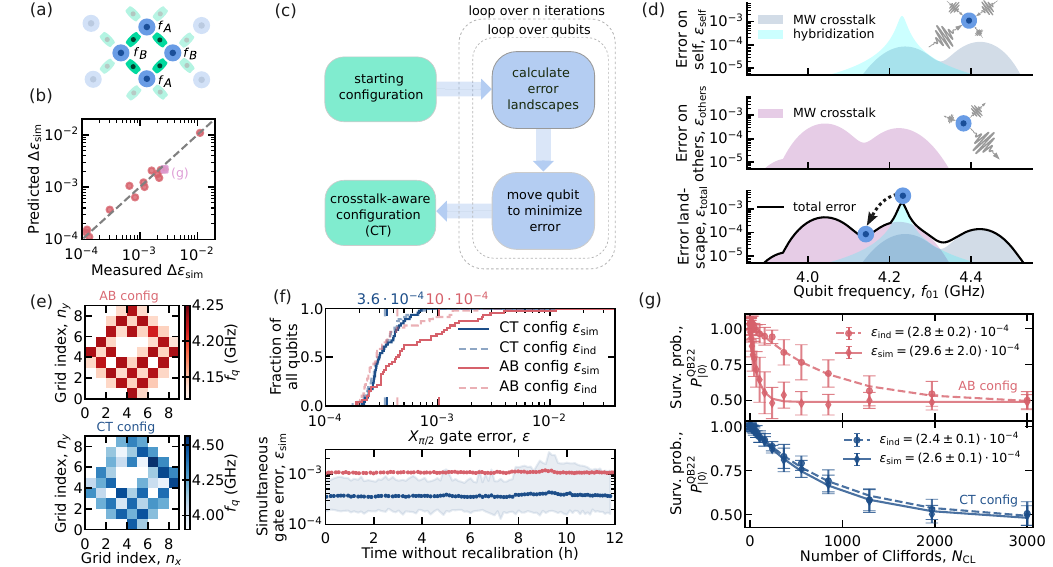}
    \caption{\textbf{Optimization of simultaneous operation of the full device using conventional pulse shapes.} \textbf{(a)} Schematic unit cell of four qubits in the device, where the qubits are set in two frequency groups $f_A$ and $f_B$, detuned by $f_A-f_B = |\alpha|/2 \approx \SI{90}{\mega\hertz}$, referred to as the reference configuration (called AB). 
    \textbf{(b)} Correlation between the measured and predicted added simultaneous error ($\Delta\varepsilon_\mathrm{sim}=\varepsilon_\mathrm{sim}-\varepsilon_\mathrm{ind}$) across the full chip in the AB configuration shown in (a). The dashed grey line indicates the two values being equal. Values below $10^{-4}$ are not shown as the uncertainty of individual gate error dominates the difference. \textbf{(c)} Schematic of the optimization algorithm (see text). \textbf{(d)} Illustration of the error landscape and qubit frequency update shown in (c). Top row: $\varepsilon_{\mathrm{self}}$, which is the error on this qubit due to the drive pulses of all other qubits (see inset illustration). Middle row: $\varepsilon_{\mathrm{others}}$, which is the error that the drive pulse of this qubit would cause on others if it was updated to have that frequency. Bottom row: the combined $\varepsilon_{\mathrm{total}}$, which is used to determine the updated qubit frequency. \textbf{(e)} Spread of qubit frequencies for the reference (AB) and optimized (CT) configuration. \textbf{(f)} Comparison of the fully simultaneous single-qubit gate fidelity in the AB (red) and CT (dark blue) configuration. Top panel: cumulative density function across the QPU, full lines indicate $\varepsilon_\mathrm{sim}$ while dashed lines indicate $\varepsilon_\mathrm{ind}$. Markers indicate mean values for $\varepsilon_\mathrm{ind}$, $\varepsilon_\mathrm{sim}$ with the same color, and the values for $\varepsilon_\mathrm{sim}$ are written above the panel. Bottom panel: 12-hour interleaved measurement of the average $\varepsilon_\mathrm{sim}$ for both configurations, shaded area indicates the minimum and maximum error for the CT configuration. \textbf{(g)} Comparison of RB traces from the data in (f) for a qubit where the simultaneous fidelity in the reference configuration was severely affected by crosstalk (magenta marker in (b)), while in the CT configuration $\varepsilon_\mathrm{sim}$ is close to $\varepsilon_\mathrm{ind}$. Error bars indicate one standard deviation over different Clifford sequences. 
    }
    \label{fig: crosstalk-optimized frequency configuration}
\end{figure*}

\section{Minimization of crosstalk-induced gate error using error landscapes}
\label{sec: CT config results}
Having established the relation between crosstalk and single-qubit gate error on a pair of qubits, we now apply the results of~\cref{sec: MW crosstalk and 1qb gate error} to mitigate crosstalk-induced errors on a large device by tuning the qubit frequencies to avoid frequency collisions. 
We first establish a reference configuration consisting of two frequency groups in the straddling regime~\cite{koch2007charge}, $A$ and $B$, detuned by $1/2\times |\alpha|/(2\pi)\approx\SI{90}{\MHz}$, see~\cref{fig: crosstalk-optimized frequency configuration}(a). This is a reasonable starting point for calibration without prior knowledge of the classical crosstalk, because it minimizes the error from nearest-neighbor hybridization crosstalk. 
Additionally, this detuning of nearest-neighbor qubits corresponds to a local minimum of error caused by microwave crosstalk as shown in~\cref{fig_main: error from ct}(c).
In the following, we use 46 qubits out of the 49 qubits on our device, because two qubits were not flux-tunable and one qubit had a low maximum qubit frequency. 

In the reference configuration, the simultaneous single-qubit gate error is significantly higher than the individual gate error due to crosstalk. In ~\cref{fig: crosstalk-optimized frequency configuration}(b), we show the measured error for all qubits with added crosstalk-induced errors $\Delta\varepsilon_{\rm sim}=\varepsilon_\mathrm{sim} - \varepsilon_\mathrm{ind} > 10^{-4}$, and compare it to the error predicted by the analytical model of~\cref{eq:mt_eps_av random phase}. To estimate the predicted error for a given target qubit, we sum up the predicted crosstalk errors independently evaluated for each of the control qubits using~\cref{eq:mt_eps_av random phase}. This is valid if the phase difference between control pulses arriving at the target qubit are sufficiently randomized. This can be either due to performing sufficiently randomized circuits, or due to varying cable-delays, microwave source phase desynchronization or a varying phase due to propagation on the device~\cite{kosen2024signal}. The strong correlation between experimental data and theory confirms the extension of the model to a larger device. In the special case of fully constructive interference of all control pulses, we note that one would have to sum the complex pulse envelopes over all control qubits before applying~\cref{eq:mt_eps_av random phase}, and the error would scale with the square of the number of qubits~$N_\mathrm{Q}^2$.  

\subsection{Crosstalk-aware optimization of the qubit frequency configuration}
\label{sec: strategy for the frequency configuration optimization}
We next utilize the model of~\cref{sec: MW crosstalk and 1qb gate error} to devise a new frequency configuration in which the error is minimized. For this purpose, we implement a minimal optimization procedure outlined in~\cref{fig: crosstalk-optimized frequency configuration}(c). The cost function for a given target qubit is given by the analytical error model in~\cref{eq:mt_eps_av random phase} summed over all control qubits. The objective is to find a qubit frequency configuration in which the cost function is minimal for a given range of qubit frequencies.  As a result of the optimization, each qubit $\mathrm{Q}_i, i \in \{1...N_\mathrm{Q}\}$ is assigned a frequency within the experimentally accessible range $f_i\in[f_\mathrm{min}, f_\mathrm{max}]$, such that the crosstalk error is minimized. To evaluate the error model, the optimizer uses experimentally measured values for the qubit frequencies $f^i_{01}$, anharmonicities $\alpha^i$, the $N_\mathrm{Q}^2$ crosstalk matrix elements $C_{ij}$ and applied pulse shapes $s^i(t)$. Throughout this section, we use conventional cosine pulse envelopes with DRAG.

In addition to the errors induced by the microwave crosstalk given by~\cref{eq:mt_eps_av random phase}, we account for errors arising from the overlap between the wavefunctions of neighboring qubits due to a residual $XX-YY$ coupling. The precise form of this error term away from resonances can be obtained from a perturbative model including two qubits and a coupling element~\cite{yan2018tunable}, but here we use a simplified heuristic Lorentzian shape centered at $f_{01}$ of the neighbouring qubits, shown in~\cref{fig: crosstalk-optimized frequency configuration}(d). We use the measured hybridization error of one pair to estimate the typical values for the Lorentzian height and width, see~\cref{ap: hybridization_error} for more details.

Finding a qubit frequency configuration that globally minimizes such an error function is non-trivial, because the search space grows exponentially with the number of qubits. The optimization landscape is additionally non-convex, as pointed out in recent works performing similar frequency configuration optimizations with different cost functions~\cite{klimov2020snake, klimov2024optimizing, jiang2025generation}. 
In practice however, there are many frequency configurations that result in a crosstalk-induced error lower than the individual gate error and it is sufficient to find a local minimum.
 Thus, to reduce the search space, we take a simplified approach and optimize line by line over qubit frequencies: 
 for a given qubit, we calculate the cost-function landscape of errors that all other qubits cause on the qubit itself, $\varepsilon_{\rm self}$, shown in the upper panel in~\cref{fig: crosstalk-optimized frequency configuration}(d). The goal is now to choose an updated frequency that satisfies our optimization target.
 Importantly, we need to ensure that the total system error is also minimized when optimizing the frequency of the given qubit. 
 Therefore, we also calculate the error the qubit causes on other qubits when moved to a certain frequency, $\varepsilon_{\rm others}$, and shown in the center panel in~\cref{fig: crosstalk-optimized frequency configuration}(d). We then combine the two errors into a total error, $\varepsilon_{\rm total}$, and find a new frequency for the qubit which satisfies the optimization criterion, indicated in the lower panel of~\cref{fig: crosstalk-optimized frequency configuration}(d).
 This ensures that the global error decreases when we update the frequency of each qubit in order to minimize the error on it. Finally, we loop over all qubits for several iterations to find a configuration with sufficiently low error, see~\cref{ap: detailed qubit frequncy optimization} for more details about the optimization algorithm.

\subsection{Measurement of simultaneous and individual gate errors for the optimized configuration}

The resulting configuration of qubit frequencies after the optimization is shown in~\cref{fig: crosstalk-optimized frequency configuration}(e) for a starting configuration where each qubit was set to its maximum frequency $f_{01, \mathrm{max}}$. The optimization target was set to find the nearest local minimum in error to $f_{01, \mathrm{max}}$. In this way, isolated gate performance was maximized for this experiment. Compared to the reference configuration, the resulting qubit frequency bandwidth, defined as the maximum spread in qubit frequencies, is larger due to the avoidance of crosstalk-induced error and additionally due to the spread in $f_{01, \mathrm{max}}$ set by junction variation in fabrication. In~\cref{ap: qpu info and statistics}, we show that despite the larger spread in qubit frequencies, we can still keep the always-on $ZZ$-interaction strength below $\SI{10}{\kilo\hertz}$ by calibrating the tunable coupler frequencies.  

We benchmark the performance after optimization by measuring the fully simultaneous single-qubit gate error $\epsilon_{\mathrm{sim}}$ for both configurations using 16-ns $X_{\pi/2}$ gates, shown in~\cref{fig: crosstalk-optimized frequency configuration}(f). We measure a mean simultaneous gate error of $3.6\times10^{-4}$, which is nearly equal to the mean individual gate error of $3.4\times10^{-4}$, see the upper panel of~\cref{fig: crosstalk-optimized frequency configuration}(f).
This is a large improvement over the reference AB configuration with a mean simultaneous gate error of $10.2\times10^{-4}$, to be compared with the mean individual gate error of $4.3\times10^{-4}$ in the same configuration. The corresponding cumulative distribution functions are shown in the upper panel of~\cref{fig: crosstalk-optimized frequency configuration}(f).
The slightly higher individual gate error in the reference configuration is attributed to increased qubit dephasing rates due to the larger distance from the maximum qubit frequencies, which increases the sensitivity to flux noise~\cite{koch2007charge}. A key observation in the cumulative distribution functions (CDF) of the simultaneous error for the two configurations is that the reference configuration contains a large tail of higher errors, caused by qubit pairs with high crosstalk. The optimization fixes these errors, while leaving most of the well performing pairs intact, strongly reducing the width of the CDF.  

We furthermore confirm the stability of the optimized configuration by measuring the simultaneous randomized benchmarking error for the two configurations interleaved over a period of 12 hours, shown in the bottom panel of~\cref{fig: crosstalk-optimized frequency configuration}(f). No recalibration was performed during this period. The strong performance gain of the frequency optimization can be seen from the fact that the maximum gate error in the optimized configuration remains below or close to the average gate error for the reference configuration during this period.

The effect of crosstalk-aware frequency optimization is shown in more detail in one of the RB traces that underlies the data of the CDF in~\cref{fig: crosstalk-optimized frequency configuration}(g). This data was taken for a qubit that was part of the tail of the CDF in the AB configuration and thus suffered from crosstalk-induced error. While the individual error rates for the qubits remain similar, the fully simultaneous error decreases by an order of magnitude in the crosstalk-aware configuration compared to the AB configuration and becomes close to the individual error rate.

\section{Crosstalk transition suppression using pulse shaping}
\label{sec: pulse shape engineering}

\begin{figure*}
    \centering
    \includegraphics{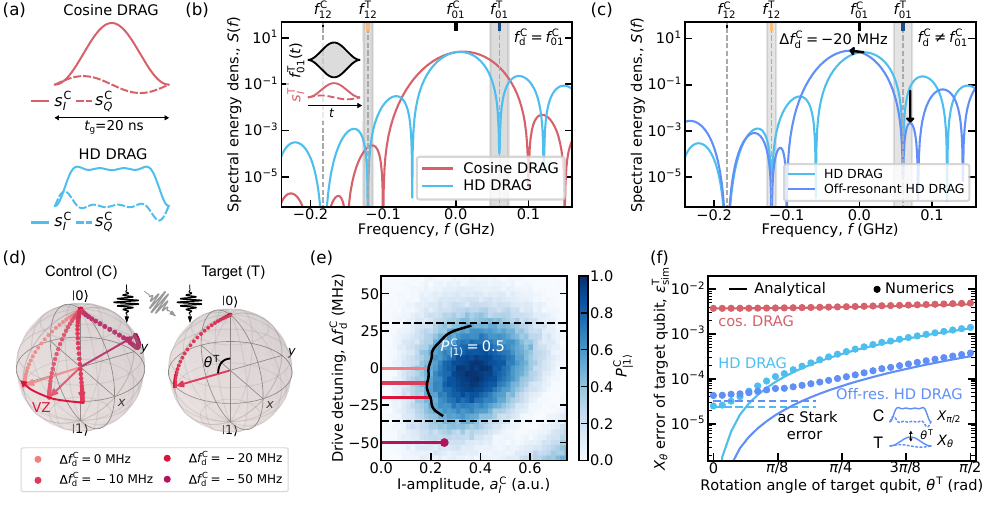}
    \caption{\textbf{Crosstalk transition suppression using higher-derivative DRAG and off-resonant drive pulses.} (a)~In-phase ($s_I^\mathrm{C}$, solid) and quadrature ($s_Q^\mathrm{C}$, dashed) envelopes of resonant 20-ns cosine DRAG (red here and later) and higher-derivative (HD) DRAG pulses (light blue here and later) applied to the control qubit. (b) Comparison of the energy spectral  density $S(f)$ for the pulse envelopes in (a). Dashed gray lines correspond to the transition frequencies $f_{01}^\mathrm{T}$, $f_{12}^\mathrm{T}$, and $f_{12}^\mathrm{C}$ suppressed by HD DRAG to mitigate crosstalk and leakage. A simultaneous drive pulse on the target qubit (inset) leads to a Rabi splitting of its $|0\rangle \leftrightarrow |1\rangle$ transition, which effectively broadens the transition frequencies $f_{01}^\mathrm{T}$ and $f_{12}^\mathrm{T}$ (shaded gray regions). (c) Energy spectral  density $S(f)$ for the resonant HD DRAG pulse and an off-resonant HD DRAG pulse (dark blue here and later) detuned by $\Delta f_\mathrm{d}^\mathrm{C}=-\SI{20}{\mega\hertz}$ with respect to the qubit frequency $f_{01}^\mathrm{C}$.  The off-resonant pulse reduces the energy spectral density around the nearest harmful transition frequency $f_{01}^\mathrm{T}$ across the frequency range corresponding to the Rabi splitting  
    (shaded gray regions).
    (d) Simulated trajectory of the state of the control qubit and the target qubit during simultaneous $X_{\pi/2}$ gates for varying off-resonant drive detunings $\Delta f_\mathrm{d}^\mathrm{C}$ of the control qubit (shades of red). A virtual $Z$ rotation can correct phase errors of the control qubit unless the drive detuning is too large (dark red trajectory). (e) Measured excited state probability of the control qubit in a Rabi experiment using a cosine DRAG pulse as a function of normalized pulse amplitude $a_I^\mathrm{C}$ and drive detuning, with the drive detunings of (d) indicated with horizontal solid lines. A 20-ns  $X_{\pi/2}$ gate can be implemented along the 
    contour with $P_{|1\rangle}^\mathrm{C} = 0.5$ for drive detunings up to $|\Delta f_\mathrm{d}^\mathrm{C}| \lesssim \SI{30}{\mega\hertz}$ (dashed lines). (f) Analytical (lines) and numerically simulated (markers) gate error of a 20-ns $X_\theta$ gate on the target qubit as a function of the target qubit rotation angle $\theta^\mathrm{T}$, with a simultaneous $X_{\pi/2}$ gate applied to the control qubit (inset) using the pulse shapes in (a)-(c). Dashed lines illustrate an approximation of the gate error due to an ac Stark shift of the target qubit induced by the drive pulse of the control qubit in the limit $\theta^\mathrm{T} \rightarrow 0$. 
    In all panels, we assume $f_{01}^\mathrm{C} - f_{01}^\mathrm{T} = -\SI{60}{\mega\hertz}$ and a crosstalk of $-\SI{13.9}{\decibel}$. }
    \label{fig: concept of higher-derivative drag}
\end{figure*}

To further reduce the impact of crosstalk and required qubit frequency bandwidth, we introduce and experimentally demonstrate a crosstalk transition suppression (CTS) protocol utilizing pulse shaping. To minimize various  terms of our analytical crosstalk error model in Eq.~\eqref{eq:mt_eps_av random phase}, the CTS protocol is designed to reduce the energy spectral density of the control qubit pulse across finite frequency bands around all relevant crosstalk-induced and leakage transitions. In practice, we implement the CTS protocol using  two complementary pulse shaping techniques, higher-derivative (HD) DRAG  and off-resonant driving.  
To simplify the problem, we first focus on crosstalk mitigation on a pair of qubits with high asymmetric microwave crosstalk schematically illustrated in Fig.~\ref{fig_main: error from ct}(a). 
The CTS protocol can then be readily extended from a single pair to the scale of the full QPU in Sec.~\ref{sec: demonstration of reduced bandwidth on full QPU using pulse shaping techniques}, because the microwave crosstalk on our QPU reduces strongly with distance and there are only a small number of high-crosstalk pairs as discussed in \cref{sec: scaling of crosstalk and error for larger devices}. 

\subsection{Errors arising from Rabi splitting during simultaneous driving}
\label{sec: rabi splitting}

First, we discuss the key requirements of pulse shaping techniques for crosstalk mitigation in multi-qubit systems in light of the analytical error model of Eq.~\eqref{eq:mt_eps_av random phase}. Based on the model, the crosstalk error of the target qubit can heavily depend on whether the target qubit is idling or driven during the control qubit pulse. Prior experimental pulse shaping demonstrations have focused on crosstalk mitigation for an idling target qubit~\cite{vesterinen2014mitigating, matsuda2025selective, wang2025suppressing}, thus largely omitting this distinction.

If the target qubit idles during the control qubit pulse, the crosstalk error model in Eq.~\eqref{eq:mt_eps_av random phase} reduces to 
\begin{equation} 
    \mathcal{E}_{\rm av, idling}
    = \frac{\lambda^2}{12}\Bigl\{
    2S[\Delta] + 3 S[\Delta + \alpha^{\rm T}] 
    \Bigr\}, 
    \label{eq:mt eps_av random phase idling}
\end{equation}
where $S[\Delta]$ and $S[\Delta + \alpha^{\rm T}]$ denote spectral energy density of the control qubit pulse $s^\mathrm{C}$ at the transition frequencies $f_{01}^\mathrm{T}$ and $f_{12}^\mathrm{T}$ of the target qubit, respectively. Thus, crosstalk-induced errors, including leakage, can be  mostly mitigated by suppressing the spectral energy density of the control qubit pulse at 
$f_{01}^\mathrm{T}$ and $f_{12}^\mathrm{T}$. For example, this can be achieved using HD DRAG pulse shaping as explained in \cref{sec: hd drag}. The analytical model in Eq.~\eqref{eq:mt eps_av random phase idling} highlights the importance of mitigating both crosstalk-activated transitions within and out of  the computational subspace  to reach optimal performance in systems based on low-anharmonicity qubits, such as transmons. Previous pulse shaping demonstrations have either mitigated transitions related to $f_{01}^\mathrm{T}$~\cite{matsuda2025selective, wang2025suppressing} or $f_{12}^\mathrm{T}$~\cite{vesterinen2014mitigating} but not both.

During simultaneous drive pulses on both qubits, the requirements for crosstalk transition suppression become more stringent since the full analytical model in Eq.~\eqref{eq:mt_eps_av random phase} must be considered. During simultaneous gates, the drive pulse on the target qubit induces a time-dependent Rabi splitting of its $|0\rangle$--$|1\rangle$ transition. The Rabi splitting gives rise to the additional error terms $S_\mathrm{c}(\Delta)$, $S_\mathrm{s}(\Delta)$, and $S_\mathrm{s\frac{1}{2}}(\Delta + \alpha^\mathrm{T})$ in Eq.~\eqref{eq:mt_eps_av random phase}. As derived in Appendix~\ref{ap: interpretation of analytical error model}, these terms can be expressed in the frequency domain using a convolution of the control qubit pulse and a kernel function  as  $S_g(\omega) = |1/(2\pi) \times (C * \mathcal{F}[g])(\omega) |^2$,
where $\mathcal{F}[g]$ denotes the Fourier transform of the kernel function $g(t)$ and $C[\omega] = \mathcal{F}[s^\mathrm{C}](\omega)$. 
Intuitively speaking, the Rabi splitting of the target qubit leads to window averaging of the Fourier transform $C[\omega]$ across finite frequency ranges around $f_{01}^\mathrm{T}$ and $f_{12}^\mathrm{T}$. The bandwidth of the averaging is proportional to the drive strength of the target qubit as given by the drive envelope $s^\mathrm{T}(t)$. For a typical 20-ns $\pi/2$-pulse, the Rabi splitting is on the order of \SI{10}{\mega\hertz}. 

Due to the Rabi splitting, it is not sufficient to suppress the control qubit pulse only at the specific transition frequencies $f_{01}^\mathrm{T}$ and $f_{12}^\mathrm{T}$ to reach optimal performance. Thus, pulse shaping strategies do not generally perform as effectively during simultaneous gates compared to the case of an idling target qubit. As a remedy, the spectral energy density should be reduced in a finite frequency band given by the Rabi splitting around the relevant transition frequencies as further discussed in \cref{sec: off-resonant control pulses}.

\subsection{Crosstalk and leakage suppression using higher-derivative (HD) DRAG pulses}
\label{sec: hd drag}

Here, we introduce an extension of our previous HD DRAG pulse shaping technique~\cite{hyyppa2024reducing}, based on theoretical ideas presented in~\cite{motzoi2013improving}, to suppress crosstalk-activated transitions within and out of the computational subspace in a multi-qubit system. The HD DRAG pulse  enables the suppression of a configurable number of transition frequencies in the frequency spectrum of the control qubit pulse. This allows us to mitigate both crosstalk-activated transition frequencies $f_{01}^\mathrm{T}$ and $f_{12}^\mathrm{T}$ as well as the leakage transition~$f_{12}^\mathrm{C}$ of the control qubit itself.  The HD DRAG pulse is also designed to ensure continuity of the in-phase and quadrature envelope functions, thus avoiding fast transients, which further mitigates leakage errors of the control qubit. Importantly, we symmetrically suppress frequencies around the center drive frequency similarly to Refs.~\cite{hyyppa2024reducing, wang2025suppressing}, thus avoiding potentially large frequency shifts arising from asymmetric spectral shaping. The symmetric suppression also simplifies experimental calibration and enables standard DRAG pulse tune-up without iterative closed-loop optimization. In~\cref{ap: maths of hd drag}, we compare HD DRAG pulse shaping to prior experimental pulse shaping techniques for crosstalk mitigation~\cite{vesterinen2014mitigating, li2024experimental, hyyppa2024reducing,  matsuda2025selective, wang2025suppressing}.

The in-phase and quadrature envelope functions of the HD DRAG pulse are given by 
\begin{align}
    s_I^\mathrm{C}(t) &= A_I^\mathrm{C} \bigg[\sum_{n=0}^K \beta_{2n} b^{(2n)} (t) \bigg], \label{eq: HD DRAG I} \\
    s_Q^\mathrm{C}(t) &= -\frac{\beta }{\alpha^\mathrm{C}} \dot{s}_I^\mathrm{C}(t) , \label{eq: HD DRAG Q}
\end{align}
where $A_I^\mathrm{C}$ is an in-phase amplitude in the units of Rabi rate, $\beta$ is the DRAG coefficient,  $\alpha^\mathrm{C}$ is the anharmonicity of the control qubit, and the dot denotes time derivative. Furthermore, $b(t)$ is a basis envelope function, and $\{\beta_{2n}\}_{n=0}^{K}$ are coefficients for the even-order derivatives with $\beta_0=1$. More details of the HD DRAG pulse shape are provided in Appendix~\ref{ap: maths of hd drag}. 

To activate CTS, we symmetrically suppress the Fourier transform of the in-phase envelope $\mathcal{F}[s_I^\mathrm{C}](f)$ at $2K$ frequencies $\{\pm f_{\mathrm{s}, j}\}_{j=1}^K$ by solving the $K$ coefficients $\{\beta_{2n}\}_{n=1}^{K}$ in Eq.~~\eqref{eq: HD DRAG I} from a system of linear equations. 
As a result, the rf drive pulse $v^\mathrm{C}(t)$ of  Eq.~\eqref{eq: CT perturbation hamiltonian} with drive frequency $f_\mathrm{d}^\mathrm{C}$ is suppressed at  $\{f_\mathrm{d}^\mathrm{C} \pm f_{\mathrm{s}, j}\}_{j=1}^K$.
In our case, $K=3$ and the set of suppressed frequencies comprises  $\{f_{\mathrm{s}, j}\} = \{|f_\mathrm{d}^\mathrm{C} - f_{01}^\mathrm{T}|, |f_\mathrm{d}^\mathrm{C} - f_{12}^\mathrm{T}|, |f_\mathrm{d}^\mathrm{C} - f_{12}^\mathrm{C}|\}$. In Figs.~\ref{fig: concept of higher-derivative drag}(a) and (b), we compare an example 20-ns HD DRAG pulse to a conventional cosine DRAG pulse in the time domain and in the frequency domain, where we also highlight the three suppressed frequencies and the relevant frequency bands for spectral suppression due to the Rabi splitting. 
In Appendix~\ref{ap: selection of suppressed frequencies for HD DRAG}, we experimentally measure the gate error and leakage on a high-crosstalk qubit pair for various combinations of suppressed transitions to confirm that all the three transition frequencies should be suppressed to achieve low  error and leakage across the straddling regime $ 0 \lesssim \Delta \lesssim - \alpha^\mathrm{T}$. 
Throughout the manuscript, we use the DRAG-L strategy ($\beta \approx 1$)~\cite{mckay2017efficient, hyyppa2024reducing} for calibrating HD DRAG pulses to minimize the leakage error of the control qubit by optimizing the DRAG coefficient $\beta$.

\subsection{Off-resonant drive pulses for crosstalk mitigation during simultaneous gates}
\label{sec: off-resonant control pulses}

To further mitigate crosstalk errors during simultaneous gates,  the Fourier transform of the control qubit pulse $C(\omega)$ should be suppressed across finite frequency bands around the transition frequencies $f_{01}^\mathrm{T}$ and $f_{12}^\mathrm{T}$ due to the Rabi splitting of the target qubit. To achieve this, we propose to use off-resonant drive pulses on the control qubit in contrast to resonant driving. Prior experimental works~\cite{chen2016measuring, wang2025suppressing} have utilized off-resonant pulses to mitigate gate errors on an individual qubit without considering the potential for crosstalk mitigation. Conveniently, off-resonant drive pulses allow us to decouple the qubit frequency, at which the qubit experiences crosstalk, from the drive frequency, at which the qubit causes crosstalk. 
In addition to off-resonant driving, wide-bandwidth suppression could potentially be also achieved using alternative pulse shaping methods, such as FAST~\cite{hyyppa2024reducing}. 

As shown in Fig.~\ref{fig: concept of higher-derivative drag}(c), we introduce a drive detuning $\Delta f_\mathrm{d}^\mathrm{C}$ for the HD DRAG pulse applied to the control qubit, which shifts the drive frequency $f_\mathrm{d}^\mathrm{C}$  away from the nearest transition frequency of the target qubit ($f_{01}^\mathrm{T}$ in this example). Furthermore, we update the suppressed frequencies and re-evaluate the envelope functions of the HD DRAG pulse to account for the change in $f_\mathrm{d}^\mathrm{C}$.  As a result, the energy spectral density of the control qubit pulse is reduced around $f_{01}^\mathrm{T}$ and $f_{12}^\mathrm{T}$ across the relevant frequency ranges determined by the Rabi splitting, see Fig.~\ref{fig: concept of higher-derivative drag}(c). This is because the energy spectral density generally falls off with increasing frequency difference from the center drive frequency. 
Consequently, the error terms $S_\mathrm{c}(\Delta)$ and $S_\mathrm{s}(\Delta)$ in Eq.~\eqref{eq:mt_eps_av random phase} and hence the gate error are reduced during simultaneous drive pulses. 

We can calibrate an $X_{\pi/2}$ gate on the control qubit across a large range of drive detunings as illustrated in \cref{fig: concept of higher-derivative drag}(d) and recently experimentally demonstrated in Ref.~\cite{wang2025suppressing} for the case of individual single-qubit gates. Namely, we can correct any remaining phase errors by applying virtual Z rotations before and after each pulse~\cite{mckay2017efficient} as long as the state vector of the qubit reaches the equator of the Bloch sphere from an initial state of $|0\rangle$. 
The use of virtual Z rotations for phase error mitigation enables us to retain the freedom in selecting the drive frequency, which is an advantage compared to approaches calibrating the drive frequency to cancel phase errors~\cite{chen2016measuring, gao2025ultra}.  

To study the available range of drive detunings $\Delta f_\mathrm{d}^\mathrm{C}$, we carry out a Rabi experiment as a function of normalized drive amplitude $a_I^\mathrm{C} \propto A_I^\mathrm{C}$ and drive detuning $\Delta f_\mathrm{d}^\mathrm{C}$ as shown in \cref{fig: concept of higher-derivative drag}(e). For simplicity, we use a conventional 20-ns cosine DRAG pulse in this experiment since the available range of drive detunings slightly depends on the exact pulse shape and thus on the qubit-qubit detuning for an HD DRAG pulse. 
In the Rabi experiment, the equator of the Bloch sphere ($P^C_{\ket{1}}=0.5$) is reached for any drive detuning satisfying $|\Delta f_\mathrm{d}^\mathrm{C}| \lesssim \SI{30}{\mega\hertz}$ with only a small or moderate increase of the drive amplitude $a_I^\mathrm{C}$ compared to the resonant case $\Delta f_\mathrm{d}^\mathrm{C}=0$. For  $|\Delta f_\mathrm{d}^\mathrm{C}| \gtrsim \SI{30}{\mega\hertz}$, the equator of the Bloch sphere is no longer reached ($P^C_{\ket{1}}<0.5$) at a gate duration of $t_\mathrm{g} = \SI{20}{\nano\second}$, and an $X_{\pi/2}$  rotation cannot be implemented. In Appendix~\ref{ap: strongly off-resonant control pulses}, we further study the limits of off-resonant driving as a function of gate duration both in experiment and in simulation to reveal a wide range of drive detunings compatible with high-fidelity $X_{\pi/2}$ gates. 

 To demonstrate the advantage of CTS combining HD DRAG pulses and off-resonant driving, we numerically simulate the gate error of the target qubit during simultaneous 20-ns pulses on both qubits 
 at a fixed qubit-qubit detuning 
 as shown in \cref{fig: concept of higher-derivative drag}(f). On the target qubit, we sweep the amplitude of a cosine DRAG pulse implementing an $X_\theta$ rotation between $\theta=0$ (no Rabi splitting)  and $\theta=\pi/2$~rad to investigate the effect of Rabi splitting. On the control qubit, we apply an $X_{\pi/2}$ rotation using either a resonant cosine DRAG pulse, a resonant HD DRAG pulse, or an off-resonant HD DRAG pulse with  $\Delta f_\mathrm{d}^\mathrm{C} = -\SI{20}{\mega\hertz}$. The  drive detuning is chosen to shift $f_\mathrm{d}^\mathrm{C}$ away from the nearest transition frequency, $f_{01}^\mathrm{T}$, while still being within the available range with some safety margin to enable robust calibration and only a small increase of pulse amplitude on the control qubit.

 If the target qubit idles, i.e., $\theta^\mathrm{T}=0$, both resonant and off-resonant HD DRAG pulses reduce the simulated gate error by two orders of magnitude compared to a resonant cosine DRAG pulse. As the rotation angle $\theta^\mathrm{T}$ of the target qubit increases and the qubits are driven simultaneously, the gate error of the target qubit grows for both resonant and off-resonant HD DRAG pulses due to the Rabi splitting.  Importantly, the off-resonant HD DRAG pulse  reduces the energy spectral density  around $f_{01}^\mathrm{T}$ and $f_{12}^\mathrm{T}$ compared to a resonant HD DRAG pulse as shown in \cref{fig: concept of higher-derivative drag}(c). Therefore, the off-resonant HD DRAG pulse reaches a lower error for simultaneous $X_{\pi/2}$-pulses ($\theta^\mathrm{T} =\pi/2$ rad) and allows an order-of-magnitude error reduction compared to a resonant cosine DRAG pulse.

 Notably,  \cref{fig: concept of higher-derivative drag}(f) establishes a good agreement between the analytical error model given by Eq.~\eqref{eq:mt_eps_av random phase} and the full numerical simulation apart from the limit $\theta^\mathrm{T} \rightarrow 0$, where the target qubit idles. In this limit, the crosstalk drive $H_\mathrm{TC}$ is no longer a small perturbation and a key assumption of the model breaks down, see Appendix~\ref{ap: analytical error model}. For an HD DRAG pulse and $\theta^\mathrm{T} \rightarrow 0$,  the error of the target qubit is dominated by the ac Stark shift induced by the control qubit pulse, which is neglected by Eq.~\eqref{eq:mt_eps_av random phase}.  In Appendix~\ref{ap: ac stark shift approximation}, we estimate the associated gate error due to the ac Stark shift, which we find to match well with the numerically evaluated error 
 in the limit $\theta^\mathrm{T} \rightarrow 0$ as shown by the dashed lines in \cref{fig: concept of higher-derivative drag}(f).  
In Appendix~\ref{ap: HD DRAG for idling error mitigation}, we experimentally confirm that the predicted error caused by the ac Stark shift also matches well with the measured crosstalk error if the target qubit idles when applying HD DRAG pulses on the control qubit.  
Additionally, these  results provide an experimental verification that HD DRAG pulse shaping reduces crosstalk errors more effectively if the target qubit idles instead of being simultaneously driven with the control qubit.

\subsection{Experimental demonstration of crosstalk transition suppression on a high-crosstalk qubit pair}
\label{sec: experimental demonstration on a high-crosstalk pair}

\begin{figure*}
    \centering
\includegraphics[width=\textwidth]
{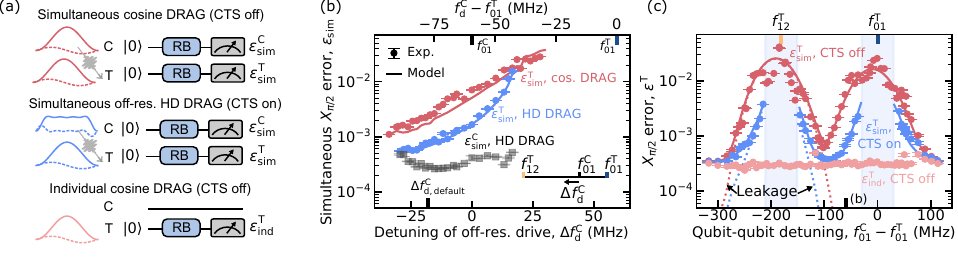}
    \caption{\textbf{Experimental demonstration of crosstalk transition suppression (CTS) for a high-crosstalk pair.}  (a) We compare the  error of simultaneous 20-ns $X_{\pi/2}$ gates measured from leakage RB with CTS on or off. For CTS on (blue here and later), we combine off-resonant driving and HD DRAG pulses on the control qubit.
    For CTS off (dark red here and later), we apply resonant cosine DRAG pulses on both qubits. As a baseline, we measure the individual gate error of the target qubit (light red here and later)  using a cosine DRAG pulse. 
    (b)~Impact of off-resonant driving on measured simultaneous $X_{\pi/2}$ error of the target qubit $\varepsilon_\mathrm{sim}^\mathrm{T}$ (markers) for the two pulse shapes. 
    Solid lines show the  analytical gate error model of Eq.~\eqref{eq:mt_eps_av random phase}.  Gray square markers denote the measured simultaneous gate error of the control qubit $\varepsilon_\mathrm{sim}^\mathrm{C}$ using an off-resonant HD DRAG pulse. The inset shows a diagram of the relevant transition frequencies  together with  the drive detuning $\Delta f_\mathrm{d}^\mathrm{C} = f_\mathrm{d}^\mathrm{C} - f_{01}^\mathrm{C}$ shifting the drive frequency away from the nearest harmful transition $f_{01}^\mathrm{T}$. The black marker on the x axis marks the magnitude of the drive detuning $\Delta f_\mathrm{d, default}^\mathrm{C}$ used by default in (c). 
    (c) Comparison of the experimental $X_{\pi/2}$ error of the target qubit $\varepsilon^\mathrm{T}$ for CTS on and off  as a function of qubit-qubit detuning $f_{01}^\mathrm{C}- f_{01}^\mathrm{T}$. 
    We denote analytical models of gate error and leakage as solid and dashed lines, respectively. For a 20-ns $X_{\pi/2}$ gate, the HD DRAG calibration may fail within $\pm\SI{30}{\mega\hertz}$ of the resonance conditions $f_{01}^\mathrm{C} =f_{01}^\mathrm{T}$ and  $f_{01}^\mathrm{C} =f_{12}^\mathrm{T}$ (shaded blue regions). The marker on the x axis denotes the qubit-qubit detuning used in (b). 
    The data in panels (b) and (c) is measured for the highest-crosstalk pair $\mathrm{Q}_5$--$\mathrm{Q}_{11}$ of our device ($C_{5, 11} = -\SI{13.9}{\decibel}$, see~\cref{ap: frequency dependence of crosstalk}). In panels (b) and (c), the error bars represent 1$\sigma$ uncertainty of the mean based on 3-6 repeated leakage RB experiments. 
    }
    \label{fig: experimental demonstration of off-resonant HD DRAG}
\end{figure*}

Next, we experimentally demonstrate that the CTS protocol reduces simultaneous single-qubit gate errors by combining off-resonant driving with HD DRAG pulse shaping. For the demonstration, we choose the qubit pair ($\mathrm{Q}_5$ -- $\mathrm{Q}_{11}$) with the highest crosstalk ($C_{5, 11} = -\SI{13.9}{\decibel}$, see~\cref{ap: frequency dependence of crosstalk}), and measure the simultaneous gate error $\epsilon_{\rm sim}^{\rm T}$ of the target qubit as a function of the off-resonant drive detuning $\Delta f_{\rm d}^{\rm C}$ of the control qubit. We repeat the experiment for two different cases, first by only using conventional cosine pulse shapes (CTS off), and then activating the crosstalk transition suppression using off-resonant HD DRAG pulses on the control qubit (CTS on), see \cref{fig: experimental demonstration of off-resonant HD DRAG}(a). For each value of the drive frequency $f_\mathrm{d}^\mathrm{C} = f_{01}^\mathrm{C} + \Delta f_\mathrm{d}^\mathrm{C}$, we update the HD DRAG envelope of the control qubit to ensure that we suppress transitions at frequencies $f_{01}^\mathrm{T}$, $f_{12}^\mathrm{T}$, and $f_{12}^\mathrm{C}$ without any further optimization of the suppressed frequencies as motivated by Appendix~\ref{ap: selection of suppressed frequencies for HD DRAG}. For both pulse shaping scenarios, we calibrate $X_{\pi/2}$ gates using relatively standard DRAG pulse tune-up experiments based on error amplification as explained in \cref{ap: single-qubit gate calibration}. 
To benchmark the average gate error $\varepsilon_\mathrm{g}$, we use leakage RB~\cite{wood2018quantification} with three-state single-shot readout after each Clifford gate sequence, see Appendix~\ref{ap: benchmarking of single-qubit gate errors} for more details.

In \cref{fig: experimental demonstration of off-resonant HD DRAG}(b), we show the measured simultaneous $X_{\pi/2}$ error of the target qubit $\varepsilon_\mathrm{sim}^\mathrm{T}$ for varying drive detuning of the control qubit $\Delta f_\mathrm{d}^\mathrm{C}$ using either a conventional cosine pulse or an HD DRAG pulse with a  gate duration of $t_\mathrm{g} = \SI{20}{\nano\second}$. For both pulse shapes, we observe that the gate error of the target qubit  monotonically decreases as the drive frequency $f_\mathrm{d}^\mathrm{C}$ is shifted away from the nearest target qubit transition frequency $f_{01}^\mathrm{T}$. By tuning the drive frequency away from the crosstalk transition $f_{01}^\mathrm{T}$, the simultaneous gate error of the target qubit reduces from $\varepsilon_\mathrm{sim}^\mathrm{T}=(16.6 \pm 0.2) \times 10^{-4}$ for a resonant HD DRAG pulse to $\varepsilon_\mathrm{sim}^\mathrm{T}=(6.5 \pm 0.1) \times 10^{-4}$ for $\Delta f_\mathrm{d}^\mathrm{C}=-\SI{20.8}{\mega\hertz}$. 
For a similar drive detuning, cosine pulse shapes also yield lower gate errors compared to the resonant driving case---validating the universality of the off-resonant driving concept---but still resulting in a factor of five higher crosstalk-induced error than for off-resonant HD DRAG  pulses.
This result verifies that the combination of HD DRAG pulse shaping and off-resonant driving forms a good basis for the CTS protocol.

In \cref{fig: experimental demonstration of off-resonant HD DRAG}(b), we also verify that the gate error of the control qubit stays approximately constant at $\varepsilon_\mathrm{sim}^\mathrm{C} \approx 3.5 \times 10^{-4}$ across the available range of drive detunings owing to  virtual Z gates correcting any phase errors. For the used gate duration of \SI{20}{\nano\second}, we can calibrate a high-fidelity gate on the control qubit for any drive detuning down to roughly $-\SI{30}{\mega\hertz}$. For off-resonant drive detunings below  $\Delta f_\mathrm{d}^\mathrm{C}<-\SI{20}{\mega\hertz}$, we, however, observe a minor increase in the leakage error of the control qubit at the $10^{-5}$ level, see Appendix~\ref{ap: supplementary fig 5 data}. Furthermore, the peak amplitude of the off-resonant HD DRAG pulse stays comparable or below that of a resonant cosine envelope down to $\Delta f_\mathrm{d}^\mathrm{C}\approx-\SI{20}{\mega\hertz}$ for the given qubit-qubit detuning. Therefore, we regard $|\Delta f_\mathrm{d}^\mathrm{C}| \sim \SI{20}{\mega\hertz}$ as a good compromise to balance crosstalk mitigation on the target qubit and robust calibration on the control qubit for $t_\mathrm{g} = \SI{20}{\nano\second}$.

In \cref{fig: experimental demonstration of off-resonant HD DRAG}(c), we further compare the error and leakage of 20-ns $X_{\pi/2}$ gates on the target qubit as a function of qubit-qubit detuning $f_{01}^\mathrm{C} - f_{01}^\mathrm{T}$ using either CTS or resonant cosine pulses. For each qubit-qubit detuning, we set the off-resonant drive detuning as detailed in Appendix~\ref{ap: strongly off-resonant control pulses} and re-evaluate the envelopes of the off-resonant HD DRAG pulse after updating the suppressed frequencies. 
Similarly to Sec.~\ref{sec: experimental relation between crosstalk and 1qb gate error}, microwave crosstalk results in a peak of gate error  around $f_{01}^\mathrm{C} \approx f_{01}^\mathrm{T}$ and a peak of leakage error around $f_{01}^\mathrm{C} \approx f_{12}^\mathrm{T}$ when implementing simultaneous $X_{\pi/2}$ gates using conventional cosine DRAG pulses. Importantly, CTS pulse shaping considerably reduces the simultaneous gate error and leakage of the target qubit across a wide range of qubit-qubit detunings. 
Using CTS, the excess simultaneous gate error $\varepsilon_\mathrm{sim}^\mathrm{T} - \varepsilon_\mathrm{ind}^\mathrm{T}$ is reduced by up to a factor of $28\pm 4 $, which occurs for $f_{01}^\mathrm{C} - f_{01}^\mathrm{T}=-\SI{81}{\mega\hertz}$. For this qubit-qubit detuning, the measured simultaneous gate error of the target qubit is $\varepsilon_\mathrm{sim}^\mathrm{T}=(3.6 \pm 0.1) \times 10^{-4}$ with CTS 
and $\varepsilon_\mathrm{sim}^\mathrm{T}=(14.2 \pm 0.1) \times 10^{-4}$ using cosine pulses. 
As a baseline, the measured individual gate error of the target qubit is $\varepsilon_\mathrm{sim}^\mathrm{T}=(3.2 \pm 0.1) \times 10^{-4}$ when no pulses are applied to the control qubit. Overall, we also observe an excellent agreement between the measured gate error and our analytical crosstalk error model in Eq.~\eqref{eq:mt_eps_av random phase}  with no free fitting parameters.

One of the limitations of the CTS protocol is that the frequency of the control qubit $f_{01}^\mathrm{C}$ must be sufficiently detuned from both $f_{01}^\mathrm{T}$ and $f_{12}^\mathrm{T}$ to enable a robust calibration of the $X_{\pi/2}$ gate despite  symmetric spectral shaping.
This is a general limitation of HD DRAG pulse shaping and other passive spectral shaping techniques~\cite{vesterinen2014mitigating, matsuda2025selective, wang2025suppressing}.  For the used gate duration $t_\mathrm{g}=\SI{20}{\nano\second}$, we achieve robust calibration down to a relatively low qubit-qubit detuning of \SI{30}{\mega\hertz}, i.e., $t_\mathrm{g} |f_{01}^\mathrm{C} - f_\mathrm{nearest}^\mathrm{T}| \gtrsim 0.6$, which corresponds to qubit-qubit detunings outside the shaded blue regions in \cref{fig: experimental demonstration of off-resonant HD DRAG}(c).

In \cref{ap: supplementary fig 5 data}, we provide supplementary experimental data for the gate error and leakage of the target qubit and the control qubit as a function of off-resonant drive detuning and gate duration. These  supplementary results demonstrate that crosstalk transition suppression using off-resonant HD DRAG pulses enables a significant reduction of crosstalk error on the target qubit, while enabling a low leakage error on both qubits and a high gate fidelity on the control qubit. The reduction of crosstalk error also unlocks faster gates and hence lower incoherent errors on both qubits compared to conventional cosine pulses. 
In this section, we have thus demonstrated crosstalk error reduction through pulse shaping for simultaneous gates that suffer from the Rabi splitting of the target qubit.

\section{Minimization of crosstalk error at reduced bandwidth using advanced pulse shaping} 
\label{sec: demonstration of reduced bandwidth on full QPU using pulse shaping techniques}

\begin{figure*}
    \centering
    \includegraphics{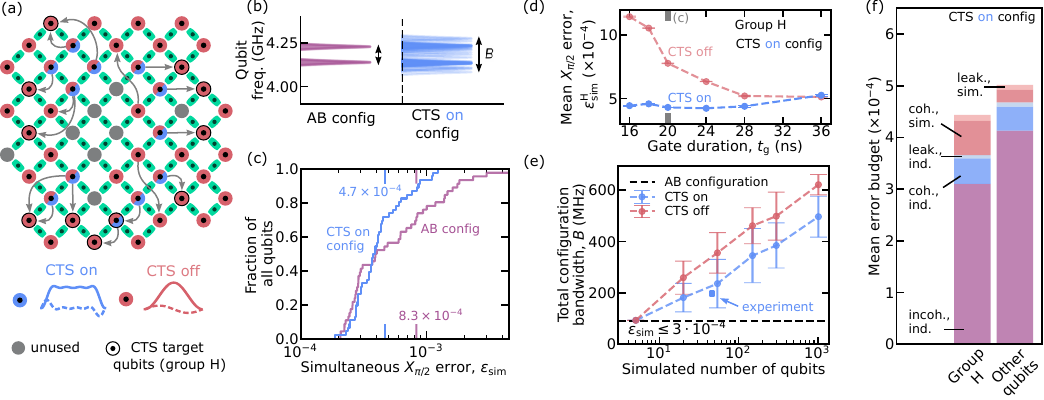}
    \caption{
    \textbf{Demonstration of CTS on the full device for high-fidelity simultaneous gates at narrow bandwidth.}
    (a)~Schematic QPU layout indicating pulse shapes applied to 46 qubits for CTS off and on. 
    For CTS off, a resonant cosine DRAG pulse (red) is applied to all qubits. For CTS on, we combine off-resonant driving and HD DRAG pulses (blue) for  12 qubits inducing high crosstalk errors on 11 target qubits forming the group H (qubits with outline). The gray arrows indicate all qubit control lines (start of the arrow) that have more than -\SI{30}{\decibel} crosstalk to another qubit (end of the arrow). 
     (b) Comparison of qubit frequency bandwidth for the reference AB configuration (purple) and a CTS-on configuration (blue). The CTS-on configuration is optimized to bound the predicted simultaneous $X_{\pi/2}$ error below $\varepsilon_\mathrm{sim} \leq 3 \times 10^{-4}$ assuming CTS on and 20-ns $X_{\pi/2}$ gates. 
    (c) Cumulative distribution function of simultaneous $X_{\pi/2}$ error from an 
    RB experiment on 46 qubits for the frequency configurations in (b) using a gate duration of \SI{20}{\nano\second}. The mean error is indicated by the same colored markers. 
    (d) Mean experimental $X_{\pi/2}$ error across group H as a function of gate duration $t_\mathrm{g}$ in the CTS-on frequency configuration using CTS on (blue) or off (red). Error bars represent the uncertainty of mean based on 5 leakage RB experiments.
    (e)  Simulated (circular markers) and experimental (square marker) bandwidth of qubit frequencies as a function of the number of qubits on the QPU based on pulse-aware optimization of qubit frequencies assuming either CTS off (red) or on (blue). Both in experiment and simulation, we target a minimal bandwidth for a predicted crosstalk-induced gate error of $\varepsilon_\mathrm{sim} \leq 3 \times 10^{-4}$ on all qubits. The simulated markers denote the average  bandwidth across 10 bootstrapped crosstalk matrices  and error bars indicate one standard deviation.
    (f) Mean error budget of 20-ns $X_{\pi/2}$ gates for the group H and the other 35 qubits using CTS at CTS-on configuration.  The error budget includes crosstalk-induced leakage (light red), crosstalk-induced errors within computational subspace (dark red), leakage error of individual gates (light blue), coherent error of individual gates (dark blue), and incoherent error (purple). In (d) and (e), dashed lines connecting the markers are a guide for the eye.
    }
    \label{fig: hd drag on bw minimized config}

\end{figure*}

In Sec.~\ref{sec: CT config results}, we demonstrated simultaneous single-qubit gate errors approaching individual gate errors using a model-based frequency optimization strategy and conventional pulse shapes. However, we did not consider the total qubit frequency bandwidth in the optimization, which is important to  enable high-fidelity elementary operations beyond single-qubit gates in large-scale QPUs. For example, a large qubit frequency bandwidth 
may pose challenges for idling the ZZ interaction at kHz precision in transmon-based architectures  \cite{marxer2023longdistance}, or necessitate more precise flux pulse predistortion for two-qubit gates.  
In this section, we address the issue by combining qubit frequency optimization with the 
crosstalk transition suppression protocol introduced in Sec.~\ref{sec: pulse shape engineering} to reduce the qubit frequency bandwidth for a given threshold of simultaneous single-qubit error.

\subsection{Bandwidth-minimized frequency configuration using crosstalk transition suppression}
\label{sec: demonstration of bW minimized configuration}

We use the same set of 46 qubits as in \cref{sec: CT config results} to demonstrate  high-fidelity simultaneous gates in a narrow-bandwidth configuration  using CTS and frequency optimization. 
To study the benefit of CTS, we compare two pulse shaping strategies denoted as CTS off and CTS on, see Fig.~\ref{fig: hd drag on bw minimized config}(a). For CTS off, we apply conventional resonant cosine DRAG pulses on all 46 qubits. For CTS on, we apply crosstalk transition suppression 
based on off-resonant HD DRAG pulses for 12 control qubits of high-crosstalk pairs. For the remaining 34 qubits, we utilize conventional cosine DRAG pulses. 
The selected high-crosstalk pairs have an asymmetric microwave crosstalk of $C_\mathrm{TC} > -\SI{30}{\decibel}$ 
and  typically a low crosstalk to other qubits. In total, there are 11 high-crosstalk target qubits labeled as the group H since two of the high-crosstalk pairs share a common target qubit ($\mathrm{Q}_{54}$). 

To achieve a narrow-bandwidth configuration with low crosstalk errors, we optimize the qubit frequency configuration using the model-based strategy introduced in~\cref{sec: CT config results} where we additionally account for the CTS pulse shaping in the error function. Here, we further adapt the optimization approach to  minimize the detuning of each qubit to a targeted AB configuration (see~\cref{sec: CT config results}), while ensuring the crosstalk error predicted by the analytical cost function remains below  an upper bound of $\varepsilon_\mathrm{sim} \leq 3\times 10^{-4}$, without including error from decoherence. Similarly to Sec.~\ref{sec: CT config results}, the cost function consists of a heuristic model for hybridization crosstalk and a pulse-shape-aware analytical model for microwave crosstalk based on Eq.~\eqref{eq:mt_eps_av random phase}. Furthermore, we incorporate a minimal detuning limit of \SI{40}{\mega\hertz} for the high-crosstalk pairs with CTS to ensure a robust calibration of 20-ns $X_{\pi/2}$ gates, see Appendix~\ref{ap: detailed qubit frequncy optimization} for further details. 

Using this optimization strategy, we obtain a qubit frequency configuration with a total bandwidth of $B=\SI{197}{\mega\hertz}$  for $t_\mathrm{g} = \SI{20}{\nano\second}$, while satisfying the given upper bound of predicted error. In the following, this optimized frequency configuration is called the CTS-on configuration. As shown in Fig.~\ref{fig: hd drag on bw minimized config}(b), the bandwidth of the CTS-on configuration is only a factor of 2 higher compared to the baseline AB configuration, 
while also being approximately equal to the typical anharmonicity $|\alpha|/(2\pi)\sim \SI{180}{\mega\hertz}$ of our device.

Subsequently, we experimentally verify that the simultaneous gate errors are low 
in the CTS-on frequency configuration. We use randomized benchmarking to characterize the error of simultaneous 20-ns $X_{\pi/2}$ gates for the optimized configuration using CTS pulse shaping and for the AB configuration using cosine pulses representing baseline performance. For the AB configuration with cosine pulses, a considerable fraction of the 46 qubits suffer from significant microwave crosstalk errors as previously discussed in Sec.~\ref{sec: CT config results}. As shown in Fig.~\ref{fig: hd drag on bw minimized config}(c), this leads to a mean gate error of $\bar{\varepsilon}_\mathrm{sim} = 8.3 \times 10^{-4}$ and  $\varepsilon_\mathrm{sim} > 10^{-3}$ for a dozen of qubits.   
In the CTS-on configuration, we manage to mitigate the tail of the error distribution caused by crosstalk, bringing the mean simultaneous gate error to $\bar{\varepsilon}_\mathrm{sim} = 4.7 \times 10^{-4}$. For clarity, the crosstalk error reduction arises partly from frequency optimization and partly from CTS pulse shaping.  Importantly, the mean simultaneous gate error approaches the mean individual gate error $\bar{\varepsilon}_\mathrm{ind} = 4.4 \times 10^{-4}$ despite the narrow frequency bandwidth below \SI{200}{\mega\hertz}.   In \cref{ap: stability of error with off-res HD DRAG }, we further demonstrate that CTS pulse shaping also provides stable gate performance over long time periods by performing repeated RB experiments for a time span of 30 hours without re-calibration.

Additionally, we show that CTS pulse shaping is necessary to achieve fast high-fidelity simultaneous gates in the optimized frequency configuration. To this end, we  measure the mean simultaneous gate error across the high-crosstalk qubits in group H as a function of gate duration $t_\mathrm{g}$  using leakage RB experiments. 
We carry out the experiments in the CTS-on configuration with or without CTS pulse shaping as shown in Fig.~\ref{fig: hd drag on bw minimized config}(d). For CTS off, the mean simultaneous gate error rapidly increases below a gate duration of $t_\mathrm{g} \lesssim \SI{30}{\nano\second}$due to increased energy spectral density at crosstalk-activated transitions. Using CTS, the mean simultaneous gate error of group H is reduced  below $5\times 10^{-4}$ down to $t_\mathrm{g}=\SI{16}{\nano\second}$. For very fast gates with $t_\mathrm{g} \lesssim \SI{16}{\nano\second}$, the gate error begins to increase on the control qubits with CTS pulses due to elevated leakage errors, see \cref{ap: error vs duration using off-res HD DRAG } for further details. 
When averaged across all 46 qubits, the gate duration of $t_\mathrm{g} = \SI{20}{\nano\second}$ used in Fig.~\ref{fig: hd drag on bw minimized config}(c) provides a good balance of low simultaneous gate error and short gate duration for our device with $|\alpha|/(2\pi) \sim \SI{180}{\mega\hertz}$ if we use CTS to suppress all the relevant crosstalk and leakage transitions.

\subsection{Scalability of the qubit frequency optimization} 
\label{sec: scalability of the qubit frequency optimization}

To investigate the scalability of the optimization strategy on larger devices and to confirm the general advantage of CTS pulse shaping, we simulate the required qubit frequency bandwidth for QPUs of various  sizes. In the simulations, we generate 10 crosstalk matrices for each selected QPU size ranging from 5 qubits to 1000 qubits using bootstrapping with replacement from a similar distribution of crosstalk 
as on the current device, see Appendix~\ref{ap: bw_simulations} for more details.  For each generated crosstalk matrix, we carry out the  qubit frequency optimization 
to estimate the resulting qubit frequency bandwidth either using CTS for qubit pairs with $C_\mathrm{TC} > -\SI{30}{\decibel}$ (CTS on) or using conventional cosine pulses on all qubits (CTS off). 
Similarly to the above experiments, we require the predicted crosstalk error to satisfy $\varepsilon_\mathrm{sim} < 3 \times 10^{-4}$ for each qubit in the optimized configuration. 

To illustrate the benefit of using CTS, we extract the average bandwidth per chip size by taking the mean of the obtained bandwidths for the 10 simulated matrices. As shown in Fig.~\ref{fig: hd drag on bw minimized config}(e), CTS pulse shaping 
reduces the mean bandwidth required for the given crosstalk error threshold by $119\pm\SI{4}{\mega\hertz}$ on average for QPU sizes of 54 qubits and greater. By adopting CTS pulse shaping, we can, for example, increase the simulated QPU size from 54 qubits to 300 qubits while retaining a similar qubit frequency bandwidth as for 54 qubits using conventional pulses.  The simulated  bandwidth using CTS also agrees within error bars  with the experimentally achieved bandwidth using 46 qubits.

In the simulations, the required qubit frequency bandwidth grows only logarithmically with the qubit count for both pulse shaping strategies. As a result, a total bandwidth of approximately \SI{500}{\mega\hertz} seems sufficient to bound the crosstalk error of simultaneous single-qubit gates below $3\times 10^{-4}$ on a 1000-qubit QPU assuming CTS pulse shaping and similar crosstalk statistics as on the current device. This suggests that for current error levels, crosstalk mitigation using the proposed techniques is feasible even on large-scale devices with 1000 qubits and beyond.   

\subsection{Detailed experimental error budget of simultaneous single-qubit gates}
\label{sec: simultaneous 1qb gate error budget}

To gain insight into the remaining error mechanisms, we estimate an experimental error budget for simultaneous 20-ns $X_{\pi/2}$ gates in the optimized configuration using CTS pulse shaping. Our error budget disentangles errors within the computational subspace into incoherent errors $\varepsilon_\mathrm{ind, incoh}$, coherent errors of individual gates $\varepsilon_\mathrm{ind, coh}$, and crosstalk-induced errors $\varepsilon_\mathrm{ct, coh}$. Furthermore, we characterize leakage errors owing to individual gates $L_\mathrm{ind}$ and crosstalk $L_\mathrm{ct}$, both of which are found to be low as required for quantum error correction at scale~\cite{miao2023overcoming, he2025experimental}. To disentangle these error sources, we perform leakage-aware purity RB experiments as well as repeated simultaneous and individual RB experiments to average over temporal coherence fluctuations as detailed in \cref{ap: error budget}. To better account for leakage errors in the estimation of coherent and incoherent errors, we extend the analysis of purity RB~\cite{wallman2015estimating, feng2016estimating} by modeling the impact of leakage on the decay rate of purity, see  Appendix~\ref{ap: benchmarking of single-qubit gate errors} for details.

In Fig.~\ref{fig: hd drag on bw minimized config}(g), we show the resulting error budget averaged separately across the 11 qubits in group H and  the 35 other qubits. In the optimized configuration with CTS pulse shaping, the mean simultaneous gate error is dominated by incoherent errors in the range of $\varepsilon_\mathrm{ind, incoh}\sim (3-4) \times  10^{-4}$ for both qubit groups. The  other error contributions sum up to roughly $1\times 10^{-4}$ since crosstalk-induced gate error and leakage in group H are reduced by roughly a factor of 5 using CTS pulse shaping. 
This also suggests that the proposed techniques could enable fast simultaneous gates with fidelities approaching $99.99\%$ if  relaxation and coherence times  are improved. 

In our measurement, the sum of average individual and crosstalk-induced leakage errors is well below $3\times 10^{-5}$, taking up only 3-4\% of the total error budget. This indicates very low leakage levels comparable to prior works on individual single-qubit gates~\cite{chen2016measuring, hyyppa2024reducing, gao2025ultra}.
Further reduction of the gate error could be achieved by mitigating the remaining coherent errors of individual gates  $\bar{\varepsilon}_\mathrm{ind, coh} \approx 5  \times 10^{-5}$, which we suspect to be dominated by uncorrected microwave pulse distortions and reflections~\cite{hyyppa2024reducing, guo2024correction}.

 \section{Discussion}
 
Here, we discuss some key requirements for the wider adoption of CTS pulse shaping on large QPUs.
 CTS pulse shaping performs well on QPUs with certain high-crosstalk pairs but low baseline crosstalk.
 As demonstrated in our experiments and simulations, CTS can be readily applied to control qubits that have high crosstalk to a single target qubit.
Furthermore, in the scenario where the drive line of the control qubit has high crosstalk to 2--3 qubits, an optimal frequency configuration may still be found, in which all the high-crosstalk qubits are parked near the spectral minima of the CTS pulse. 
However, for devices with a larger fraction of high crosstalk pairs, CTS pulse shaping may reduce the crosstalk error on  chosen target qubits but increase the error on other qubits.   
The strong decay of crosstalk with distance of \SI{-6.4}{\decibel}/\SI{}{\milli\meter} measured in this work helps in this regard to limit the number of high crosstalk elements per control line, and it would be important to maintain or improve upon this when scaling to larger devices.

An important advantage of the CTS technique is that it requires only a small number of pulse parameters. 
The HD DRAG pulse is parametrized in terms of  suppressed frequencies that can be directly assigned based on qubit transition frequencies. Additionally, there exists a relatively broad range of values for the off-resonant drive detuning $\Delta f_\mathrm{d}^\mathrm{C}$ that provides a significant crosstalk error reduction while supporting robust calibration. This enables efficient model-based optimization of the qubit frequencies, because the pulse shapes depend deterministically on the frequency configuration and can thus be quickly updated during the optimization loop. 
 Furthermore, the pulse shaping techniques introduced in this work are complementary to active crosstalk cancellation \cite{nuerbolati2022canceling}, which offers interesting possibilities for combining these approaches for scalable crosstalk error mitigation on devices with high crosstalk.

The analytical error model in Eq.~\eqref{eq:mt_eps_av random phase} and the experimental results in \cref{fig_main: error from ct}(c) show that leakage errors can become the dominant error mechanism of simultaneous single-qubit gates in the presence of microwave crosstalk. 
To enable deep algorithms, or quantum error correction, leakage errors need to be carefully mitigated~\cite{mcewen2021removing, he2025experimental}.  In~\cref{sec: demonstration of bW minimized configuration}, we demonstrated a mean simultaneous leakage error of $\bar{L}_\mathrm{sim} =2 \times 10^{-5}$ on 46 qubits, while retaining fast gates with $t_\mathrm{g}=\SI{20}{\nano\second}$ and a low spread in qubit frequencies of \SI{197}{\mega\hertz}. 
Additionally, the pulse calibration used to minimize the error of fully simultaneous gates can also be applied in cases where one needs partially simultaneous gates, such as syndrome measurements, without requiring recalibration. 
These observations highlight that 
a combination of qubit frequency optimization and CTS pulse shaping holds potential to support quantum error correction at scale.

\section{Conclusions}

In this work, we introduced techniques to model and mitigate crosstalk errors in simultaneous single-qubit gates on large-scale superconducting quantum processors. By combining model-based optimization of qubit frequencies and pulse shaping for crosstalk transition suppression,  
we experimentally demonstrated fast, simultaneous single-qubit gates on 46 qubits in a narrow-band configuration with errors approaching individually operated gates. This constitutes an experimental demonstration of pulse shaping techniques for crosstalk mitigation on a large superconducting QPU, advancing beyond previous demonstrations on small-scale systems with 2-3 qubits~\cite{vesterinen2014mitigating, berger2024dimensionality, wang2025suppressing, matsuda2025selective}.
The proposed approach was further shown to effectively mitigate crosstalk-induced single-qubit gate error in simulations with up to at least 1000 qubits, assuming the same spatial scaling and statistical distribution of crosstalk as on the current device.  
Looking forward, the optimization method is also compatible with a holistic optimization of all other required elementary operations, such as readout and two-qubit gates~\cite{klimov2024optimizing, marxer2025999}.
Overall, the demonstrated techniques provide an important step towards scaling up to large-scale quantum processors with high-fidelity simultaneous operations.

\section{Acknowledgement}
This work was supported by Business Finland through project CfoQ (787/31/2025). Parts of this work are included in patent  applications filed by IQM Finland Oy. We are grateful to all technical teams at IQM Quantum Computers for their support in enabling this work.

\section{Author contributions}
E.H. and J.W. performed the experiments.
J.W., E.H., and A.V. conceived the experiment.
J.An and J.T. created the analytical error model. J.An, A.H., M.Pap., and J.T. provided theory support.
E.H., J.W., and J.An. performed the data analysis. 
J.W., E.H., J.An., C.F.C., A.Ge., and A.V., wrote the manuscript with input from all coauthors.
S.D., S-G.K., A.K., M.L., T.L., K.M., A.P., S.S., V.S., and W.L. contributed to device fabrication and packaging. 
 A.Gu., K.J., A.La., and C.O.K. contributed to device design. 
K.H., R.K., O.K., J.K., A.Li., L.O., S.P, D.V., P.V., and M.Sar. contributed to the measurement hardware and system operation.
S.J, M.Par., I.S., I.T., F.T.,  and  C.F.C.  contributed to initial device calibration.
E.H., J.W., J.Ad., R.B., V.B.,  S.D.F., A.G.F., Z.G., D.G., T.H.,  S.I., J.I., S.J., J.L., N.L.,  P.L., F.M., J.M.,  M.Par.,  M.R., J.R., I.S., M.Sav., I.T.,  B.T., F.T., J.V., N.W.,  C.F.C., and A.V. contributed to the calibration software. 
J.He., H-S.K.,  F.D., and  J.Ha. enabled the project. 
C.F.C, A.Ge., and A.V. supervised  the project.

\bibliography{main}
\pagebreak
\clearpage

\onecolumngrid
\title{Supplementary information to: Mitigating crosstalk errors for simultaneous single-qubit gates on a superconducting quantum processor}

\maketitle

\clearpage

\appendix
\newcolumntype{C}{>{\centering\arraybackslash} m{1.4cm} }

\renewcommand{\thefigure}{S\arabic{figure}}
\renewcommand{\theHfigure}{S\arabic{figure}}
\renewcommand{\bibnumfmt}[1]{[S#1]}
\newcommand{\note}[1]
  {\begingroup{\color{blue}[NOTE: \textit{#1}]}\endgroup}

\setcounter{equation}{0}
\setcounter{figure}{0}
\setcounter{table}{0}
\setcounter{page}{1}
\setcounter{section}{0}

\section{Quantum processing unit}
\label{ap: qpu info and statistics}

The QPU consists of two chip tiers. The top tier contains the transmon qubits, tunable couplers, readout resonators and drive lines. The bottom tier is used for routing and includes the flux lines, readout lines, through-silicon vias (TSVs).
The two tiers are integrated using indium bump flip-chip bonding.

\begin{figure}
    \centering
    \includegraphics{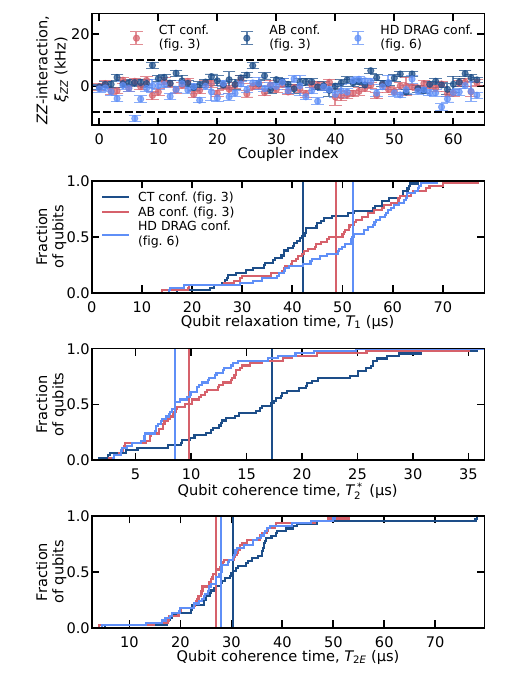}
    \caption{\textbf{Quality metrics for the configurations discussed in main text. } From top to bottom, measurements of the residual $ZZ$ interaction $\xi$, $T_1$, $T_2^*$ and $T_2^E$ for the three  configurations discussed in the main text. Vertical lines indicate median values (see text). Horizontal dashed lines in the upper panel indicate \SI{10}{\kilo\hertz} residual $ZZ$ interaction. 
    }
    \label{fig: qpu_statistics}
\end{figure}

An overview of qubit and coupler parameters of the device is given in~\cref{tab:design_params}. Device statistics are shown in~\cref{fig: qpu_statistics}, measured on 46 qubits in the three main configurations discussed in the main text. To confirm that the optimized configuration did not induce significant $ZZ$ interactions during idling of the qubits, we measured the $ZZ$ interaction strength using a conditional frequency Ramsey experiment also known as joint amplification of ZZ (JAZZ) \cite{ku2020suppression, shirai2023all} that  applies an echo $\pi$-pulse to the target qubit to increase the signal to noise ratio. In all three configurations, the residual $ZZ$ interaction $\xi_{ZZ}$, remained below \SI{10}{\kilo\hertz} for nearly all qubits as shown in the upper panel of~\cref{fig: qpu_statistics}.
Additionally, we measured $T_1$, $T_2^*$, obtained from a Ramsey experiment, and $T_2^E$ obtained from a standard Hahn echo experiment. We found median values $T_1 = \{\SI{42.0}{\micro\second}, \SI{48.7}{\micro\second}, \SI{52.2}{\micro\second}\}$, $T_2^*=\{\SI{17.2}{\micro\second}, \SI{9.8}{\micro\second}, \SI{8.6}{\micro\second}\}$ and $T_2^E=\{\SI{30.3}{\micro\second}, \SI{27.0}{\micro\second}, \SI{28.0}{\micro\second}\}$ for the crosstalk optimized configuration of~\cref{fig: crosstalk-optimized frequency configuration}, the AB configuration of ~\cref{fig: crosstalk-optimized frequency configuration} and the optimized configuration using CTS pulse shaping of~\cref{fig: hd drag on bw minimized config}, respectively. In this data, we excluded one qubit from the reported medians due to a fitting error.  
The $T_2^*$ was higher for the crosstalk optimized configuration because there the target frequency for the optimization was set to the maximum qubit frequency, reducing the sensitivity to flux noise.

\begin{table}[ht]
\begin{tabular}{l|r}
\hline
\textbf{Parameter}  & \textbf{Values} \\
 Readout freq. &$f_\mathrm{ro}\in$ [\SI{5.0}{\giga\hertz}, \SI{5.98}{\giga\hertz}] \\
Maximum qubit freq.  &  $f^q_\mathrm{max}\in $ [\SI{4.15}{\giga\hertz}, \SI{4.56}{\giga\hertz}] \\
Qubit anharmonicities  &  $|\alpha|/(2\pi)\in $ [\SI{170}{\mega\hertz}, \SI{183}{\mega\hertz}]\\ 
Target max. coupler freq. &  $f^c_\mathrm{max} =\SI{6.5}{\giga\hertz}$  \\
Target coupler idling freq.   &  $f^c_\mathrm{\xi=0}=\SI{6.3}{\giga\hertz}$\\
Target qubit-coupler coupling &  $g_{qc}/(2\pi)=\SI{63}{\mega\hertz}$ \\
\end{tabular}
\caption{\label{tab:design_params} \textbf{Summary of component parameters}. Shown ranges represent the spread as measured on the device. Individual values for the couplers represent targeted design parameters for all couplers. 
}
\end{table}

\section{Crosstalk matrix and measurement methods.}
\label{ap: crosstalk matric measurement}
\begin{figure}
    \centering
    \includegraphics{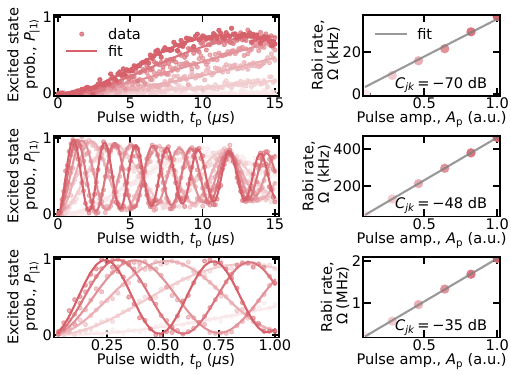}
    \caption{\textbf{Measured cross-Rabi rate data for varying crosstalk strength.} From top to bottom are shown measurements of the cross-Rabi rate for $\mathrm{Q}_{37}$ and control lines that have \SI{-70}{\decibel}, \SI{-48}{\decibel} and \SI{-35}{\decibel} crosstalk, respectively. Measurements consist of a square pulse with varying amplitude and width before readout is performed. The measured excited state probability data (red shaded dots) is shown with a fit to~\cref{eq:ap_cross_rabi_rate_fit} (similarly colored lines) in the left panels for various amplitudes. Right panels indicate the extracted Rabi rate from the fits for each pulse amplitude (same shade red dots as right panel). A linear fit (grey line) is used to reliably estimate the Rabi rate at unit pulse amplitude.  }
    \label{fig: crosstalk individual traces}
\end{figure}

The pair-wise crosstalk $C_{kj}$ as defined in~\cref{eq: crosstalk definition}, is characterized using a combination of diagonal- and cross-Rabi rate measurements. The Rabi rate is extracted by varying both the time and amplitude of a square $\pi$-pulse~\cite{spring2022high}, see ~\cref{fig: crosstalk individual traces}. The rates are measured in the AB configuration to reduce the impact of hybridization crosstalk on the extracted cross-Rabi rates. 
For each pulse amplitude, we fit a decaying sinusoidal function to the resulting Rabi oscillation of the excited state population $P_{\ket{1}}$ given by
\begin{equation}
     P_{\ket{1}}(t) = -\frac{1}{2} \cos\left(\Omega t\right)e^{(-t / T_\mathrm{exp}) - (t / T_\mathrm{gauss})^2} + \frac{1}{2}
     \label{eq:ap_cross_rabi_rate_fit}
\end{equation}
where $\Omega$ is the extracted Rabi rate during the square pulse, and the exponential decay time-constant $T_\mathrm{exp}$ and Gaussian decay time-constant $T_\mathrm{gauss}$ are left as free fit parameters to allow for a robust extraction of $\Omega$ for longer pulses where incoherent processes start affecting the populations. 
We then perform a linear fit to the extracted Rabi rates for the various amplitudes $A_p$ and extract the slope such that we can extract the Rabi rate for a given qubit--control line pair $j,k$ at unit pulse amplitude $\Omega_{jk} = \Omega_{|A_p=1}$. 

\begin{figure*}
    \centering
    \includegraphics{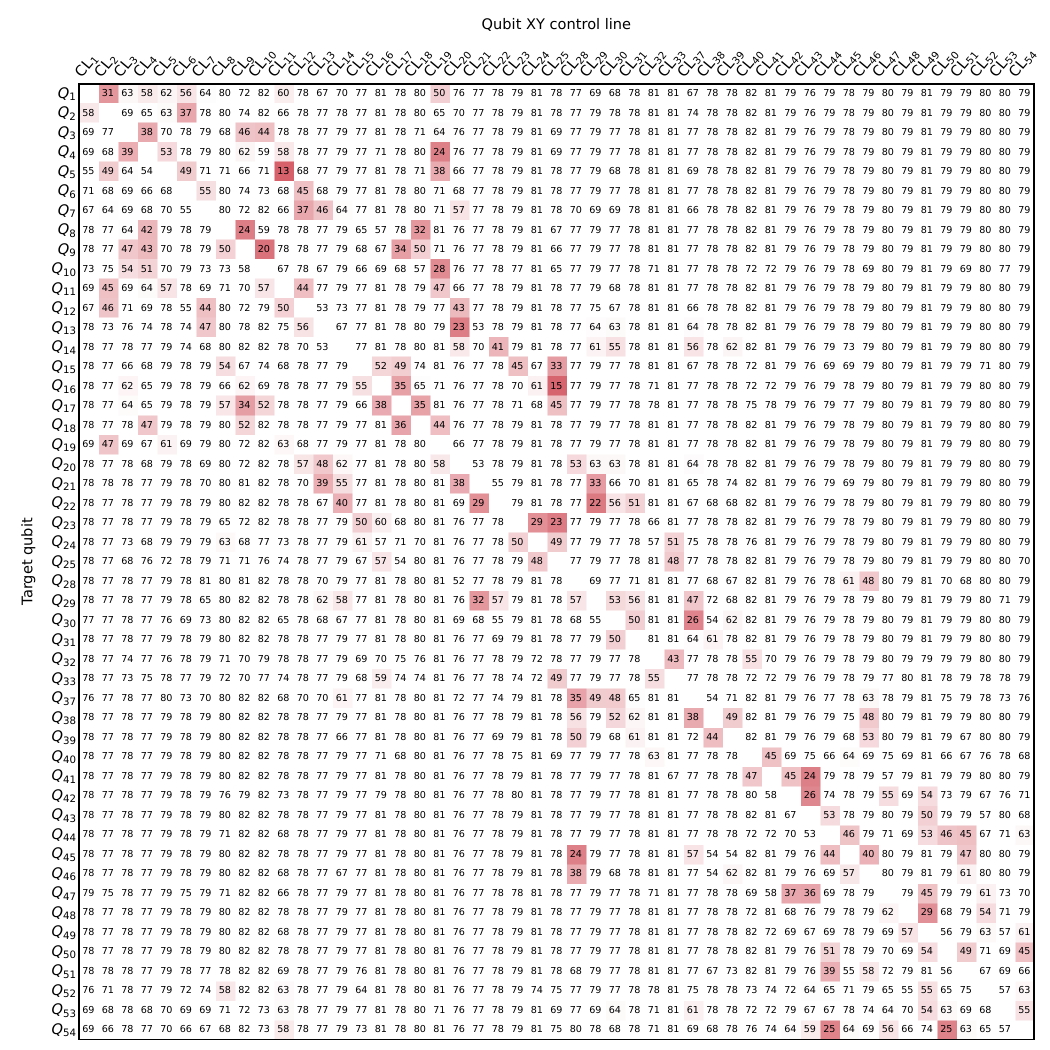}
    \caption{\textbf{Full crosstalk matrix.} All measured crosstalk matrix elements $C_{kj}$  that underlie the histogram shown in~\cref{fig_main: device and crosstalk}(c). The matrix elements are based on ~\cref{eq: crosstalk definition} and expressed in dB. For visual clarity, the minus sign is omitted in the elements. High crosstalk elements are highlighted with varying shades of red. For each control line, there are typically less than two high-crosstalk elements. This facilitates crosstalk-induced error suppression using pulse shaping since the required number of suppressed frequencies in the pulse spectrum remains low. As qubits are labeled sequentially across the device, the concentration of high-crosstalk elements near the diagonal indicates the local nature of the crosstalk on the device as illustrated in~\cref{fig_main: error from ct}(e). Elements, for which crosstalk could be measured directly using pulse widths of up to \SI{15}{\micro\second}, are displayed in black, while gray elements represent upper bounds estimated using the methods described in~\cref{ap: crosstalk matric measurement}.}.
    \label{fig: crosstalk matrix}
\end{figure*}

As the diagonal Rabi and cross-Rabi rates can vary over several orders of magnitude, i.e. from tens of kHz to hundreds of MHz, the stability of the fits depends strongly on the chosen ranges of pulse width, amplitude and number of points. 
We therefore measured the data with varying ranges in pulse widths and amplitudes, while also implementing some safeguards against failed fits.  If the excited state probability for a given range of pulse widths and amplitude is $P_{\ket{1}}<0.5$, corresponding to a $\pi/2$ rotation in absence of decoherence, the oscillation rate is deemed too small to be measured with those ranges and we estimate an upper bound on the cross-Rabi rate
$\Omega_\mathrm{lim}:=\arccos\left(1 - 2P_{\ket{1},\mathrm{threshold}}\right) / (2\pi t_\mathrm{max})
$. If this is not the case, we proceed to the linear fitting of extracted rates and implement outlier rejection to mitigate cases where the highest amplitude is too high for the given rate, causing undersampling of the oscillation, or the lowest $A_p$ are too low, causing high fit uncertainty in the extracted rates. Finally, if the r-squared values of the fit are below a set threshold, indicating there is significant deviation from a linear fit, the extraction is deemed failed for that range of amplitudes and times and remeasured with a different range. For example, in cases of neighbouring qubits that are close in frequency, the extracted crosstalk can be dominated by wavefunction hybridization and scale non-linear with the pulse amplitude~\cite{spring2022high}.  The full matrix with all extracted crosstalk elements is shown in~\cref{fig: crosstalk matrix}.

\subsection{Frequency dependence of the crosstalk magnitude.}
 \label{ap: frequency dependence of crosstalk}
\begin{figure}
    \centering
    \includegraphics{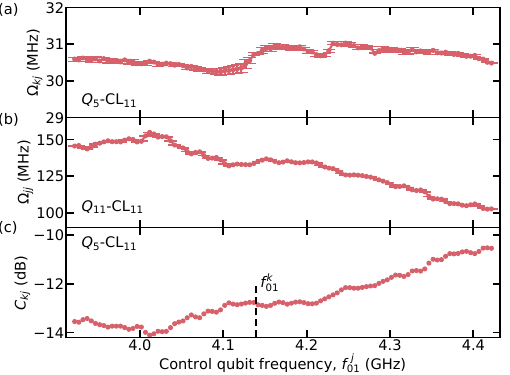}
    \caption{\textbf{Variation in crosstalk due to a frequency-dependent transfer function of the control lines. } 
    (a) Cross-Rabi rate versus control qubit ($\mathrm{Q}_{11}$) frequency for the highest-crosstalk pair $\mathrm{Q}_5$--$\mathrm{Q}_{11}$ of the device. (b) Corresponding diagonal Rabi rate of $\mathrm{Q}_{11}$. (c) Resulting crosstalk of qubit pair $\mathrm{Q}_5$--$\mathrm{Q}_{11}$, demonstrating that crosstalk may vary by about \SI{3}{\decibel} over the considered frequency ranges due to the transfer function of the drive line. Black dashed line indicates $f_{01}$ of the target qubit ($\mathrm{Q}_5$). Horizontal bars for points in upper panels indicate the fit uncertainty of the Rabi rate extraction.}
    \label{fig: crosstalk versus detuning}
\end{figure}

Throughout this work we have neglected the frequency dependence of crosstalk, which may arise due to a frequency dependent transfer function of the control lines. In~\cref{fig: crosstalk versus detuning}, we quantify the violation of this assumption for the highest crosstalk pair $\mathrm{Q}_{5}$--$\mathrm{Q}_{11}$ considered in Figs.~\ref{fig_main: error from ct}(c,d) and \ref{fig: experimental demonstration of off-resonant HD DRAG}. 
We sweep the frequency $f_{01}^{j}|_{j=\mathrm{Q}_{11}}$ of the control qubit ($\mathrm{Q}_{11}$), while keeping the frequency of the target qubit ($\mathrm{Q}_5$) fixed and measure both the diagonal Rabi rate $\Omega_{jj}(f_{01}^j)_{|j=\mathrm{Q}_{11}}$ and the cross-Rabi rate from the control line of $\mathrm{Q}_{11}$ to $\mathrm{Q}_5$. 
The diagonal Rabi rate $\Omega_{jj}(f_{01}^j)_{|j=\mathrm{Q}_{11}}$ varies from $\SI{100}{\mega\hertz}$ to $\SI{150}{\mega\hertz}$ due to a frequency dependence of the line transfer function and electronics output, while the cross-Rabi rate (measured at the fixed frequency of $\mathrm{Q}_5$) remains approximately fixed at \SI{30.5}{\mega\hertz}. The variation in Rabi rate results in roughly \SI{3}{\decibel} variation in the measured magnitude of the crosstalk over a range of roughly $\SI{0.5}{\giga\hertz}$. Thus, we estimate that neglecting the frequency dependence of the crosstalk matrix in the model-based optimization can lead to a misestimate of crosstalk-induced error by roughly 40\%. In order to obtain a good match between theory and data in~\cref{fig_main: error from ct}(b,c) and ~\cref{fig: experimental demonstration of off-resonant HD DRAG}, which were taken at different frequency configurations, we re-measured the crosstalk in each configuration.

\section{Estimate of nearest neighbour hybridization error.}\label{ap: hybridization_error}
\begin{figure}
    \centering
    \includegraphics{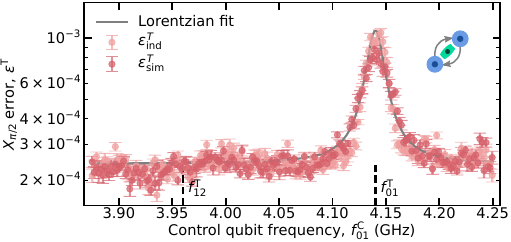}
    \caption{\textbf{Nearest neighbour hybridization crosstalk error versus qubit-qubit detuning.} 
    Simultaneous and individual randomized benchmarking error on a target qubit ($\mathrm{Q}_{15}$) while sweeping a neighbouring control qubit frequency ($\mathrm{Q}_{23}$). Solid line indicates an empirical fit using a Lorentzian function of~\cref{eq:ap_lorentzian}, yielding $A=25\times10^3$ and $\gamma=9.4\times 10^6~\SI{}{\hertz}$. The classical crosstalk for the pair was measured to be below \SI{45}{\decibel} in both directions in the AB configuration, so the error is dominated by hybridization error due to wavefunction overlap. For each data point, we calibrated standard single-qubit gate parameters for both qubits after changing the control qubit frequency followed by RB. The coupler frequency was fixed at the calibrated $\xi_{ZZ}=0$ point for the control qubit at \SI{4.25}{\giga\hertz}. The error on the control qubit (not shown) was of similar strength due to the symmetric nature of hybridization errror. }
    \label{fig: hybridization crosstalk versus detuning}
\end{figure}

As part of calibrating a certain frequency configuration, we set the tunable couplers to a frequency that eliminates the residual longitudinal $ZZ$ interaction between the connected pair of qubits. However, this frequency does not necessarily cancel the transverse interaction and a residual wavefunction hybridization between the qubits may remain~\cite{yan2018tunable,valles-sanclemente2025optimizinga}. This hybridization causes two types of errors during single-qubit gates. The first  applies to simultaneous gates, creating an effective symmetric crosstalk that scales non-linearly with the amplitude of the drive~\cite{spring2022high} and which can be estimated using the methods described in~\cref{ap: crosstalk matric measurement}. The second is an additional contribution to the individual gate error due to part of the wavefunction residing on the neighbouring qubit. As this is a transverse interaction, both of these errors scale strongly with the detuning between the qubit $f_{01}$ transition frequencies.

In~\cref{fig: hybridization crosstalk versus detuning}, we illustrate an example of hybridization error for a pair of qubits with negligible  classical microwave crosstalk. There, we measure the error of a $X_{\pi/2}$ gate on the target qubit  when sweeping the control qubit frequency using both simultaneous and individual RB. We observe a peak in both individual and simultaneous gate fidelity of the target qubit, a key signature of hybridization error distinguishing it from classical crosstalk-induced error.
An empirical Lorentzian function was fitted to the measured error of the form
\begin{equation}\label{eq:ap_lorentzian}
    \varepsilon=\frac{A}{\pi}  \frac{\gamma}{(f - f_0)^2 + \gamma^2}+\varepsilon_0,
\end{equation}
where $\gamma, A, \epsilon_0$ are free fit parameters. These fitted values were used as estimates for the error in the crosstalk-aware frequency optimization.  

Additionally,  we generally expect increased hybridization error for a collision between neighbouring $f_{01}\leftrightarrow f_{12}$ frequencies due to residual non-zero coupling between $\ket{11}\leftrightarrow\ket{02}$ states at the coupler frequency corresponding to $\xi_{ZZ}=0$. For this work, we found that we could already reach a frequency configuration with sufficient gate fidelity by avoiding only  $f_{01}\leftrightarrow f_{01}$ collisions without explicitly considering $f_{01}\leftrightarrow f_{12}$ collisions. This is also visible in the data shown in~\cref{fig: hybridization crosstalk versus detuning}, where for the calibrated coupler frequency, 
the $f_{01}\leftrightarrow f_{12}$ collision did not show an increase in error above the coherence limited error of roughly $ 2\times10^{-4}$.

\section{Analytical error model for crosstalk-induced single-qubit gate error}
\label{ap: analytical error model}

Here, we describe the analytical model of~\cref{sec: experimental relation between crosstalk and 1qb gate error} in more detail. We derive the single-qubit gate error of the target qubit caused by the crosstalk from the drive line of the control qubit during simultaneous single-qubit gates, and obtain~\cref{eq:mt_eps_av random phase}. We model the target qubit as a driven transmon qubit with the Hamiltonian
\begin{align}\label{target_hamiltonian suppl}
    \hat{H}_{\rm T} &= \hbar \omega_{01}^\mathrm{T}\hat{a}^\dagger \hat{a} + \frac{\hbar \alpha^{\rm T}}{2}\hat{a}^\dagger \hat{a}^\dagger \hat{a} \hat{a} \nonumber \\
    &+ \ii \hbar [s_I^{\rm T}(t) \cos(\omega_{\rm d}^\mathrm{T}t + \phi^{\rm T}) \nonumber \\ 
    &+ s_Q^{\rm T}(t)\sin(\omega_{\rm d}^\mathrm{T}t + \phi^{\rm T})](\hat{a}^\dagger - \hat{a}),
\end{align}
where $\omega_{\rm T}$ is the angular frequency of target qubit, $\alpha_{\rm T}$ is the corresponding anharmonicity,  $s_I^\mathrm{T}$ and $s_Q^\mathrm{T} $ are the in-phase and quadrature envelope functions of the drive pulse applied to the target qubit, $\omega_{\rm d}^\mathrm{T}$ is the angular drive frequency of the target qubit, and $\phi^\mathrm{T}$ is the drive phase. Drive crosstalk coming from a simultaneous drive of the control qubit is given by the Hamiltonian
\begin{align}\label{control drive term suppl}
    \hat{H}_{\rm CT} &= \ii \lambda \hbar [s_I^\mathrm{C}(t) \cos(\omega_\mathrm{d}^\mathrm{C}t + \phi^{\rm C})\nonumber \\  &+ s_Q^\mathrm{C}(t)\sin(\omega_\mathrm{d}^\mathrm{C}t + \phi^{\rm C})](\hat{a}^\dagger - \hat{a}),
\end{align}
where $\lambda$ is a dimensionless crosstalk parameter related to the magnitude of drive crosstalk  as $\lambda^2 =C_\mathrm{TC}$, $\omega_\mathrm{d}^\mathrm{C}$ is the angular drive frequency of the control qubit, $\phi_{\rm C}$ is the drive phase of the control qubit, and $s_I^\mathrm{C}(t)$ and $s_Q^\mathrm{C}(t)$ are the in-phase and quadrature envelope functions of the drive pulse applied to the control qubit.

Let us make a transformation to the rotating frame of the bare target qubit with
\begin{align}
    \hat{U}(t) = \text{exp}(-\ii (\omega_{01}^{\rm T}\hat{a}^\dagger \hat{a} + \frac{ \alpha^{\rm T}}{2}\hat{a}^\dagger \hat{a}^\dagger \hat{a} \hat{a})t).
\end{align}
We make the rotating wave approximation to the drive of the target qubit and truncate the Hamiltonian subspace to the three lowest eigenstates. The target qubit Hamiltonian is then given by
\begin{align}
    \hat{H}_{\rm T} \hat{\approx} \frac{\ii \hbar}{2}\begin{pmatrix}
        0 & -s^\mathrm{T}(t)e^{\ii \phi^\mathrm{T}} & 0 \\
        s^{\mathrm{T}}(t)^*e^{-\ii \phi^\mathrm{T}} & 0 & -\sqrt{2}s^\mathrm{T}(t)e^{\ii \gamma^{\rm T}} \\
        0 & \sqrt{2}s^{\mathrm{T}}(t)^*e^{-\ii \gamma^{\rm T}} & 0
    \end{pmatrix},
\end{align}
 where we assume $\omega_{\rm d}^{\rm T} = \omega_{01}^{\rm T}$ and denote $\gamma^{\rm T} =\phi^{\rm T} - \alpha^{\rm T}t$ and $s^\mathrm{T} = s_I^\mathrm{T}(t) - \ii s_Q^\mathrm{T}(t)$.
In this frame, the crosstalk term is written as
\begin{align}
    \hat{H}_{\rm CT}
    \hat{\approx} &\frac{\ii \lambda \hbar}{2}  
    \begin{pmatrix}
       0 & - s^\mathrm{C}(t)e^{\ii\delta } & 0\\
       s^\mathrm{C}(t)^*e^{-\ii\delta } & 0 & - \sqrt{2}s^\mathrm{C}(t)e^{\ii\gamma^{\rm C} } \\
       0 & \sqrt{2}s^\mathrm{C}(t)^*e^{-\ii\gamma^{\rm C}} &0
    \end{pmatrix},
\end{align}
where $\gamma^{\rm C} = \phi^{\rm C} - (\Delta+\alpha^{\rm T})t$, $\delta = \phi^{\rm C}- \Delta t$, $s^\mathrm{C}(t) = s_I^\mathrm{C}(t) - \ii s_Q^\mathrm{C}(t)$ and $\Delta = \omega_{01}^{\rm T} - \omega_{\rm d}^{\rm C}$ is the detuning between the target qubit and the drive of the control qubit.

We are only interested in the error introduced by the drive of the control qubit so we assume that the drive of the target qubit does not cause any leakage to the $f$-state during the gate. The time-evolution of the target qubit is then given by the ideal unitary

\begin{align}
    \hat{U}_{\rm ideal} \hat = \begin{pmatrix}
        \cos\frac{\theta(t)}{2} & -e^{\ii \phi^{\rm T}}\sin\frac{\theta(t)}{2} & 0 \\
        e^{-\ii \phi^{\rm T}}\sin\frac{\theta(t)}{2} & \cos\frac{\theta(t)}{2} & 0 \\
        0 & 0 & 1
    \end{pmatrix},
\end{align}
where the rotation angle $\theta(t)$ of the target qubit is given by an integral of its in-phase envelope function, i.e, $\theta(t) = \int_0^t \text{d}t's_I^\mathrm{T}(t')$ at any time $0 \leq t \leq t_\mathrm{g}$, where $t_\mathrm{g}$ is the gate duration.

We make a transformation to the interaction picture with $\hat{U}_{\rm ideal}(t)$ and write the von Neumann equation as 
\begin{align}
    \dt{\check{\rho}} = -\frac{\ii}{\hbar}[\check{H}_{\rm CT}, \check{\rho}],
\end{align}
where the transformed operator $\hat O$ is denoted with $\check{O} = \hat{U}^\dagger_{\rm ideal}(t)\hat{O}\hat{U}_{\rm ideal}(t)$.
We obtain the density-operator dynamics by expanding the solution of the von Neumann equation perturbatively in the crosstalk amplitude $\lambda$ ($0\le \lambda \ll 1$). Retaining terms up to second order in $\lambda$ yields
\begin{align}
    \check{\rho}(t_\mathrm{g})
    &\approx \check{\rho}(0)
    -\frac{\ii}{\hbar}\int_0^{t_\mathrm{g}} \mathrm{d}t\,\big[\check{H}_{\rm CT}(t), \check{\rho}(0)\big] \nonumber\\
    &- \frac{1}{\hbar^2}\int_0^{t_\mathrm{g}} \mathrm{d} t \int_0^t \mathrm{d} t_1\,\big[\check{H}_{\rm CT}(t), \big[\check{H}_{\rm CT}(t_1), \check{\rho}(0)\big]\big].
\end{align}
We transform the density operator back into the Schrödinger picture and obtain 
\begin{align}
    \hat{\rho}(t_\mathrm{g}) &\approx \hat{U}_{\rm ideal}(t_\mathrm{g})\check{\rho}(0)\hat{U}_{\rm ideal}^\dagger(t_\mathrm{g}) \nonumber \\
    &-\frac{\ii}{\hbar}\int_0^{t_\mathrm{g}} \text{d}t\hat{U}_{\rm ideal}(t_\mathrm{g})[\check{H}_{\rm CT}(t), \check{\rho}(0)]\hat{U}_{\rm ideal}^\dagger(t_\mathrm{g}) \nonumber \\
    &-\frac{1}{\hbar^2}\int_0^{t_\mathrm{g}} \text{d}t \int_0^t \text{d} t_1\hat{U}_{\rm ideal}(t_\mathrm{g}) \nonumber \\ &\quad[\check{H}_{\rm CT}(t),[\check{H}_{\rm CT}(t_1) \check{\rho}(0)]]\hat{U}_{\rm ideal}^\dagger(t_\mathrm{g}).
\end{align}
Given that $\hat{\rho}_{\rm ideal}(t_\mathrm{g})=\hat{U}_{\rm ideal}(t_\mathrm{g}) \hat{\rho}(0) \hat{U}_{\rm ideal}^\dagger(t_\mathrm{g})$ and $\check{\rho}(0) = \hat{\rho}(0)$, we compute the fidelity under crosstalk as 
\begin{align}
    \mathcal{F} &= \Tr{[\hat{\rho}_{\rm ideal}(t_\mathrm{g})\hat{\rho}(t_\mathrm{g})]} \nonumber \\
    &= 1 - \frac{\ii}{\hbar}\int_0^{t_\mathrm{g}} \text{d}t\Tr{[\check{H}_{\rm CT}(t), \hat{\rho}(0)]\hat{\rho}(0)} \nonumber \\
    &-\frac{1}{\hbar^2}\int_0^{t_\mathrm{g}} \text{d}t\int_0^t\text{d}t_1 \Tr{[\check{H}_{\rm CT}(t), [\check{H}_{\rm CT}(t_1), \hat{\rho}(0)]]\hat{\rho}(0)}.
\end{align}
Since the trace operation is invariant under cyclic permutations, the first order term is zero. The error due to crosstalk is then written as
\begin{align}\label{eq:eps}
\mathcal{E} &\equiv 1 - \mathcal{F} \nonumber\\
&= \frac{1}{2\hbar^2}\int_{0}^{t_\mathrm{g}}\!\mathrm{d}t \int_{0}^{t_\mathrm{g}}\!\mathrm{d}t_1\,
\Bigl[
  \langle \check{H}_{\rm CT}(t)\check{H}_{\rm CT}(t_1)\rangle \\
&
+ \langle \check{H}_{\rm CT}(t_1)\check{H}_{\rm CT}(t)\rangle 
- 2 \langle \check{H}_{\rm CT}(t)\hat{\rho}(0)\check{H}_{\rm CT}(t_1)\rangle
\Bigr], \nonumber
\end{align}
where 
\begin{align}
        \langle \check{H}_{\rm CT}(t_i)\check{H}_{\rm CT}(t_j)\rangle &= \Tr{\check{H}_{\rm CT}(t_i)\check{H}_{\rm CT}(t_j)\hat{\rho}(0)}, \nonumber \\
        \langle \check{H}_{\rm CT}\hat{\rho}(0)(t)\check{H}_{\rm CT}(t_1)\rangle &= \Tr{\check{H}_{\rm CT}(t)\hat{\rho}(0)\check{H}_{\rm CT}(t_1)\hat{\rho}(0)}
\end{align}
and we used the fact that the integrand in the above definition of the crosstalk error is symmetric with respect to the integration variables.

We assume that the target qubit can be in any pure state in the Hilbert space spanned by the two lowest states. Consequently, the initial density operators are written as
\begin{align}
    \hat{\rho}(0) = \frac{1}{2}\left[\hat{I} + \cos\vartheta\hat{\Sigma}_z+ \sin\vartheta(\cos\varphi\hat{\Sigma}_x + \sin\varphi\hat{\Sigma}_y)\right],
\end{align}
where $\hat{I}$ is the $3\times 3$ identity operator, $\vartheta \in [0, \pi]$, $\varphi \in [0, 2\pi]$, and $\Sigma_i$ for $i \in [x, y, z]$ are Pauli operators extended into three dimensions as 
\begin{align}
    \Sigma_x &\equiv \begin{pmatrix}
        0 & 1 & 0 \\
        1 & 0 & 0 \\
        0 & 0 & 0 
    \end{pmatrix}, \\
    \Sigma_y &\equiv \begin{pmatrix}
        0 & -\ii & 0 \\
        \ii & 0 & 0 \\
        0 & 0 & 0 
    \end{pmatrix}, \\
    \Sigma_z &\equiv \begin{pmatrix}
        1 & 0 & 0 \\
        0 & -1 & 0 \\
        0 & 0 & 0 
    \end{pmatrix}.
\end{align}
Since we want to compute the average fidelity, we write
\begin{align}
    \hat{\rho}_{\rm av}(0) = \frac{1}{4\pi}\int_0^\pi\text{d}\vartheta \sin\vartheta \int_0^{2\pi} \text{d}\varphi \hat{\rho}(0) = \frac{1}{2}\hat{I}.
\end{align}
Calculating the average  of~\eqref{eq:eps} over all initial density operators and simplifying using trigonometric identities then yields
\begin{equation}
\begin{aligned}
    \mathcal{E}_{\rm av} &= \frac{1}{4\hbar^2}\int_0^{t_\mathrm{g}} \text{d}t\int_0^{t_\mathrm{g}} \text{d}t_1 \nonumber \\
    &\times \Bigl[\Tr{\check{H}_{\rm CT}(t)\check{H}_{\rm CT}(t_1)} \\
    &- \frac{1}{3}\Bigl(\Tr{\check{H}_{\rm CT}(t)\Sigma_z\check{H}_{\rm CT}(t_1)\Sigma_z} \\ 
    &+ \Tr{\check{H}_{\rm CT}(t)\Sigma_x\check{H}_{\rm CT}(t_1)\Sigma_x} \\
    &+ \Tr{\check{H}_{\rm CT}(t)\Sigma_y\check{H}_{\rm CT}(t_1)\Sigma_y}\Bigr)\Bigr],
\end{aligned}
\end{equation}

We observe that we need to compute the time-integrals of the perturbation Hamiltonian $\check H_{\rm CT}(t)$. We can do this since we previously assumed that the ideal unitary $\hat{U}_{\rm ideal}$ describes the time-evolution of the target qubit at all times $0\leq t \leq t_\mathrm{g}$. We write the time-integrals as 
\begin{align}
    \bar{H}_{\rm CT} \equiv \int_0^{t_\mathrm{g}} \text{d}t \check{H}_{\rm CT}(t).
\end{align}
The average error arising from the crosstalk is then given by
\begin{equation}
\begin{aligned}
\mathcal{E}_{\rm av}
&= \frac{1}{4\hbar^2}\Bigl\{
\operatorname{Tr}\!\bigl[\bar{H}_{\rm CT}\bar{H}_{\rm CT}\bigr]
-\frac{1}{3}\bigl(
\operatorname{Tr}\!\bigl[\bar{H}_{\rm CT}\hat{\Sigma}_z\bar{H}_{\rm CT}\hat{\Sigma}_z\bigr] \\
&+\operatorname{Tr}\!\bigl[\bar{H}_{\rm CT}\hat{\Sigma}_x\bar{H}_{\rm CT}\hat{\Sigma}_x\bigr] +\operatorname{Tr}\!\bigl[\bar{H}_{\rm CT}\hat{\Sigma}_y\bar{H}_{\rm CT}\hat{\Sigma}_y\bigr]
\bigr)
\Bigr\} \\
&= \frac{\lambda^2}{12}\Bigl\{
|C[\Delta]|^2 + |C_{\rm c}[\Delta]|^2 + |C_{\rm s}[\Delta]|^2 \\
&+ 3\bigl|C_{{\rm c}\frac{1}{2}}[\Delta + \alpha^{\rm T}]\bigr|^2 
 + 3\bigl|C_{{\rm s}\frac{1}{2}}[\Delta + \alpha^{\rm T}]\bigr|^2 \\
&
+ \operatorname{Re}\!\bigl(
e^{2\ii(\Delta\phi)}\,\{C[\Delta]^2
- C_{\rm c}[\Delta]^2
- C_{\rm s}[\Delta]^2\}
\bigr)
\Bigr\},
\end{aligned}
\label{eq:eps_av intermediate}
\end{equation}
where we have defined
\begin{align} \label{eq: C definitions}
    C[\omega] &= \int_{-\infty}^\infty \text{d}te^{-\ii \omega t}s^{\rm C}(t),\nonumber \\
    C_{\rm s}[\omega] &= \int_{-\infty}^\infty \text{d}te^{-\ii \omega t}s^{\rm C}(t)\sin\theta(t),\nonumber \\
    C_{\rm c}[\omega] &= \int_{-\infty}^\infty \text{d}te^{-\ii \omega t}s^{\rm C}(t)\cos\theta(t),\\
    C_{{\rm s}\frac{1}{2}}[\omega] &= \int_{-\infty}^\infty \text{d}te^{-\ii \omega t}s^{\rm C}(t)\sin\frac{\theta(t)}{2},\nonumber \\
    C_{{\rm c}\frac{1}{2}}[\omega] &= \int_{-\infty}^\infty \text{d}te^{-\ii \omega t}s^{\rm C}(t)\cos\frac{\theta(t)}{2},\nonumber
\end{align}
and $\Delta \phi = \phi^{\rm T} - \phi^{\rm C}$. We define corresponding spectral densities as $S_i[\omega] = |C_i[\omega]|^2$ and write the final expression for the error as
\begin{equation} 
    \begin{aligned}
    \mathcal{E}_{\rm av}
    &= \frac{\lambda^2}{12}\Bigl\{
    S[\Delta] + S_{\rm c}[\Delta] + S_{\rm s}[\Delta] \\
    & + 3 S_{{\rm c}\frac{1}{2}}[\Delta + \alpha^{\rm T}]
    + 3 S_{{\rm s}\frac{1}{2}}[\Delta + \alpha^{\rm T}] \\
    &
    + \operatorname{Re}\!\bigl(
    e^{2\ii(\Delta\phi)}\,\{C[\Delta]^2
    + C_{\rm c}[\Delta]^2
    + C_{\rm s}[\Delta]^2
    \}\bigr)
    \Bigr\},
    \end{aligned}
\label{eq:eps_av}
\end{equation}
which completes the derivation of Eq.~\eqref{eq:mt_eps_av random phase}.

Assuming a randomized drive phase  of the control qubit, such as approximately in simultaneous RB, we further compute the average error over the phase difference $\Delta \phi$. Consequently, the last term in Eq.~\eqref{eq:eps_av} is zero, which yields
\begin{equation} 
    \begin{aligned}
    \bar{\mathcal{E}}_{\rm av}
    &= \frac{\lambda^2}{12}\Bigl\{
    S[\Delta] + S_{\rm c}[\Delta] + S_{\rm s}[\Delta] \\
    &+ 3 S_{{\rm c}\frac{1}{2}}[\Delta + \alpha^{\rm T}]
    + 3 S_{{\rm s}\frac{1}{2}}[\Delta + \alpha^{\rm T}] 
    \Bigr\}.
    \end{aligned}
\label{eq:eps_av random phase}
\end{equation}
The expression in Eq.~\eqref{eq:eps_av random phase} further  simplifies if the target qubit idles. In that case, the kernel functions are constant and the average gate error on the target qubit due to the pulse from the control qubit reduces to
\begin{equation} 
    \mathcal{E}_{\rm av, idling}
    = \frac{\lambda^2}{12}\Bigl\{
    2S[\Delta] + 3 S[\Delta + \alpha^{\rm T}] 
    \Bigr\}. 
    \label{eq:eps_av random phase idling}
\end{equation}

Importantly, the phase-dependent error term $\mathcal{E}_\phi = \lambda^2/12 \times \operatorname{Re}\!\bigl(
    e^{2\ii(\Delta\phi)}\{C[\Delta]^2
    + C_{\rm c}[\Delta]^2
    + C_{\rm s}[\Delta]^2\}
    \bigr)$  in Eq.~\eqref{eq:eps_av} is bounded by the error terms of the phase-averaged error formula in Eq.~\eqref{eq:eps_av random phase} since
\begin{align}
    |\mathcal{E}_\phi| &= \frac{\lambda^2}{12}|\operatorname{Re}\bigl(
    e^{2\ii(\Delta\phi)}\{e^{\ii 2\gamma'}|C[\Delta]|^2
    + e^{\ii 2\gamma_\mathrm{c}'}|C_{\rm c}[\Delta]|^2 \nonumber \\
    &+ e^{\ii 2\gamma_\mathrm{s}'}  |C_{\rm s}[\Delta]|^2 \}
    \bigr)| \nonumber \\
    &\leq \frac{\lambda^2}{12} \bigl( |C[\Delta]|^2 + |C_{\rm c}[\Delta]|^2 + |C_{\rm s}[\Delta]|^2 \bigr) \nonumber \\
    &=\frac{\lambda^2}{12} \bigl(  S[\Delta] + S_{\rm c}[\Delta] + S_{\rm s}[\Delta]  \bigr),
\end{align}
where $C[\Delta] = |C[\Delta]|e^{\ii \gamma'}$ and analogously for $C_{\rm c}[\Delta]$ and $C_{\rm s}[\Delta]$. Due to this upper bound, the magnitude of the phase-dependent error term $\mathcal{E}_\phi$ is also reduced when using crosstalk mitigation techniques that minimize the error terms of the phase-averaged error formula in Eq.~\eqref{eq:eps_av random phase}. For this reason, the crosstalk transition suppression techniques considered in this work, such as HD DRAG and off-resonant driving, also help to mitigate crosstalk errors in a scenario, in which the phase difference is not averaged. This argument generalizes also to a larger number of control qubits, which can be motivated by considering $s^{\rm C}(t)$ as a sum of the drive pulses and noting that the Fourier transform is a linear operation.

To validate the analytical model, we compare the excess gate error of the target qubit given by the model against QuTiP~\cite{lambert2025qutip5quantumtoolbox} simulations. In the simulations, we model the time evolution of the target qubit under drive crosstalk using the Hamiltonian in Eq.~\eqref{target_hamiltonian suppl} and the drive crosstalk term in Eq.~\eqref{control drive term suppl}. Here, we assume $\omega_{01}^{\rm T} / (2\pi) = \omega_{\rm d}^{\rm T} / (2\pi) = \SI{4.074}{\giga\hertz}$, and $\alpha^{\rm T} / (2\pi) \approx -\SI{182}{\mega\hertz}$. The details of the simulation methods, gate calibration procedure, and the average gate fidelity calculation follow those presented in the supplementary Sec. C of Ref.~\cite{hyyppa2024reducing}. For the target qubit, we calibrate an $X_{\pi/2}$ gate based on a cosine DRAG-L pulse. For the control qubit, we calibrate either a cosine DRAG-L $\pi/2$-pulse or an HD DRAG-L $\pi/2$-pulse. To calculate the excess gate error of the target qubit, $\epsilon_{\rm sim}^{\rm T} - \epsilon_{\rm ind}^{\rm T}$, we first compute the average gate error in the absence of crosstalk ($\lambda$ = 0) to obtain the individual gate error $\epsilon_{\rm ind}^{\rm T}$. We then compute the simultaneous gate error $\epsilon_{\rm sim}^{\rm T}$ by repeating the calculation with the assumed crosstalk parameter $\lambda$ related to the magnitude of the crosstalk as $\lambda^2 = C_\mathrm{TC}$. 
Since the crosstalk error depends on the drive phase of the control qubit according to Eq.~\eqref{eq:eps_av}, we compute the average simultaneous fidelity over the phase difference $\Delta\phi \in [0, 2\pi]$.

\begin{figure}[ht]
    \centering
\includegraphics{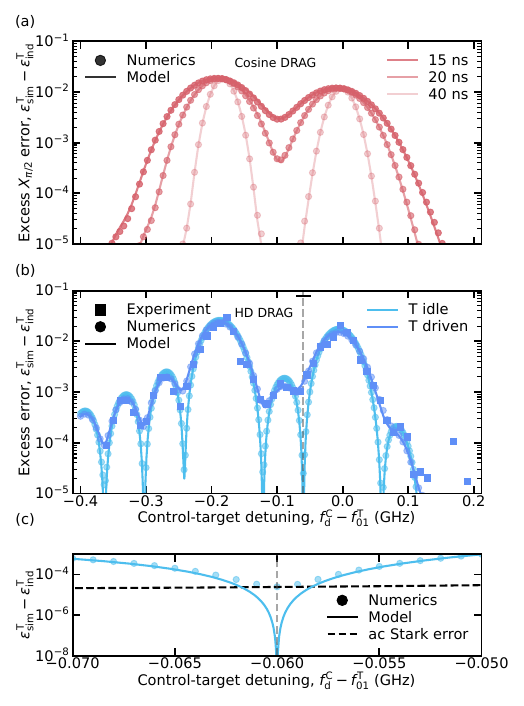}
    \caption{\textbf{Comparison of the analytical crosstalk error model against QuTiP simulations.} (a) Excess $X_{\pi/2}$ gate error of the target qubit as a function of control-target detuning for the analytical model in Eq.~\eqref{eq:eps_av random phase} (lines) and for a full numerical simulation (circular markers) assuming cosine DRAG $\pi/2$-pulses on both qubits and $C_{\rm TC} = -\SI{15}{\decibel}$. Different pulse durations are denoted by different shades of red. (b) Excess error of the target qubit  for the analytical model in Eq.~\eqref{eq:eps_av random phase} (lines) and for a numerical simulation (circular markers) assuming $C_{\rm TC} = -\SI{13.85}{\decibel}$ and a 20-ns HD DRAG $\pi/2$-pulse on the control qubit in two scenarios: i) the target qubit is idling (light blue), and ii) the target qubit is driven with a 20-ns cosine DRAG $\pi/2$-pulse (dark blue).  We determine the suppressed frequencies of the HD DRAG pulse for $f_\mathrm{d}^\mathrm{C} - f_{01}^\mathrm{T} = -\SI{60}{\mega\hertz}$ (dashed vertical gray line), and keep them fixed across $f_\mathrm{d}^\mathrm{C}$.  Square markers show experimental simultaneous gate error for a driven target qubit. (c) Zoom-in of (b) highlighting the difference of the numerical simulation and the analytical model near the spectral minimum at  $f_\mathrm{d}^\mathrm{C} - f_{01}^\mathrm{T} = -\SI{60}{\mega\hertz}$ due to the ac Stark error (dashed black line) given by Eq.~\eqref{ac stark error}.}
    \label{fig: model_vs_simulation}
\end{figure}

In Fig.~\ref{fig: model_vs_simulation}(a), we show the excess $X_{\pi/2}$ gate error of the target qubit for different gate durations over a range of control-target detunings when using cosine DRAG pulses on both the target and control qubit, and $C_{\rm TC} = -\SI{15}{\decibel}$. We observe that the analytical error model is in excellent agreement with numerical simulations across a wide range of detunings and magnitudes of error. In Fig~\ref{fig: model_vs_simulation}(b), we show the excess gate error of the target qubit when applying a 20-ns HD DRAG $\pi/2$-pulse on the control qubit and the target qubit is assumed to either idle or be  driven by a 20-ns cosine DRAG $\pi/2$-pulse. For the HD DRAG pulse, we choose the suppressed frequencies  $\{f_{\mathrm{s}, j}\} = \{|f_\mathrm{d}^\mathrm{C} - f_{01}^\mathrm{T}|, |f_\mathrm{d}^\mathrm{C} - f_{12}^\mathrm{T}|, |f_\mathrm{d}^\mathrm{C} - f_{12}^\mathrm{C}|\}$ for a control-target detuning of $f_\mathrm{d}^\mathrm{C} - f_{01}^\mathrm{T}=-\SI{60}{\mega\hertz}$, and keep them fixed over the drive detuning sweep. We note that the analytical model is able to capture the excess gate error of the target qubit in both the idle and driven scenarios for almost the entire range of drive detunings. In the case of an idling the target qubit, the analytical error model, however, underestimates the gate error compared to the full numerical simulation across a narrow range of frequencies around the minima $S[\omega] = 0$ as shown in panel (c) of  \cref{fig: model_vs_simulation}. This is because the HD DRAG pulse perfectly cancels all the error terms in the analytical model resulting in $\mathcal{E}_{\rm av} = 0$. However, the analytical model in Eq.~\eqref{eq:eps_av random phase} neglects the ac Stark shift of the target qubit caused by the control qubit pulse. The full numerical simulation captures this ac Stark shift of the target qubit, resulting in a non-zero gate error at $S[\omega] = 0$.  Furthermore, we observe that both the analytical error model and the simulations are in excellent agreement with corresponding experimental results across the entire range of detunings for a driven target qubit.

\subsection{Approximation of the ac Stark error caused by the control qubit pulse}
\label{ap: ac stark shift approximation}
If the target qubit is not driven, we can approximate the ac Stark shift of its $\ket{0}\leftrightarrow\ket{1}$ transition induced by the control-qubit pulse using second-order perturbation theory. Following Ref.~\cite{Veps_l_inen_2018}, we obtain
\begin{align}\label{ac stark shift}
    \delta_{01}^{\rm ac}(t)
    = -\frac{|\lambda s^{\rm C}(t)|^2}{2\Delta}
    + \frac{|\lambda s^{\rm C}(t)|^2}{2(\Delta - \alpha^{\rm T})}
    = -\frac{\lambda^2 \alpha^{\rm T}|s^{\rm C}(t)|^2}{2\Delta(\alpha^{\rm T} - \Delta)},
\end{align}
where $\lambda s^{\rm C}(t)$ is the control-qubit drive envelope scaled by the crosstalk amplitude onto the target qubit.

Due to the ac Stark shift, the target qubit accumulates a phase $\phi_{\rm ac}^{\rm T}$ over the duration of the control pulse, which manifests as a coherent $Z$ error on the target qubit. A coherent phase error of angle $\phi_{\rm ac}^{\rm T}$ results in an average infidelity
\begin{align}\label{ac stark error}
    \mathcal{E}_{\rm ac}^{\rm T} = \frac{1 - \cos{\phi_{\rm ac}^{\rm T}}}{3} \approx \frac{(\phi_{\rm ac}^{\rm T})^2}{6},
\end{align}
where the latter approximation assumes $|\phi_{\rm ac}^{\rm T}|\ll 1$. The expression in Eq.~\eqref{ac stark error} follows from the standard closed-form formula for the average fidelity between two unitaries~\cite{Pedersen_2007},
\begin{align}
    F_{\rm avg}(\hat{U},\hat{V})=\frac{d+|\mathrm{Tr}(\hat{U}^\dagger \hat{V})|^2}{d(d+1)},
\end{align}
evaluated for $\hat{U}=\hat{I}$, $\hat{V}=e^{-\ii \hat{\sigma}_z \phi_{\rm ac}^{\rm T}/2}$, and $d=2$.

From Eqs.~\eqref{ac stark shift} and \eqref{ac stark error}, the ac-Stark-induced phase satisfies $\phi_{\rm ac}^{\rm T}\propto \lambda^2$, implying an infidelity scaling $\mathcal{E}_{\rm ac}^{\rm T}\propto \lambda^4$. This is parametrically distinct from cross-driving errors captured by the analytical model in Eq.~\eqref{eq:eps_av random phase}, which scale as $\mathcal{E}_{\rm av}^{\rm T}\propto \lambda^2$. Moreover, in the large-detuning limit $|\Delta|\gg |\alpha^{\rm T}|$, Eq.~\eqref{ac stark shift} yields $\delta_{01}^{\rm ac}(t)\propto 1/\Delta^2$ and therefore $\mathcal{E}_{\rm ac}^{\rm T}\propto 1/\Delta^4$. In contrast, cross-driving errors can be strongly suppressed with detuning when the control pulse has rapidly decaying spectral weight at the target frequency. Consequently, while $\mathcal{E}_{\rm ac}^{\rm T}$ is typically small for experimentally relevant parameters, it may become the dominant residual contribution at sufficiently large detunings.

The perturbative expression for the ac Stark shift in Eq.~\eqref{ac stark shift} is not applicable when the target qubit is simultaneously driven. In the simultaneous-drive regime considered here, we nevertheless find the corresponding ac-Stark-induced contribution to be typically orders of magnitude smaller compared to the error predicted by the analytical model in Eq.~\eqref{eq:eps_av random phase}.

\section{Interpretation of the analytical error model}
\label{ap: interpretation of analytical error model}

Here, we provide further discussion on the interpretation of the error terms  $S[\Delta]$, $S_{\rm c}[\Delta]$, $S_{\rm s}[\Delta]$, $S_{{\rm c}\frac{1}{2}}[\Delta + \alpha^\mathrm{T}]$, and $S_{{\rm s}\frac{1}{2}}[\Delta + \alpha^\mathrm{T}]$ in Eq.~\eqref{eq:eps_av random phase}.

\subsection{Energy spectral density}
\label{ap: error terms spectral density}

The error terms proportional to $S[\Delta]$, $S_{\rm c}[\Delta]$, and $S_{\rm s}[\Delta]$ in Eq.~\eqref{eq:eps_av random phase} correspond to gate errors within the computational subspace $\{|0\rangle, |1\rangle \}$ of the target qubit. We attribute these terms to the off-resonant cross driving of the target qubit within its computational subspace due to the crosstalk-mediated control qubit pulse. Intuitively, the first term $S[\Delta] = |\mathcal{F}[s^\mathrm{C}](\Delta)|^2 = |\int_{-\infty}^\infty \mathrm{d} t e^{-\ii \Delta t} s^\mathrm{C}(t)|^2$ 
is equal to the energy spectral density of the control qubit pulse $s^\mathrm{C}(t)$ evaluated at the detuning $\Delta$  between the target qubit and the drive of the control qubit. 
Thus, the errors related to  $S[\Delta]$ can be mitigated by shaping the control qubit pulse to have a root of its Fourier transform  at $\Delta$, i.e., $\mathcal{F}[s^\mathrm{C}](\Delta) = 0$. This can be achieved using existing pulse shaping techniques, such as DRAG \cite{motzoi2009simple} or higher-derivative extensions of DRAG \cite{motzoi2013improving, hyyppa2024reducing, li2024experimental, wang2025suppressing, gao2025ultra}. As illustrated in \cref{fig: explaining rabi splitting}(a), such pulse shaping techniques do not, however, necessarily suppress the error terms $S_{\rm c}[\Delta]$ and $S_{\rm s}[\Delta]$ that account for the Rabi splitting of the target qubit as discussed in the following subsection. In the special case of an idling target qubit with $s^\mathrm{T}(t) = 0$, there is no Rabi splitting and a drive pulse designed to suppress $S[\Delta]$ also mitigates $S_{\rm c}[\delta]$ and $S_{\rm s}[\Delta]$ since   $S_{\rm c}[\Delta]= S[\Delta]$ and $S_{\rm s}[\Delta]= 0$. 

\subsection{Rabi splitting}
\label{ap: error terms rabi splitting}

\begin{figure*}
    \centering
\includegraphics{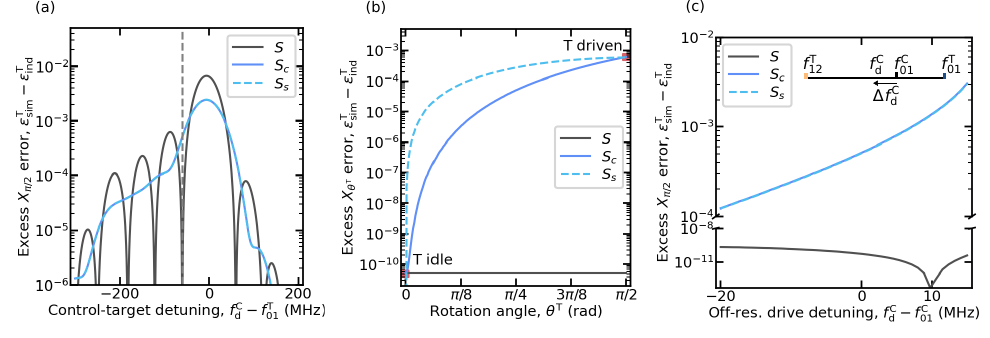}
    \caption{\textbf{Rabi splitting.} (a) Excess $X_{\pi/2}$ gate error of the target qubit owing to the individual error terms $S$, $S_{\rm c}$, and $S_{\rm s}$ of the analytical model as a function of control-target detuning assuming an HD DRAG $\pi/2$-pulse on the control qubit and a cosine DRAG $\pi/2$-pulse on the target qubit. The vertical dashed line corresponds to the control-target detuning $f_\mathrm{d}^\mathrm{C} - f_{01}^\mathrm{T} = -\Delta$ used to determine the suppressed frequencies of the HD DRAG pulse, thus ensuring $S[\Delta] = 0$. Note that $S_{\rm c}$ and $S_{\rm s}$ are essentially smoothed variants of $S$. (b) Excess $X_{\theta}$ error  of the target qubit owing to $S$, $S_{\rm c}$, and $S_{\rm s}$ as a function of the rotation angle $\theta^{\rm T}$ of the single-qubit gate applied to the target qubit when simultaneously applying  an HD DRAG $\pi/2$-pulse on the control qubit. The red squares highlight the special cases of an idling target qubit (left) and a target qubit driven with a $\pi/2$-pulse (right). (c)  Excess $X_{\pi/2}$ error  of the target qubit owing to $S$, $S_{\rm c}$, and $S_{\rm s}$ as a function of off-resonant drive detuning of the control qubit assuming an HD DRAG $\pi/2$-pulse on the control qubit and a cosine DRAG $\pi/2$-pulse on the target qubit. Throughout the panels, we assume a control-target detuning of $f_\mathrm{01}^\mathrm{C} - f_{01}^\mathrm{T} = -\SI{60}{\mega\hertz}$ and a crosstalk of $C_\mathrm{TC}=-\SI{13.9}{\decibel}$ similarly to Fig.~\ref{fig: concept of higher-derivative drag}.}
    \label{fig: explaining rabi splitting}
\end{figure*}

We can write the error terms $S_{\rm c}[\Delta]$ and $S_{\rm s}[\Delta]$ in Eq.~\eqref{eq:eps_av random phase} based on a convolution of the Fourier transform $\mathcal{F}[s^\mathrm{C}]$ and a kernel function. The bandwidth of the kernel function turns out to be determined by the Rabi splitting of the target qubit due to its own drive pulse $s^\mathrm{T}(t)$.

To explain this, we first note that multiplication in the time domain is equivalent to convolution in the frequency domain
\begin{equation}
    \int_{-\infty}^{\infty} \mathrm{d}t e^{-\mathrm{i} \omega t} f(t) g(t) = \frac{1}{2\pi} (\mathcal{F}[f] * \mathcal{F}[g])[\omega],
\end{equation}
where $\mathcal{F}[f](\omega)= \int_{-\infty}^\infty \mathrm{d}t \exp(-\mathrm{i} \omega t) f(t) $ denotes the Fourier transform of $f(t)$, and $*$ denotes the convolution operator. This allows us to re-write $C_{\rm s}[\omega]$ and $ C_{\rm c}[\omega]$ in Eq.~\eqref{eq: C definitions} in terms of $\mathcal{F}[s^\mathrm{C}](\omega)=C[\omega]$ as 
\begin{align} 
    C_{\rm s}[\omega] &= \frac{1}{2\pi} (C * \mathcal{F}[\sin{\theta}])[\omega] , \label{eq: Cs definitions as convolution} \\
    C_{\rm c}[\omega] &= \frac{1}{2\pi} (C * \mathcal{F}[\cos{\theta}])[\omega], \label{eq: Cc definitions as convolution} 
\end{align}
where  $\mathcal{F}[\sin{\theta}]$ and $ \mathcal{F}[\cos{\theta}]$ are the kernel functions in the frequency domain. The above equations  signify that $C_{\rm s}[\omega]$ and $C_{\rm c}[\omega]$ correspond to  convolutions of the Fourier transform $C(\omega)$. 
However, they do not directly reveal the bandwidth of the kernel functions in Eqs.~\eqref{eq: Cs definitions as convolution} and \eqref{eq: Cc definitions as convolution}. Namely, the kernel functions are not uniquely defined as their exact form depends on the chosen continuation of $\theta(t)$ outside of the time interval $t \in [0, t_\mathrm{g}]$.

To gain further insight into the relevant bandwidth of the kernel functions, we re-write $C_{\rm s}[\omega]$ and $ C_{\rm c}[\omega]$ in Eq.~\eqref{eq: C definitions} in an alternative way
\begin{align} 
    C_{\rm s}[\omega] &= \int_{-\infty}^\infty \text{d}te^{-\ii \omega t}s^\mathrm{C}(t) \frac{e^{\ii \theta(t)} - e^{-\ii \theta(t)}}{2\ii} \nonumber \\
    &= \frac{1}{2\ii} \bigg [ \int_{-\infty}^\infty \text{d}t \bigg ( e^{-\ii ( \int_0^t \omega  + s^\mathrm{T}(t')\mathrm{d}t' )}s^\mathrm{C}(t) \nonumber \\
    &~~~~~~~~~~~~~~~~- e^{-\ii ( \int_0^t \omega  - s^\mathrm{T}(t')\mathrm{d}t' )}s^\mathrm{C}(t)  \bigg ) \bigg ], \label{eq: Cs shifted Fourier transform} \\
        C_{\rm c}[\omega] &= \int_{-\infty}^\infty  \text{d}te^{-\ii \omega t}s^\mathrm{C}(t) \frac{e^{\ii \theta(t)} + e^{-\ii \theta(t)}}{2} \nonumber \\
    &= \frac{1}{2} \bigg [ \int_{-\infty}^\infty \text{d}t \bigg ( e^{-\ii ( \int_0^t \omega  + s^\mathrm{T}(t')\mathrm{d}t' )}s^\mathrm{C}(t) \nonumber \\
    &~~~~~~~~~~~~~~~~+ e^{-\ii ( \int_0^t \omega  - s^\mathrm{T}(t')\mathrm{d}t' )}s^\mathrm{C}(t)  \bigg ) \bigg ], \label{eq: Cc shifted Fourier transform}
\end{align}
where we have used $\theta(t) = \int_{0}^t s^\mathrm{T}(t') \mathrm{d} t'$. The above equations can be interpreted as modified Fourier transforms, in which the angular frequency $\omega$ is dynamically shifted  by the drive strength $\pm s^\mathrm{T}(t)$ of the target qubit. The drive strength $s^\mathrm{T}(t)$ is also equal to the instantaneous Rabi splitting of the target qubit due to its own drive under a two-level approximation. Thus, the Rabi splitting  $s^\mathrm{T}(t)$ essentially determines the relevant bandwidth of the smoothing for the error terms $S_{\rm c}[\Delta]$ and $S_{\rm s}[\Delta]$. For this reason, we use the term \textit{Rabi splitting terms} to denote $S_{\rm c}[\Delta]$ and $S_{\rm s}[\Delta]$. 

For further intuition, we briefly consider the special case of a rectangular drive pulse on the target qubit, i.e., $s^\mathrm{T}(t) = s^\mathrm{T} = \theta / t_\mathrm{g}$. 
Using Eqs.~\eqref{eq: Cs shifted Fourier transform} and \eqref{eq: Cc shifted Fourier transform}, we see that the terms $C_{\rm s}[\omega]$ and $ C_{\rm c}[\omega]$ simplify as 
\begin{align} 
    C_{\rm s}[\omega] &= \frac{1}{2\ii} \left ( C[\omega + \theta/t_\mathrm{g}] - C[\omega - \theta/t_\mathrm{g}] \right ),  \label{eq: Cc for rectangular pulse} \\
    C_{\rm c}[\omega] &= \frac{1}{2} \left ( C[\omega + \theta/t_\mathrm{g}] + C[\omega - \theta/t_\mathrm{g}] \right ),  \label{eq: Cs for rectangular pulse}
\end{align}
which corresponds to the Fourier transform $C[\omega]$ shifted by the Rabi splitting $\pm s^\mathrm{T}= \pm \theta/t_\mathrm{g}$. For the square pulse, it thus holds  that
\begin{align}
S_\mathrm{s}[\omega] + S_\mathrm{c}[\omega] &= \frac{|C[\omega + \theta/t_\mathrm{g}]|^2 + |C[\omega - \theta/t_\mathrm{g}]|^2}{2} \nonumber \\
&\approx |C[\omega]|^2 = S[\omega], \label{eq: Sc + Ss}
\end{align}
for sufficiently small $\theta$ or away from spectral minima, where the energy spectral density behaves smoothly. A similar approximation is valid also for the conventional cosine pulse shape away from spectral minima.

The smoothing of $S_{\rm c}[\Delta]$ and $S_{\rm s}[\Delta]$ by the Rabi splitting has important consequences for the mitigation of asymmetric microwave crosstalk when shaping the spectrum of the control qubit pulse using, e.g., HD DRAG.  Only in the absence of a drive pulse on the target qubit, there is no Rabi splitting and all the error terms $S[\Delta]$, $S_{\rm c}[\Delta]$ and $S_{\rm s}[\Delta]$ are perfectly minimized using an HD DRAG pulse on the control qubit. In the presence of a drive pulse on the target qubit, the Rabi splitting terms $S_{\rm c}[\Delta]$ and $S_{\rm s}[\Delta]$ are non-zero at $\Delta$ as shown in Fig.~\ref{fig: explaining rabi splitting}(a) even though the HD DRAG pulse has been shaped to ensure $S[\Delta]=0$. The error terms $S_{\rm c}[\Delta]$ and $S_{\rm s}[\Delta]$ typically increase monotonically with increasing drive amplitude on the target qubit as shown in Fig.~\ref{fig: explaining rabi splitting}(b).

As shown in Fig.~\ref{fig: explaining rabi splitting}(c), the Rabi splitting terms can be significantly reduced by shifting the drive frequency of the control qubit away from the nearest target qubit transition using the off-resonant driving technique introduced in Sec.~\ref{sec: off-resonant control pulses} and further discussed in Sec.~\ref{ap: strongly off-resonant control pulses}. As illustrated in Fig.~\ref{fig: concept of higher-derivative drag}(c) of the main text, the off-resonant driving helps to reduce the magnitude of the Fourier transform $\mathcal{F}[s^\mathrm{C}]$ around the nearest target qubit transition across the frequency range associated with the Rabi splitting, which consequently decreases $S_{\rm c}[\Delta]$ and $S_{\rm s}[\Delta]$.

\subsection{Leakage error}
\label{sec: leakage}

The error terms proportional to $S_{{\rm c}\frac{1}{2}}[\Delta + \alpha^\mathrm{T}]$ and $S_{{\rm s}\frac{1}{2}}[\Delta + \alpha^\mathrm{T}]$ in Eq.~\eqref{eq:eps_av random phase} correspond to leakage errors to the second excited state $|2\rangle$ of the target qubit due to the off-resonant cross driving of its $|1\rangle \leftrightarrow|2\rangle$ transition by the control qubit pulse. Similarly to the Rabi splitting terms $S_{\rm c}[\Delta]$ and $S_{\rm s}[\Delta]$,  the leakage terms  $S_{{\rm c}\frac{1}{2}}[\Delta + \alpha^\mathrm{T}]$ and $S_{{\rm s}\frac{1}{2}}[\Delta + \alpha^\mathrm{T}]$ essentially correspond to a smoothed energy spectral density of the control qubit pulse $s^\mathrm{C}(t)$ evaluated at a detuning of $\Delta + \alpha^\mathrm{T}$, which  corresponds to the $f_{12}^\mathrm{T}$ transition of the target qubit.  Thus, these terms are maximized if the drive frequecy of the control qubit matches the $|1\rangle \leftrightarrow |2\rangle $ transition frequency of the target qubit, i.e., $f_\mathrm{d}^\mathrm{C} = f_{12}^\mathrm{T}$. For these leakage  terms, the bandwidth of  the kernel function is half of the bandwidth corresponding to the Rabi splitting terms $S_{{\rm c}}[\Delta]$ and $S_{{\rm s}}[\Delta]$. Intuitively speaking, the $|1\rangle \leftrightarrow |2\rangle$ transition only attains half of the shift compared to $|0\rangle \leftrightarrow |1\rangle$ since the Rabi splitting affects the levels $|0\rangle$ and $|1\rangle$.

\section{Optimization of the qubit frequency configuration}
\begin{algorithm}[t]
\caption{\textbf{Qubit frequency configuration optimization}. Summary of the steps performed in the optimization loop shown in~\cref{fig: crosstalk-optimized frequency configuration}(c, d). The input parameters are a general list of qubit parameters $\vec{p}$ and all qubit frequencies $\vec{f_q}$. Then for each qubit $i$, a one-dimensional error landscape $\epsilon_{\mathrm{others}}$ is calculated for an array of frequencies of that qubit $\vec{f^i}$. The pulse shapes for all qubits $\vec{s}$ are either constant (cosine DRAG) or determined by the qubit frequencies $\vec{f^q},\vec{f^i}$ (off-resonant HD DRAG).  Finally, a new frequency is chosen for qubit $i$ by minimizing the landscape and updated in the vector of qubit frequencies $\vec{f_q}$. }\label{alg: optimization}

\begin{algorithmic}
\State \textbf{input} $\vec{p}, \vec{f_q}$
\For{$n \in 1..N_\mathrm{it}$} 
    \For{$i \in 1\dots N_\mathrm{Q}$}
        \State $\vec{f}\gets [f_{min}, f_{max}^i]$
        \State $\vec{s}\gets \mathrm{eval\_pulse\_shapes}(\vec{f}, \vec{f}_q, \vec{p})$
        \State $\epsilon_\mathrm{self}\gets \mathrm{ct\_self}(i,\vec{f},\vec{f}_q,
        \vec{s}, \vec{p})+  \mathrm{hybridization}(i, \vec{f}, \vec{f}_q,  \vec{p})$
        \State $\epsilon_\mathrm{others}$ $\gets$ ct\_others($i,\vec{f}, \vec{f}_q, \vec{s}$, $\vec{p}$)
        \State $\epsilon_\mathrm{total}$ $\gets$ $\max(\epsilon_\mathrm{self}$,$\epsilon_\mathrm{others}$)
        \State $\vec{f}_q[i] \gets $minimize($\vec{f}$,$f^i_{target}$, $\epsilon_\mathrm{total}$)
        
        \EndFor
       
\EndFor
\end{algorithmic}
\end{algorithm}
\label{ap: detailed qubit frequncy optimization}

We now provide a more detailed description of the minimal 
line-by-line optimization algorithm described in~\cref{sec: strategy for the frequency configuration optimization}. A summary is shown in~\cref{alg: optimization}. 
For each qubit, an array of frequencies $\vec{f^i}$ is determined, to which the qubit can be moved in order to minimize the global error cost function $C = \sum_i^{N_\mathrm{Q}} \epsilon_\mathrm{self}^i(\vec{f}^i, \vec{f_q},\vec{s},  \vec{p})$, where $\epsilon_\mathrm{self}^i$ is the \textit{error on self} for qubit $i$ as a function of pulse shapes used on all qubits $\vec{s}$, current frequencies of other qubits $\vec{f_q}$, and calibrated qubit parameters $\vec{p}$.
The frequency array $\vec{f^i}$ spans $[f_\mathrm{min}, f_\mathrm{max}^i]$, where $f_\mathrm{min}$ was set to $\SI{3.6}{\giga\hertz}$ and $f^i_\mathrm{max}\in [\SI{4.15}{\giga\hertz},\SI{4.56}{\giga\hertz}]$ are the maximum frequencies of qubit $i$ set by device fabrication. We then estimate $\epsilon_\mathrm{self}^i(\vec{f}^i, \vec{f_q}, \vec{s}, \vec{p})$ by evaluating the model in~\cref{eq:mt_eps_av random phase} for the array  $\vec{f^i}$ and summing error contributions from all other qubits at their currently set frequency. Here, pulse shapes of all qubits are taken into account. The hybridization error is then also calculated using the Lorentzian approximation discussed in~\cref{ap: hybridization_error}, centered at $\ket{0}\leftrightarrow\ket{1}$ frequencies of the neighbouring qubits to avoid $f^i_{01}-f^j_{01}$ collisions. As discussed in~\cref{ap: hybridization_error}, we generally expect that a similar term should be included to avoid $f^i_{01}-f^j_{12}$ collisions. However, we found that was not necessary for the typical idling frequencies of the couplers in this device. 

Then, we evaluate the \textit{error on others} $\epsilon_\mathrm{others}^i(\vec{f}^i,\vec{f_q}, \vec{s}, \vec{p})$ for the same frequencies $\vec{f^i}$ by evaluating the model in~\cref{eq:mt_eps_av random phase} assuming we move the current qubit to that frequency and summing the error contributions it causes on all other qubits at their currently set frequency. Hybridization crosstalk is symmetric between control and target qubits, so there is no need to include it in $\epsilon_\mathrm{others}^i$. The \textit{total error} $\epsilon_\mathrm{total}$ is then computed by taking the maximum of the two error landscapes at each frequency in $\vec{f}$. Including the  $\epsilon_\mathrm{others}^i(\vec{f}^i,\vec{f_q}, \vec{s}, \vec{p})$ is a key part of the algorithm that ensures that  $C$ will not increase too much when we move a qubit to minimize the error on itself. 
The efficiency of the error model evaluation can be improved due to the sparsity of the crosstalk matrix. By ignoring qubit pairs that have negligible crosstalk, we can strongly reduce the number of qubits to include in the summation of $\epsilon_\mathrm{others}^i$ and  $\epsilon_\mathrm{self}^i$.

Having established the error landscapes, we then use a minimization procedure to move the qubit to a frequency close to a desired target configuration $\vec{f}_\mathrm{target}$ that satisfies the optimization criterion, i.e. $\Delta\varepsilon_\mathrm{sim} < \varepsilon_\mathrm{target}$.
  This procedure is subsequently repeated for each qubit, looping $n$ times, making sure the optimization target is reached for all $N_\mathrm{Q}$ as shown by the dashed boxes in~\cref{fig: crosstalk-optimized frequency configuration}(c).
Note that this optimization procedure does not guarantee to find a global minimum and thus the resulting configuration depends on the starting configuration and ordering of the components that are optimized. Nevertheless, there exist many solutions to the optimization that in practise satisfy  $\Delta\varepsilon_\mathrm{sim} < \varepsilon_\mathrm{target}$ for realistic error targets. 

 \subsection{Optimization for crosstalk-optimized configuration}
 We now discuss specific settings for the data shown in the main text. For the minimization of crosstalk error in~\cref{sec: CT config results}, $\vec{f}_\mathrm{target}$ was set to the maximum qubit frequencies $f_{max}^i$ and we chose the nearest local minimum to a target frequency, that satisfied a predicted $\Delta\varepsilon_\mathrm{sim}$ below a threshold of $3\times10^{-4}$. To make sure a minimum would exist, we included a small slope ($\epsilon_\mathrm{slope} < 10^{-5}$) from low to high frequency.

\subsection{Optimization for CTS-on configuration with pulse shaping}
To optimize qubit frequencies for the CTS-on configuration minimizing the bandwidth in~\cref{sec: demonstration of bW minimized configuration}, we now include the pulse shapes as an additional degree of freedom. However, instead of adding a costly optimization over pulse shapes, we determine the off-resonant HD DRAG pulse shapes completely by the set of suppressed frequencies, which in turn are fully determined by the qubit frequencies and their anharmonicities. Thus, we only specify on which control qubits we apply off-resonant HD DRAG pulses and which target qubits they aim to suppress. Then, we keep the rest of the optimization the same as in~\cref{alg: optimization}, only choosing a new frequency in each step for a qubit and updating the error landscapes and pulse shapes accordingly.

We applied CTS using off-resonant HD DRAG pulses for pairs with high crosstalk ($C_{kj}>-\SI{30}{\decibel}$) to a single other qubit. We found that this was roughly optimal to minimize the total frequency spread as the HD DRAG pulse shapes require a minimal detuning to its target qubit for the calibration to succeed. Hence, most gains are obtained for pairs with high enough crosstalk that would otherwise be detuned beyond the set minimal detuning. 
To model this, we included a phenomenological minimal detuning in the $\epsilon_\mathrm{self}^i$ that limits $|f_{01}^\mathrm{T} - f_{01}^\mathrm{C}|$ and $|f_{12}^\mathrm{T} - f_{01}^\mathrm{C}|$ above $\SI{40}{\mega\hertz}$ for HD DRAG pairs to ensure a robust calibration of 20-ns $X_{\pi/2}$ gates.

\begin{figure}
    \centering
     \includegraphics{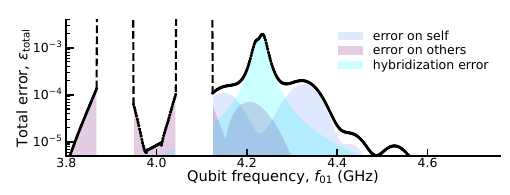}
    \caption{\textbf{Example error landscape using CTS.} Shown is the total error versus possible frequencies to place a qubit during one step of the qubit frequency optimization. The total error consists of a combination of hybridization and microwave crosstalk errors. Here, CTS pulse shaping was on for the qubit under consideration, and thus the shape of the major contribution to $\epsilon_\mathrm{others}$ follows \cref{fig: experimental demonstration of off-resonant HD DRAG}(c). The vertical dashed lines denotes boundaries of the minimum detuning of the qubit to its CTS target transition frequencies $f_{01}^\mathrm{T}, f_{12}^\mathrm{T}$. Additionally, there are contributions to $\epsilon_\mathrm{self}$ that come from another qubit using cosine DRAG pulses and thus follow the double peaked Gaussian error shape. The landscape is shown for $Q_{37}$ during the first optimization step starting from the AB configuration and including errors to and from the other 45 qubits.}
    \label{fig: cts_on_error_landscape}
\end{figure}

To limit the spread of qubit frequencies, $f_\mathrm{target}$ was set to the AB configuration, and we picked the frequency closest to $f_\mathrm{target}$ that still satisfied $\Delta\varepsilon_\mathrm{sim} < 3\times 10^{-4}$ during minimization along each line.
After the optimization, we further shifted the frequency of each qubit by $\SI{5}{\mega\hertz}$ in the direction of reducing predicted error, to account for a slight frequency dependence of crosstalk and reduce the error of qubits with a steep slope of error. 
An example  error landscape combining HD DRAG pulse shapes, hybridization error and the minimal detuning limit is shown in Fig.~\ref{fig: cts_on_error_landscape}. 
After preparing the bandwidth-minimized configuration on the QPU, we mitigate near-resonant two-level systems (TLS) on three qubits which impacted $\epsilon_{ind}$ by adjusting the qubit frequencies by 10--\SI{20}{\mega\hertz} in a direction satisfying  $\varepsilon_\mathrm{sim} \leq 3\times 10^{-4}$.

\section{Simulation of CTS optimization on larger devices using bootstrapping}
\label{ap: bw_simulations}
We now discuss in more detail the simulations performed for~\cref{fig: hd drag on bw minimized config}(e). We first establish the measured crosstalk dataset as a distribution from which we sample, using bootstrapping with replacement. For this purpose, we bin all measured pairwise crosstalk data from the full matrix shown in~\cref{fig: crosstalk matrix} into bins dependent on the inter-pair distance, where we rounded the distance to \SI{0.5}{\milli\meter} to ensure enough data was inside each bin. This gave a discrete set of bins with on average 84 elements per bin leading up to a maximum of \SI{17}{\milli\meter} total distance. A key approximation made here is that the crosstalk is \textit{only} dependent on the inter-pair distance. 

We then generate chip-topologies with the number of qubits ranging from 5 to 1000, which have the same pitch in millimeters for each few-qubit unit-cell as in our actual device. For each of these topologies, we generate a new pairwise distance matrix. 
We then convert those pairwise distance matrices to crosstalk matrices by sampling one of the experimentally measured crosstalk values from the above explained set of bins, with replacement, for each distance value in the matrix. Given that above $\sim$\SI{10}{\milli\meter} the measured crosstalk is already nearly always below the measurement limit, we assume that any distance larger than the largest distance of \SI{17}{\milli\meter} on the measured device yields the measurement limit of \SI{-76}{\decibel}. We then repeat this bootstrapping procedure for each chip size 10 times and save the resulting crosstalk matrices.  

After establishing the crosstalk matrices for each chip size, we run a similar optimization of qubit frequencies and pulse shapes as performed on the actual device for each matrix, setting the AB configuration as the target frequency configuration and using a fixed anharmonicity of $\alpha/(2\pi)=\SI{180}{\mega\hertz}$. 
Here, we did not consider $f^q_\mathrm{max}$ for simplicity. As a fixed rule, we apply CTS (off-resonant driving and HD DRAG) only on control qubits with crosstalk higher than $\SI{-30}{\decibel}$, and use cosine DRAG for all other qubits similar to the experiments. Thus in the optimization, optimal pulse shapes were evaluated in each optimization step for control qubits with CTS, suppressing both $f_{01}$ and $f_{12}$ for the current target qubit frequency.  We additionally used the same \SI{20}{\nano\second} gate length, minimum detuning limit of \SI{40}{\mega\hertz},  
and procedure to set the off-resonant driving frequency. The resulting bandwidth per chip size using CTS is then calculated as the difference of the highest and lowest qubit frequency, averaged over the 10 runs per chip size. We additionally performed the optimization on the same crosstalk matrices, turning off CTS 
and using cosine DRAG for all qubits to serve as the reference data in~\cref{fig: hd drag on bw minimized config}(e). .

\section{Implementation and calibration of CTS pulse shaping using off-resonant HD DRAG}
\label{ap: implementation and calibration of off-resonant HD DRAG}

In this section, we provide more details on the CTS pulse shaping technique employed to mitigate microwave crosstalk errors in Sections~\ref{sec: pulse shape engineering} and \ref{sec: demonstration of reduced bandwidth on full QPU using pulse shaping techniques}. First, we present a mathematical description of  HD DRAG pulse shaping in Section~\ref{ap: maths of hd drag} and discuss the implementation and limits of off-resonant driving for crosstalk error mitigation in Section~\ref{ap: strongly off-resonant control pulses}. Subsequently, Section~\ref{ap: selection of suppressed frequencies for HD DRAG} motivates our choice to suppress three transition frequencies in the spectra of HD DRAG pulses. We discuss the gate calibration protocol for off-resonant HD DRAG pulses in Section~\ref{ap: single-qubit gate calibration}, which mostly follows conventional DRAG pulse calibration using error amplification experiments. In Sec.~\ref{ap: HD DRAG for idling error mitigation}, we provide an experimental demonstration of more effective crosstalk mitigation using HD DRAG pulses in a case where the target qubit is idling instead of being simultaneously driven with the control qubit as is the case in the main text. The experimental results provide further verification of the analytical model and ac Stark error approximation as shown in Appendix~\ref{ap: analytical error model}, and highlight that the gate error of an idling target qubit is not representative of the simultaneous gate error when using spectral shaping techniques. 

\subsection{Mathematical definition of HD DRAG pulses and comparison to prior literature}
\label{ap: maths of hd drag}

Here, we provide mathematical details on the implementation of HD DRAG pulses discussed in Sec.~\ref{sec: hd drag}, extending Supplementary Sec. B of our earlier work \cite{hyyppa2024reducing}. We also compare  HD DRAG pulses to previous experimental demonstrations of pulse shaping for crosstalk mitigation.  

\subsubsection{Derivation of the HD DRAG pulse shape}

The in-phase and quadrature envelopes of an HD DRAG pulse can be written as 
\begin{align}
        s_I^\mathrm{C}(t) &= A_I^\mathrm{C} \bigg[\sum_{n=0}^K \beta_{2n} b^{(2n)} (t) \bigg], \label{eq: HD DRAG I suppl} \\
    s_Q^\mathrm{C}(t) &= -\frac{\beta }{\alpha^\mathrm{C}} \dot{ s}_I^\mathrm{C}(t) , \label{eq: HD DRAG Q suppl}
\end{align}
where $b(t)$ is a basis envelope, and $\{\beta_{2n}\}_{n=0}^{K}$ are the coefficients of the even-order derivatives with $\beta_0=1$. We solve the $K$ coefficients $\{\beta_{2n}\}_{n=1}^{K}$ to suppress the Fourier transform of the in-phase envelope $\mathcal{F}[s_I^\mathrm{C}](f)$ symmetrically at $2K$ frequencies $\{\pm f_{\mathrm{s}, j} \}_{j=1}^{K}$,  equivalent to suppressing the rf drive pulse with drive frequency $f_\mathrm{d}^\mathrm{C}$ at frequencies $\{f_\mathrm{d}^\mathrm{C} \pm f_{\mathrm{s}, j} \}_{j=1}^{K}$. 

This gives us the following system of equations
\begin{equation}
    \mathcal{F}[s_I^\mathrm{C}](f_{\mathrm{s}, j}) = 0,~j=1, \dots, K. \label{seq: HD FT of I zero}
\end{equation}
Using the relation $\mathcal{F}[b^{(n)}](f) = (\ii 2\pi f)^n \mathcal{F}[b](f)$, we can simplify the expression of the Fourier transform  $\mathcal{F}[s_I^\mathrm{C}](f)$  as
\begin{equation}
    \mathcal{F}[s_I^\mathrm{C}](f) = A_I^\mathrm{C} \mathcal{F}[b](f) \bigg [\sum_{n=0}^K \beta_{2n} (-1)^n (2\pi f)^{2n} \bigg  ]. \label{eq: Omega I hat polynomial}
\end{equation}

Thus, the system of equations in Eq.~\eqref{seq: HD FT of I zero} reduces to a set of linear equations
\begin{equation}
    \sum_{n=0}^K \beta_{2n} (-1)^n (2\pi f_{\mathrm{s}, j})^{2n} = 0,~j=1, \dots, K,  \label{eq: equation of betas for HD DRAG}
\end{equation}
from which the coefficients $\{\beta_{2n}\}_{n=1}^{K}$ can be straightforwardly solved if the suppressed frequencies $\{f_j\}_{j=1}^K$ are distinct. In our implementation, we select the suppressed frequencies as $\{f_{\mathrm{s}, j}\} = \{|f_\mathrm{d}^\mathrm{C} - f_{01}^\mathrm{T}|, |f_\mathrm{d}^\mathrm{C} - f_{12}^\mathrm{T}|, |f_\mathrm{d}^\mathrm{C} - f_{12}^\mathrm{C}|\}$ to avoid exciting the $|0\rangle \leftrightarrow |1\rangle$ and $|1\rangle \leftrightarrow |2\rangle$ transitions of the target qubit as well as the $|1\rangle \leftrightarrow |2\rangle$ transition of the control qubit itself. Furthermore, the DRAG coefficient $\beta$ is calibrated using the DRAG-L technique to further reduce leakage to the second excited state $|2\rangle$ of the control qubit. 

The basis function $b(t)$ needs to be chosen with care to ensure the continuity of $s_I^\mathrm{C}(t)$ and $s_Q^\mathrm{C}(t)$ and to avoid high-frequency components in their Fourier transforms. To ensure continuity when suppressing $K$ frequencies, we utilize a cosine series with $K+1$ terms as the basis envelope
\begin{equation}
    b(t) = \sum_{k=1}^{K + 1} d_k[1 - \cos(2\pi k t/t_\mathrm{p})], \label{seq: b cosine series} 
\end{equation}
where $t_\mathrm{p}$ is the pulse duration that together with any zero padding adds up to the total gate duration $t_\mathrm{g}$. Furthermore, $\{d_k\}_{k=1}^{K+1}$ denote the coefficients to be solved to satisfy the continuity requirements $b(0)=b^{(1)}(0)=\dots=b^{(2K+1)}(0)=b(t_\mathrm{p})=b^{(1)}(t_\mathrm{p})=\dots=b^{(2K+1)}(t_\mathrm{p})=0$. Since $b(t)$ is expressed as a cosine series, the odd derivatives automatically evaluate to zero at $t=0$ and $t=t_\mathrm{p}$. The continuity conditions for the even derivatives can be satisfied by solving $\{d_k\}_{k=1}^{K+1}$ from the following system of linear equations
\begin{align}
    \sum_{k=1}^{K+1} d_k (-1)^{n+1} k^{2n} &= 0, ~n=1, \dots, K, \label{eq: basis function coefficient eq 1} \\
    \sum_{k=1}^{K+1} d_k &= 1, \label{eq: basis function coefficient eq 2}
\end{align}
where the latter equation normalizes the sum of the coefficients to an arbitrarily chosen value.

\subsubsection{Comparison of HD DRAG to prior literature}
 
The presented implementation of the HD DRAG pulse is an extension of our previous work~\cite{hyyppa2024reducing}, where we demonstrated leakage mitigation for sub-10-ns single-qubit gates on an individual qubit by suppressing the spectrum of the in-phase envelope at the $ef$-transition  using $K=1$. Furthermore, we note that HD DRAG pulses can be viewed as a special case of recursive multi-derivative DRAG pulses~\cite{li2024experimental}. In this context, the HD DRAG pulse suppresses the leakage transition $f_{12}^\mathrm{C}$ and $2K$ additional one-photon transitions symmetrically located around the center drive frequency. The symmetric spectral suppression of the in-phase envelope also bears resemblance to the dual-DRAG pulse recently demonstrated in Ref. \cite{wang2025suppressing} with the goal to mitigate hybridization crosstalk. The dual-DRAG pulse reduces errors caused by hybridization crosstalk near the resonance  $f_{01}^\mathrm{C} \approx f_{01}^\mathrm{T}$ by suppressing the in-phase envelope at frequencies $\pm |f_{01}^\mathrm{T} - f_{01}^\mathrm{C}|$. Leakage errors near the resonance $f_{01}^\mathrm{C} \approx f_{12}^\mathrm{T}$ were not considered.

In prior literature, Wah-Wah pulses~\cite{schutjens2013single, vesterinen2014mitigating} and selective excitation pulses~\cite{matsuda2025selective} have also been demonstrated to mitigate the impact of crosstalk on a target qubit by suppressing multiple transition frequencies. However, these drive pulses suffer from discontinuities in the time domain, leading to increased leakage error of the control qubit due to elevated spectral energy at high frequencies.

Overall, the HD DRAG pulse considered in our work goes beyond previous experimental pulse shaping demonstrations \cite{vesterinen2014mitigating, matsuda2025selective, wang2025suppressing} by suppressing both transition frequencies $f_{01}^\mathrm{T}$ and $f_{12}^\mathrm{T}$ of the target qubit and also the leakage transition $f_{12}^\mathrm{C}$ as discussed in \cref{sec: rabi splitting}. Based on our experiments, all of these three transition frequencies need to be considered to reach a low gate error and leakage across a wide range of qubit-qubit detunings, see Fig.~\ref{fig: hd drag with varying number of suppressed frequencies} in Appendix~\ref{ap: selection of suppressed frequencies for HD DRAG}.

\subsection{Off-resonant drive pulses}
\label{ap: strongly off-resonant control pulses}

\begin{figure}
    \centering
\includegraphics{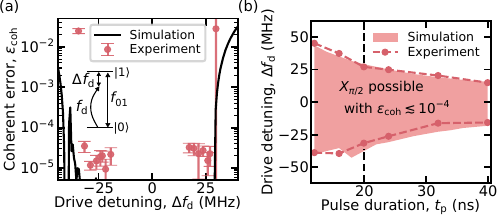}
    \caption{\textbf{Available range of drive detunings for off-resonant drive pulses.} Coherent error $\varepsilon_\mathrm{coh}$ measured from purity RB (circular markers) together with simulated coherent error (black line) as a function of drive detuning $\Delta f_\mathrm{d}$ used for a cosine DRAG pulse implementing an $X_{\pi/2}$ gate. The error bars denote $1\sigma$ uncertainty of the fit. (b) Upper and lower bound of measured drive detunings (circular markers) enabling $X_{\pi/2}$ gates with $\varepsilon_\mathrm{coh} \lesssim 10^{-4}$ as a function of pulse duration. The shaded red region shows the corresponding range based on simulations. }
    \label{fig: limits of off-res drive}
\end{figure}

Here, we provide further experimental results on the available range of drive detunings for off-resonant drive pulses. We also describe our strategy for choosing the drive detunings for the experiments discussed in Secs.~\ref{sec: pulse shape engineering} and \ref{sec: demonstration of reduced bandwidth on full QPU using pulse shaping techniques}. 

\subsubsection{Available range of drive detunings for high-fidelity gates}

In  Sec.~\ref{sec: off-resonant control pulses}, we experimentally demonstrated that off-resonant drive pulses enable Rabi oscillations reaching the equator of the Bloch sphere $P_{|1\rangle} = 0.5$ for a large range of drive detunings. For a typical a 20-ns pulse duration, this was possible for drive detunings up to $|\Delta f_\mathrm{d}^\mathrm{C}| \lesssim \SI{30}{\mega\hertz}$ as shown in Fig.~\ref{fig: concept of higher-derivative drag}(e).
Here, we further perform full gate calibration and benchmarking to confirm that the coherent error of $X_{\pi/2}$ gates can indeed be made minimal  using virtual Z rotations across this range of detunings. In Fig.~\ref{fig: limits of off-res drive}(a), we show the coherent error measured from a purity RB experiment (see Appendix~\ref{ap: benchmarking of single-qubit gate errors}) together with the simulated coherent error as a function of drive detuning $\Delta f_\mathrm{d}$ for a 20-ns cosine DRAG pulse. For simplicity, we apply the cosine DRAG pulse shape in our experiments and simulations since the available range of drive detunings for HD DRAG pulses slightly depends on the specific choice of suppressed frequencies, see Fig.~\ref{fig: supplementary data on single-pair detuning sweep}.  As in Appendix~\ref{ap: analytical error model}, we carry out the simulations using QuTiP based on the Hamiltonian of a driven transmon qubit in Eq.~\eqref{target_hamiltonian suppl} truncated to four levels.  For $t_\mathrm{p} = \SI{20}{\nano\second}$, we observe that a coherent error below $\varepsilon_\mathrm{coh} < 10^{-4}$ can be experimentally reached  for drive detunings $-\SI{31.5}{\mega\hertz} \lesssim \Delta f_\mathrm{d} \lesssim \SI{26.9}{\mega\hertz}$. Outside of this range, the coherent error rapidly increases both in experiments and in simulations as the qubit state does not reach the equator of the Bloch sphere or barely reaches it, thus hampering the gate calibration.

In Fig.~\ref{fig: limits of off-res drive}(b), we further study the available range of drive detunings for $X_{\pi/2}$ gates based on cosine DRAG pulses as a function of pulse duration in both experiments and simulations. As expected, we observe that the available range of drive detunings increases with reduced gate duration roughly as  $\propto 1 / t_\mathrm{p}$. Overall, the experimentally characterized limits of drive detuning  agree well with those predicted by simulations of coherent error.  For fast gates with $t_\mathrm{p} < \SI{20}{\nano\second}$, the lower bound of the drive detuning range becomes limited by elevated leakage errors since a negative drive detuning shifts the drive frequency towards the own $f_{12}$ transition of the driven qubit. 

\subsubsection{Selection of drive detuning}

In the experiments of Secs.~\ref{sec: pulse shape engineering} and \ref{sec: demonstration of reduced bandwidth on full QPU using pulse shaping techniques}, we choose the off-resonant drive detuning $\Delta f_\mathrm{d}^\mathrm{C}$ of the control qubit before gate calibration experiments. We aim to select a drive detuning that shifts the drive frequency away from the nearest target qubit transition to mitigate crosstalk errors on the target qubit, while still allowing the implementation of high-fidelity $X_{\pi/2}$ gates on the control qubit. In practice, we set the drive detuning as $\Delta f_\mathrm{d}^\mathrm{C} = \gamma \times |\Delta f_\mathrm{d}^\mathrm{C}|$, where the sign $\gamma$ and magnitude $|\Delta f_\mathrm{d}^\mathrm{C}|$ are given by 
\begin{align}
    \gamma &= (f_{01}^\mathrm{C} - f_\mathrm{nearest}^\mathrm{T}) / |f_{01}^\mathrm{C} - f_\mathrm{nearest}^\mathrm{T}|, \label{eq: off-res. drive sign} \\
    |\Delta f_\mathrm{d}^\mathrm{C}| &= \begin{cases} \Delta f_\mathrm{d, default}^\mathrm{C},~~|f_{01}^\mathrm{C} - (f_{01}^\mathrm{T} + f_{12}^\mathrm{T}) / 2| > \Delta f_\mathrm{d, default}^\mathrm{C}, \\
    0.9 \times |f_{01}^\mathrm{C} - (f_{01}^\mathrm{T} + f_{12}^\mathrm{T}) / 2|, \textrm{ otherwise}, 
    \end{cases}
\end{align}
where $f_\mathrm{nearest}^\mathrm{T} \in \{f_{01}^\mathrm{T}, f_{12}^\mathrm{T}\}$ is the nearest target qubit transition frequency with respect to $f_{01}^\mathrm{C}$, and $\Delta f_\mathrm{d, default}^\mathrm{C}$ is the default magnitude of the drive detuning used if $f_{01}^\mathrm{C}$ is not too close to the mean of $f_{01}^\mathrm{T}$ and  $f_{12}^\mathrm{T}$. If the control qubit frequency $f_{01}^\mathrm{C}$ is  close to the middle of the straddling regime $(f_{01}^\mathrm{T} + f_{12}^\mathrm{T}) / 2$, we prefer to set the off-resonant drive frequency near $(f_{01}^\mathrm{T} + f_{12}^\mathrm{T}) / 2$ corresponding to a local minimum of crosstalk error. The factor of $0.9$ is used to avoid coalescing suppressed frequencies.

We choose the default magnitude of the drive detuning $\Delta f_\mathrm{d, default}^\mathrm{C}$ to enable a significant reduction of crosstalk-induced error on the target qubit, while still allowing robust calibration of high-fidelity $X_{\pi/2}$ gates on the control qubit. As a good rule of thumb, we advice to use a default drive detuning of $\Delta f_\mathrm{d, default}^\mathrm{C} \in [15, 20]~\mathrm{MHz} \times \SI{20}{\nano\second} / t_\mathrm{p}$. In our experiments, we use $\Delta f_\mathrm{d, default}^\mathrm{C}=\SI{18}{\mega\hertz}$ in \cref{fig: experimental demonstration of off-resonant HD DRAG}(c),   $\Delta f_\mathrm{d, default}^\mathrm{C}=\SI{22}{\mega\hertz} \times 20~\mathrm{ns} / t_\mathrm{p}$ in \cref{fig: supplementary data on single-pair duration sweep}, and $\Delta f_\mathrm{d, default}^\mathrm{C}=\SI{15}{\mega\hertz}$ in \cref{fig: hd drag on bw minimized config}. We chose a rather conservative value of the drive detuning for the demonstration on the full QPU to ensure a robust calibration across all qubits. Namely,  some pairs using CTS pulse shaping had a relatively low qubit-qubit detuning of around \SI{40}{\mega\hertz} in the optimized configuration.

\subsection{Selection of suppressed frequencies for HD DRAG}
\label{ap: selection of suppressed frequencies for HD DRAG}

Throughout the experiments in Secs.~\ref{sec: pulse shape engineering} and \ref{sec: demonstration of reduced bandwidth on full QPU using pulse shaping techniques}, we use three suppressed frequencies for HD DRAG pulses and we directly set the suppressed frequencies  based on the measured transition frequencies $f_{01}^\mathrm{T}$, $f_{12}^\mathrm{T}$, and $f_{12}^\mathrm{C}$ as $\{f_{\mathrm{s}, j}\} = \{|f_\mathrm{d}^\mathrm{C} - f_{01}^\mathrm{T}|, |f_\mathrm{d}^\mathrm{C} - f_{12}^\mathrm{T}|, |f_\mathrm{d}^\mathrm{C} - f_{12}^\mathrm{C}|\}$. Here, we motivate these choices by first establishing the approximate optimality of the chosen suppressed frequencies $\{f_{\mathrm{s}, j}\}$ followed by a comparison of HD DRAG pulses with a varying number of suppressed frequencies. 

\begin{figure}
    \centering    \includegraphics{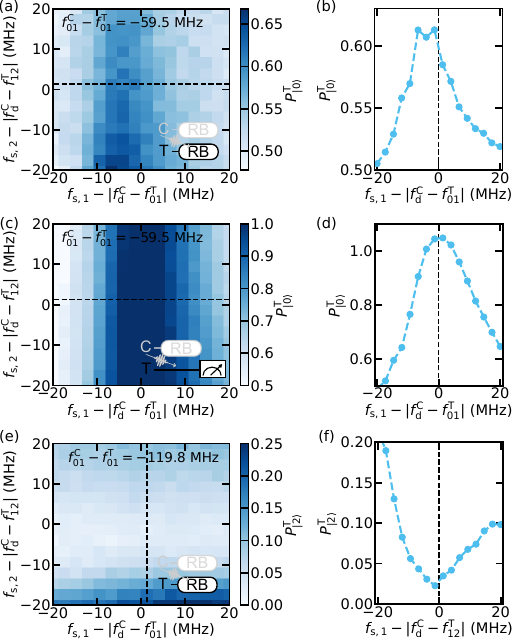}
    \caption{\textbf{Optimality of suppressed frequencies for resonant HD DRAG pulses.} (a) Ground-state probability of the target qubit after a fixed-length RB sequence applied to both qubits as a function of suppressed frequencies of the control qubit pulse targeting $f_{01}^\mathrm{T}$ and $f_{12}^\mathrm{T}$. Throughout the panels, we apply  resonant HD DRAG pulses to the control qubit and use 200 Clifford gates and 15 different randomizations for the single-qubit RB sequence. (b) Line cut corresponding to the horizontal dashed line in (a). (c) Same as (a) but applying the fixed-length RB sequence  only to the control qubit without any pulses on the target qubit. (d) Line cut corresponding to the horizontal dashed line in (c). (e) $f$-state probability of the target qubit after a fixed-length RB sequence  applied to both qubits as a function of suppressed frequencies of the control qubit pulse targeting $f_{01}^\mathrm{T}$ and $f_{12}^\mathrm{T}$. (f) Line cut corresponding to the dashed line in (e). In (a)-(d), we used the qubit pair $\mathrm{Q}_5$--$\mathrm{Q}_{11}$ with a crosstalk of $-\SI{13.9}{\decibel}$ and a qubit-qubit detuning of $f_{01}^\mathrm{C}- f_{01}^\mathrm{T} = -\SI{59.5}{\mega\hertz}$, whereas in (e)-(f) we used the qubit pair $\mathrm{Q}_{16}$-$\mathrm{Q}_{25}$ with a crosstalk of $-\SI{15.9}{\decibel}$ and a qubit-qubit detuning of $f_{01}^\mathrm{C}- f_{01}^\mathrm{T} = -\SI{119.8}{\mega\hertz}$. In (a)-(b) and (e)-(f), cosine DRAG pulses were applied to the target qubit. The vertical dashed lines in (b), (d), and (f) show the default suppressed frequencies chosen by our gate calibration procedure based on the measured transition frequencies. 
    }
    \label{fig: 2d param sweep of hd drag}
\end{figure}

\begin{figure*}
    \centering
    \includegraphics{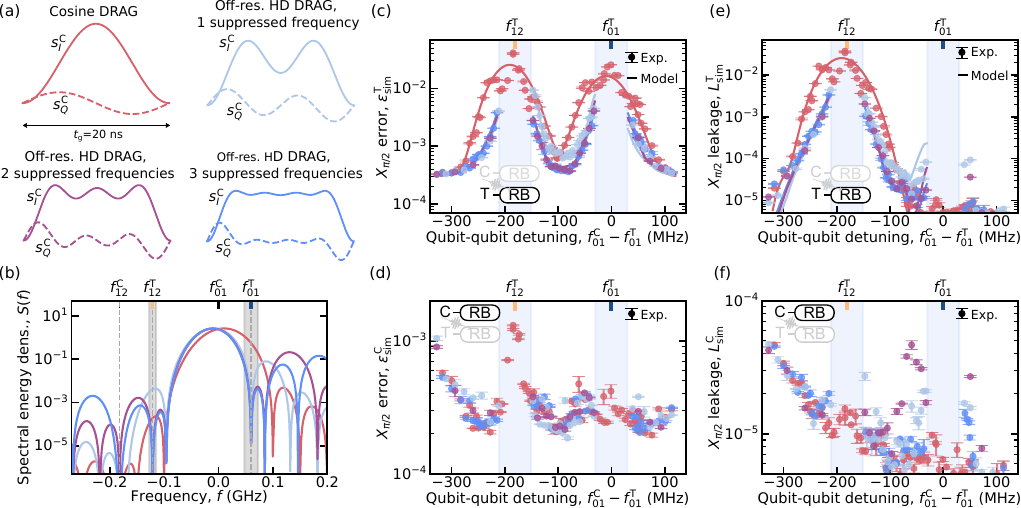}
    \caption{\textbf{Comparison of HD DRAG pulses with 1--3 suppressed frequencies as a function of qubit-qubit detuning.} (a) Example in-phase (solid line) and quadrature (dashed) envelopes for the studied pulse shapes of the  control qubit, including resonant cosine DRAG (red), and off-resonant HD DRAG suppressing 1-3 nearest transitions (light blue, purple, or blue) in the set $\{f_{01}^\mathrm{T}, f_{12}^\mathrm{T}, f_{12}^\mathrm{C}\}$. (b) Spectral energy density $S(f)$ for the pulses in (a), with the harmful transition frequencies $\{f_{01}^\mathrm{T}, f_{12}^\mathrm{T}, f_{12}^\mathrm{C}\}$ denoted by vertical dashed gray lines and the typical Rabi splitting of the target qubit for 20-ns $X_{\pi/2}$ pulses shown with  shaded gray regions. (c) Experimental simultaneous $X_{\pi/2}$ error of the target qubit $\varepsilon_\mathrm{sim}^\mathrm{T}$  measured from leakage RB (markers)  as a function of qubit-qubit detuning $f_{01}^\mathrm{C} - f_{01}^\mathrm{T}$ when applying the pulse shapes in (a) for the control qubit and a conventional cosine DRAG pulse on the target qubit. Solid lines show the corresponding prediction of the analytical gate error model in~\cref{eq:mt_eps_av random phase} without any fitting parameters. (d) Same as (c) but showing the simultaneous gate error of the control qubit  $\varepsilon_\mathrm{sim}^\mathrm{C}$. (e) Experimental simultaneous leakage error of the target qubit $L_\mathrm{sim}^\mathrm{T}$ for $X_{\pi/2}$ gates measured from leakage RB (markers) as a function of qubit-qubit detuning $f_{01}^\mathrm{C} - f_{01}^\mathrm{T}$ for the same pulse shapes. Solid lines show the corresponding prediction of the analytical leakage error model without any fitting parameters. (f) Same as (e) but showing the simultaneous leakage error of the control qubit $L_\mathrm{sim}^\mathrm{C}$.   Throughout the panels, the data is collected for the highest-crosstalk pair $\mathrm{Q}_5$--$\mathrm{Q}_{11}$ of our device, with a crosstalk of $-\SI{13.9}{\decibel}$ using $X_{\pi/2}$ pulses with $t_\mathrm{g} = t_\mathrm{p} = \SI{20}{\nano\second}$. For the off-resonant HD DRAG pulses, we use a default drive detuning of $\Delta f_\mathrm{d, default}^\mathrm{C}=\SI{18}{\mega\hertz}$. In panels (c)-(f), the shaded blue regions show the range of qubit-qubit detunings, where the calibration of 20-ns HD DRAG pulses may not be robust.  
    }
    \label{fig: hd drag with varying number of suppressed frequencies}
\end{figure*}

In Fig. \ref{fig: 2d param sweep of hd drag}, we sweep the suppressed frequencies $f_{\mathrm{s}, 1}$ and $f_{\mathrm{s}, 2}$ targeting $f_{01}^\mathrm{T}$ and $f_{12}^\mathrm{T}$ to gain insight on the optimal values and sensitivity of these parameters. For simplicity, we apply a resonant 20-ns HD DRAG pulse on the control qubit, and a resonant 20-ns cosine DRAG pulse on the target qubit. We first study the error on the qubit pair $\mathrm{Q}_5$--$\mathrm{Q}_{11}$ with a crosstalk of $-\SI{13.9}{\decibel}$ and a qubit-qubit detuning of $f_{01}^\mathrm{C}- f_{01}^\mathrm{T} = -\SI{59.5}{\mega\hertz}$, indicating that $f_{01}^\mathrm{T}$ is the nearest harmful transition  and the gate error of the target qubit is dominated by errors within the computational subspace. Hence, we measure the ground-state probability of the target qubit after simultaneous fixed-length RB sequences with 200 Clifford gates on both qubits to obtain a proxy for the RB gate error. In the two-dimensional sweep of suppressed frequencies $f_{\mathrm{s}, 1}$ and $f_{\mathrm{s}, 2}$, we re-evaluate the pulse envelopes of the control qubit for each pair of suppressed frequencies, while ensuring that the in-phase envelope integrates to a given value and the DRAG coefficient stays constant. Based on the 2D and 1D parameter sweeps shown in \cref{fig: 2d param sweep of hd drag}(a) and (b), the gate error of the target qubit is significantly reduced by suppressing the spectrum near $f_{01}^\mathrm{T}$ and   the minimum of the gate error is observed approximately at $f_{\mathrm{s}, 1} = |f_{01}^\mathrm{T} - f_\mathrm{d}^\mathrm{C}|$. This motivates our choice of setting the suppressed frequencies directly based on the measured transition frequencies, without further calibration experiments.

Interestingly, we observe that the HD DRAG pulse on the control qubit mitigates gate errors of the target qubit much more effectively if the target qubit is idling during the fixed-length RB sequence on the control qubit as shown in \cref{fig: 2d param sweep of hd drag}(c) and (d). Setting $f_{\mathrm{s}, 1} = |f_{01}^\mathrm{T} - f_\mathrm{d}^\mathrm{C}|$ provides nearly perfect mitigation of excitation errors of the idling target qubit since there is no Rabi splitting in the absence of a drive pulse on the target qubit. See Sec.~\ref{ap: HD DRAG for idling error mitigation} for further discussion.

We also study the optimal choice of $f_{\mathrm{s}, 2}$ targeting $f_{12}^\mathrm{T}$ to mitigate leakage errors of the target qubit. To this end, we carried out experiments on the qubit pair $\mathrm{Q}_{16}$-$\mathrm{Q}_{25}$ with a crosstalk of $-\SI{15.9}{\decibel}$ and a qubit-qubit detuning of $f_{01}^\mathrm{C}- f_{01}^\mathrm{T} = -\SI{119.8}{\mega\hertz}$, indicating that $f_{12}^\mathrm{T}$ is the nearest harmful transition and the gate error of the target qubit is dominated by leakage errors. Hence, we measure the $f$-state probability of the target qubit after simultaneous fixed-length RB sequences with 200 Clifford gates on both qubits to obtain a proxy of the leakage error.  Based on the 2D and 1D parameter sweeps shown in \cref{fig: 2d param sweep of hd drag}(e) and (f), the gate error of the target qubit is significantly reduced by suppressing the spectrum near $f_{12}^\mathrm{T}$, with the minimum of the leakage error indeed observed approximately at $f_{\mathrm{s}, 2} = |f_{12}^\mathrm{T} - f_\mathrm{d}^\mathrm{C}|$. This provides further support for our choice of selecting the suppressed frequencies directly based on the measured transition frequencies.

In Fig.~\ref{fig: hd drag with varying number of suppressed frequencies}, we investigate the gate error of the target qubit and the control qubit when varying the number of  suppressed frequencies in the HD DRAG pulses applied to the control qubit. In addition to a cosine DRAG pulse and an HD DRAG pulse with three suppressed frequencies considered in the main text, we also compare the gate error for HD DRAG pulses with either 1 or 2 suppressed frequencies as visualized in Fig.~\ref{fig: hd drag with varying number of suppressed frequencies}(a). In these cases, we choose the suppressed frequencies to mitigate the nearest transition frequencies depending on the qubit-qubit detuning as illustrated by the example energy spectral densities in Fig.~\ref{fig: hd drag with varying number of suppressed frequencies}(b). Based on Fig.~\ref{fig: hd drag with varying number of suppressed frequencies}(c), we observe that a single suppressed frequency is not sufficient to mitigate crosstalk-induced errors in the straddling regime $\alpha^\mathrm{T}/(2\pi) < f_{01}^\mathrm{C} - f_{01}^\mathrm{T} < 0$. In the straddling regime, the spectrum of the control qubit pulse should be suppressed at both $f_{01}^\mathrm{T}$ and $f_{12}^\mathrm{T}$ to efficiently mitigate crosstalk errors of the target qubit especially near the middle of the straddling regime. Neglecting the suppression of $f_{12}^\mathrm{T}$ leads to elevated leakage errors of the target qubit, see Fig.~\ref{fig: hd drag with varying number of suppressed frequencies}(e). 

By further investigating the simultaneous gate error and leakage of the control qubit in Figs.~\ref{fig: hd drag with varying number of suppressed frequencies}(d) and (f), we observe that the leakage error of the control qubit is elevated from $L_\mathrm{sim}^\mathrm{C} \sim 10^{-5}$ to $L_\mathrm{sim}^\mathrm{C} \sim 5 \times 10^{-5}$ if we only suppress $f_{01}^\mathrm{T}$ and $f_{12}^\mathrm{T}$ while neglecting $f_{12}^\mathrm{C}$. Note that we apply the DRAG-L strategy to choose the DRAG coefficient to minimize leakage in both cases. Hence, already for gate fidelities approaching 99.99\%, it is important to also mitigate leakage errors of the control qubit by suppressing  $f_{12}^\mathrm{C}$. For these reasons, we choose to suppress all three transition frequencies $f_{01}^\mathrm{T}$, $f_{12}^\mathrm{T}$, and $f_{12}^\mathrm{C}$ to mitigate crosstalk-induced errors and leakage across a wide range of qubit-qubit detunings in the experiments considered in the main text. In certain special cases, such as outside of the straddling regime, it may be sufficient to only suppress the nearest harmful transition frequency, which may enable the application of HD DRAG pulses for shorter gate durations, while still limiting the leakage error of the control qubit. 

Increasing the number of suppressed frequencies beyond three may lead to rapidly oscillating envelope functions or high peak amplitudes for fast gates and densely packed suppressed frequencies. Relaxing the assumption of symmetrically suppressed frequencies or increasing either the gate duration or the suppressed frequencies $f_{\mathrm{s}, j}$ may enable the suppression of a higher number of frequencies, while avoiding high peak amplitudes and rapidly oscillating components. Furthermore, the FAST DRAG technique introduced in \cite{hyyppa2024reducing} may enable efficient crosstalk and leakage mitigation at shorter gate durations than possible with HD DRAG thanks to configurable pulse regularization. However, this comes at the cost of additional pulse hyperparameters that need to be set either heuristically or with numerical optimization.

\subsection{Calibration of single-qubit gate parameters}
\label{ap: single-qubit gate calibration}

\begin{figure*}
    \centering
\includegraphics{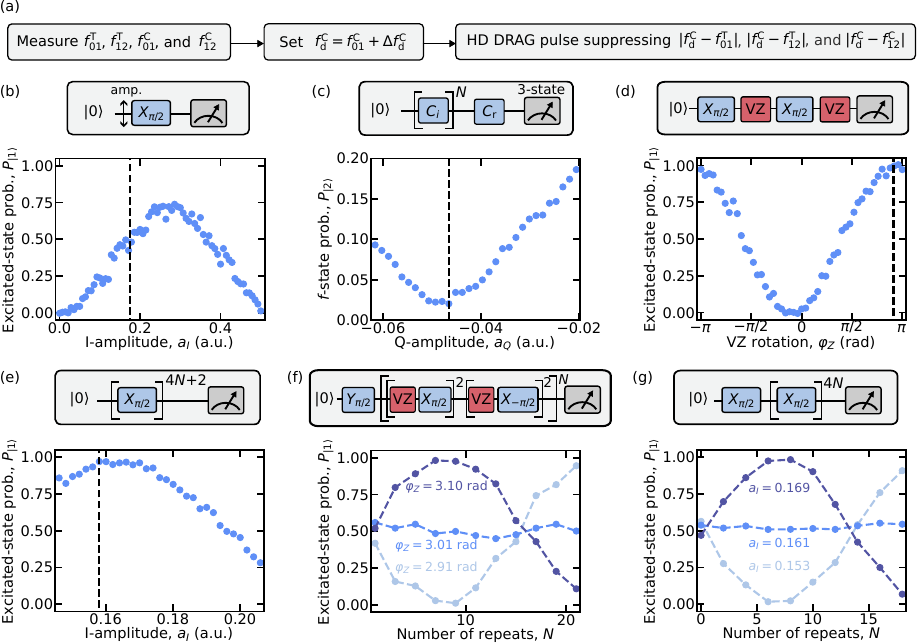}
    \caption{\textbf{Calibration of off-resonant $X_{\pi/2}$ pulses.} (a) For a high-crosstalk qubit pair, we first determine the drive frequency and suppressed frequencies of the HD DRAG pulse applied to the control qubit by measuring the transition frequencies $f_{01}^\mathrm{T}$, $f_{12}^\mathrm{T}$, $f_{01}^\mathrm{C}$, and $f_{12}^\mathrm{C}$. (b) Initial in-phase amplitude $a_I$ calibration based on an amplitude Rabi experiment. (c) A sweep of fixed-point leakage RB over the quadrature envelope amplitude $a_Q \propto \beta$ to minimize leakage errors. (d) Initial calibration of virtual Z angle $\varphi_z$. (e) Rough error amplification experiment to mitigate overrotation errors.  (f) Error amplification measurement to mitigate phase errors. (g) More precise error amplification measurement to mitigate overrotation errors. Some of the experiments  were iterated 1-3 times to reduce coherent errors to negligible levels. }
    \label{fig: supplementary x90 calibration}
\end{figure*}

Here, we describe our procedure for calibrating the parameters of an off-resonant control pulse implementing a single-qubit $X_{\pi/2}$-gate. We assume the control pulse to take the form $v(t) = s_I (t) \cos(\omega_\mathrm{d} t + \varphi) +  s_Q (t) \sin(\omega_\mathrm{d} t + \varphi) $, where $s_I(t) = \Omega_{ii} \times a_I \times s_I ^\mathrm{norm}(t)$ with $a_I$ being the normalized in-phase amplitude and $\Omega_{ii}$ being the Rabi rate per unit amplitude,  $s_Q(t) = -\beta\dot{s}_I(t)/\alpha$ with $\beta$ being the DRAG coefficient, $\omega_\mathrm{d}=2\pi f_\mathrm{d}$ is the angular drive frequency, and the phase $\varphi$ is advanced by $\varphi_z/2$ before and after each gate to implement virtual Z rotations~\cite{mckay2017efficient}. In the experiments discussed in Secs.~\ref{sec: pulse shape engineering} and \ref{sec: demonstration of reduced bandwidth on full QPU using pulse shaping techniques}, we use the DRAG-L calibration strategy described in Appendix G of Ref.~\cite{hyyppa2024reducing} with slight modifications as described below to ensure robust calibration for off-resonant control pulses.

We summarize the key steps of our calibration in \cref{fig: supplementary x90 calibration} using a single high-crosstalk pair as an example.  First, we experimentally determine the transition frequencies $f_{01}^\mathrm{T}$, $f_{12}^\mathrm{T}$, $f_{01}^\mathrm{C}$, and $f_{12}^\mathrm{C}$ as shown in \cref{fig: supplementary x90 calibration}(a). Based on the measured transition frequencies, we select an appropriate off-resonant drive detuning for the control qubit $\Delta f_\mathrm{d}^\mathrm{C}$ and the associated drive frequency $f_\mathrm{d}^\mathrm{C} = f_{01}^\mathrm{C} + \Delta f_\mathrm{d}^\mathrm{C}$ as explained in Sec.~\ref{ap: strongly off-resonant control pulses}. For the target qubit, the drive frequency is set as the qubit frequency $f_{01}^\mathrm{T}$ unless the target qubit acts as a control qubit in another high-crosstalk pair. If using an HD DRAG pulse on the control qubit, the suppressed frequencies of the in-phase envelope are chosen as $\{f_{\mathrm{s}, j} \}_{j=1}^{3} = \{|f_\mathrm{d}^\mathrm{C} - f_{01}^\mathrm{T}|, |f_\mathrm{d}^\mathrm{C} - f_{12}^\mathrm{T}|, |f_\mathrm{d}^\mathrm{C} - f_{01}^\mathrm{C}|  \}$ without any further calibration experiments as motivated in Sec.~\ref{ap: selection of suppressed frequencies for HD DRAG}.

For both qubits, the pulse parameters $\{a_I, \beta, \varphi_z \}$ are calibrated such that the amplitude $a_I$ controls the rotation angle of the gate, the DRAG coefficient $\beta$ minimizes leakage errors, and the virtual Z angle $\varphi_z$ counteracts any accumulated phase errors during the $X_{\pi/2}$ gate. We use similar gate calibration experiments for cosine DRAG and HD DRAG pulse shapes. First, a rough estimate of the in-phase amplitude $a_I$ is obtained from an amplitude Rabi experiment, in which we search for the smallest amplitude resulting in $P_{|1\rangle} = 0.5$ as illustrated in \cref{fig: supplementary x90 calibration}(b). Subsequently, an initial estimate for the DRAG coefficient is obtained in a fixed-point leakage RB experiment, in which we sweep the amplitude $a_Q\propto \beta$ of the quadrature envelope to minimize the  leakage population of the second excited state $|2\rangle$ after a fixed number of Clifford gates, see \cref{fig: supplementary x90 calibration}(c). Next, an initial estimate for the virtual Z angle is obtained using the circuit depicted in  \cref{fig: supplementary x90 calibration}(d), which is similar to that used in Ref.~\cite{wang2025suppressing}.

Subsequently, the parameter estimates are refined using iterated error amplification experiments. To improve the estimate of the in-phase amplitude $a_I$, we carry out error amplification experiments with repeated $X_{\pi/2}$-gates to amplify overrotation errors. For strongly off-resonant drive pulses, the initial calibration of $\beta$ and $\varphi_z$ may affect the optimal value of the amplitude $a_I$. As a result, we first apply a rough amplitude error amplification experiment as  shown in \cref{fig: supplementary x90 calibration}(e), which is followed by a more precise error amplification experiment in subsequent iterations as shown  \cref{fig: supplementary x90 calibration}(g). Leakage errors are further minimized by an additional round of the fixed-point leakage RB sweep with a finer sweep range. Phase errors are mitigated by refining the virtual Z angle $\varphi_\mathrm{z}$ using an error amplification experiment based on the circuit depicted in \cref{fig: supplementary x90 calibration}(f). The initial $Y_{\pi/2}$ and $X_{\pi/2}$ gates in Figs.~\ref{fig: supplementary x90 calibration}(f) and (g) render the excited state probability $P_{|1\rangle}$ linearly sensitive to phase errors and overrotation errors, respectively, which enables more precise calibration~\cite{hyyppa2024reducing, gao2025ultra}.  We iterate the error amplification experiments for $\{a_I, \beta, \varphi_z \}$  a couple of times to ensure low coherent errors and a robust calibration even with relatively large drive detunings $\Delta f_\mathrm{d}^\mathrm{C}$. 
For resonant control pulses, some of the iterations and initial calibration experiments, such as those in Figs.~\ref{fig: supplementary x90 calibration}(c) and (d), may be omitted since approximate parameter values may be reliably inferred prior to the calibration.

\subsection{Mitigation of crosstalk errors for an idling target qubit using HD DRAG}
\label{ap: HD DRAG for idling error mitigation}

\begin{figure}
    \centering
    \includegraphics{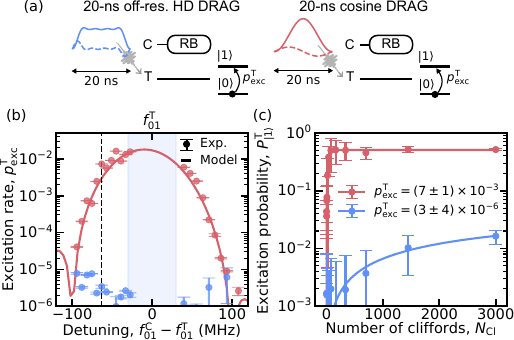}
    \caption{\textbf{Reduction of excitation rate for an idling target qubit using HD DRAG.} (a) We measure the excitation rate $p_\mathrm{exc}^\mathrm{T}$ of an idling target qubit when applying a single-qubit RB sequence on the control qubit based on either 20-ns off-resonant HD DRAG pulses (blue) or 20-ns cosine DRAG pulses (red). (b) Measured excitation rate $p_\mathrm{exc}^\mathrm{T}$ of the target qubit (markers) as a function of qubit-qubit detuning, with the error bars denoting the $1\sigma$ uncertainty of mean based on 3 repeated experiments. We show the prediction of an analytical model for cosine DRAG (solid red line)  based on the first three terms of Eq.~\eqref{eq:mt_eps_av random phase}. For HD DRAG, the model  predicts no excitation error for ideally calibrated suppressed frequencies. The shaded blue region shows a range of qubit-qubit detunings, for which the calibration of the 20-ns HD DRAG pulse may not be robust. The dashed lines shows the example qubit-qubit detuning $f_{01}^\mathrm{C} - f_{01}^\mathrm{T} = -\SI{63}{\mega\hertz}$ used in (c).  (c) Measured excitation probability $P_{|1\rangle}^\mathrm{T}$ of the target qubit (markers) as a function of the number of Clifford gates in the RB sequence. The error bars denote the standard deviation of the excited state probability across different random Clifford sequences.  Solid lines show exponential fits used to extract the excitation rate~$p_\mathrm{exc}^\mathrm{T}$. The data is for the same high-crosstalk pair as in \cref{fig: experimental demonstration of off-resonant HD DRAG} with a crosstalk of -\SI{13.9}{\decibel}.
    } 
    \label{fig: excitation rate}
\end{figure}

During simultaneous gates, the effectiveness of HD DRAG pulse shaping is hampered by the Rabi splitting of the target qubit as discussed in Section~\ref{sec: off-resonant control pulses} and Appendix~\ref{ap: interpretation of analytical error model}.
For an idling target qubit, HD DRAG pulses can ideally suppress all the terms of the analytical model in Eq.~\eqref{eq:mt_eps_av random phase}, leaving only a minor error induced by the ac Stark shift of the target qubit.
Therefore, our analytical and numerical results in Figs.~\ref{fig: concept of higher-derivative drag}(f) and \ref{fig: model_vs_simulation}(b) suggest that HD DRAG pulse shaping could provide a considerably larger reduction of crosstalk error if the target qubit is idling instead of driven. Such a scenario is relevant for many quantum circuits that apply simultaneous single-qubit gates only to a subset of the qubits, while the rest of the qubits  are idling. In this section, we  experimentally investigate the effectiveness of HD DRAG pulse shaping for an idling target qubit. 

\begin{figure*}
    \centering
    \includegraphics{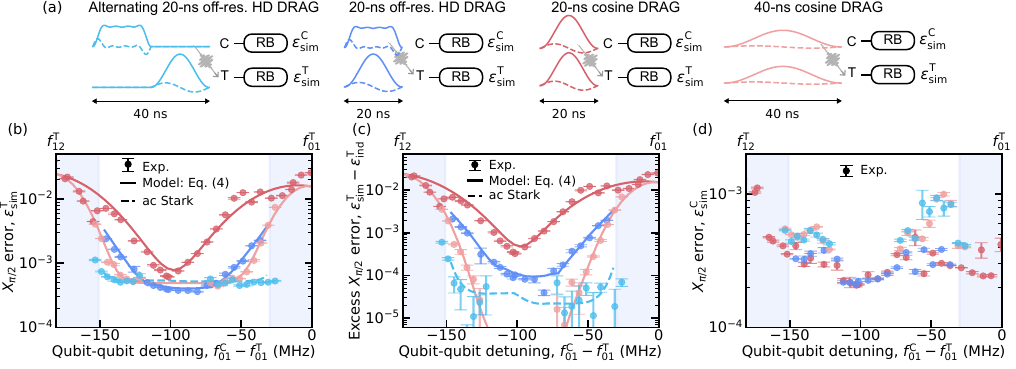}
    \caption{\textbf{Comparison of $X_{\pi/2}$ error for an idling and driven target qubit using HD DRAG.} (a) We use RB to benchmark crosstalk errors for an idling target qubit by applying a 20-ns off-resonant HD DRAG pulse on the control qubit and a 20-ns cosine DRAG pulse on the target qubit in an alternating manner (light blue), resulting in an effective total gate duration of \SI{40}{\nano\second}. This is compared against three alternatives of fully simultaneous gates, including a 20-ns off-resonant HD DRAG pulse on the control qubit and a 20-ns cosine DRAG pulse on the target qubit (dark blue), 20-ns cosine DRAG pulses on both qubits (dark red), and 40-ns cosine DRAG pulses on both qubits (light red). (b) Measured $X_{\pi/2}$ error of the target qubit (markers) based on leakage RB experiments as a function of qubit-qubit detuning for the pulse shaping scenarios of (a). The solid lines show the analytical model in Eq.~\eqref{eq:mt_eps_av random phase} for the three fully simultaneous scenarios. For 
    alternating HD DRAG and cosine DRAG pulses, we present the  approximate error caused by the ac Stark shift of the target qubit based on Eq.~\eqref{ac stark error} (dashed light blue line). The measured individual error has been added to the analytical model and to the approximation of the error induced by the ac Stark shift.  (c) Same as (b) but showing the excess crosstalk-induced $X_{\pi/2}$ error $\varepsilon_\mathrm{sim}^\mathrm{T} - \varepsilon_\mathrm{ind}^\mathrm{T}$ of the target qubit. (d) Measured $X_{\pi/2}$ error of the control qubit as a function of qubit-qubit detuning. In panels (b)-(d), the shaded blue regions denote qubit-qubit detunings, for which the calibration of the 20-ns HD DRAG pulse is not necessarily robust. The error bars present the $1\sigma$ uncertainty of the mean based on 3--6 repeated leakage RB experiments. The data is for the same high-crosstalk pair as in \cref{fig: experimental demonstration of off-resonant HD DRAG} with a crosstalk of -\SI{13.9}{\decibel}.
    } 
    \label{fig: alternating hd drag }
\end{figure*}

\subsubsection{Effect of HD DRAG pulse shaping on the excitation rate of an idling target qubit}

First, we study the impact of pulse shaping on the excitation rate of an idling target qubit using the highest-crosstalk pair $\mathrm{Q}_5$--$\mathrm{Q}_{11}$ of our device with $C_{5, 11} = -\SI{13.9}{\decibel}$. Similarly to Refs.~\cite{wang2025suppressing} and \cite{matsuda2025selective},  we apply random single-qubit RB sequences to the control qubit and measure the excitation rate of the target qubit as illustrated in Fig.~\ref{fig: excitation rate}(a). To estimate the excitation rate per gate $p_\mathrm{exc}^\mathrm{T}$, we first measure the excited state probability $P_{|1\rangle}^\mathrm{T}$ of the target qubit as a function of the number of Clifford gates in the RB sequence. 
Subsequently, we fit an exponential model $P_{|1\rangle}(N_\mathrm{Cl}) = A_\mathrm{e} + B_\mathrm{e} \lambda_\mathrm{e}^{N_\mathrm{Cl}}$ and compute the excitation rate per gate as $p_\mathrm{exc}^\mathrm{T} = P_{|1\rangle}(N_\mathrm{Cl} = 1) / N_\mathrm{g} = -B_\mathrm{e}(1 - \lambda_\mathrm{e})/N_\mathrm{g}$, where $N_\mathrm{g}=2.167$ is the average number of $\pi/2$-pulses per Clifford in our decomposition. 

In  Fig.~\ref{fig: excitation rate}(b), we compare the measured excitation rate for 20-ns HD DRAG and cosine DRAG pulses as a function of qubit-qubit detuning around the $f_{01}^\mathrm{C} \approx f_{01}^\mathrm{T}$ resonance. For the HD DRAG pulse, we use an off-resonant drive detuning of $|\Delta f_\mathrm{d}^\mathrm{C}| = \SI{10}{\mega\hertz}$ on the control qubit, though a resonant drive would suffice such as well in case the target qubit is idling. Near the qubit-qubit resonance, HD DRAG reduces the excitation rate by over three orders of magnitude as shown in Fig.~\ref{fig: excitation rate}(b). Figure~\ref{fig: excitation rate}(c) illustrates an example measurement of the excitation rate  for  $f_{01}^\mathrm{C} - f_{01}^\mathrm{T} \approx -\SI{63}{\mega\hertz}$, where the excitation rate is reduced from $p_\mathrm{exc}^\mathrm{T} = (7 \pm 1) \times 10^{-3}$ using cosine DRAG to $p_\mathrm{exc}^\mathrm{T} = (3 \pm 4) \times 10^{-6}$ using HD DRAG. For cosine DRAG pulses, the excitation rate  reduces to the level  $p_\mathrm{exc}^\mathrm{T} \lesssim 10^{-5}$  only for qubit-qubit detunings approaching \SI{100}{\mega\hertz}. Thus, HD DRAG pulse shaping enables us to reduce the qubit-qubit detuning  by a factor of three without increasing the excitation rate.  We also observe that the analytical model based on the first three terms of Eq.~\eqref{eq:mt_eps_av random phase} provides a good agreement with the measured values of the excitation rate for cosine DRAG. For HD DRAG, the analytical model predicts no errors for ideally chosen suppressed frequencies, though the model ignores the ac Stark shift of the target qubit due to the control qubit pulse.

\subsubsection{Effect of HD DRAG pulse shaping on the gate error of an idling target qubit}

When using spectral shaping techniques, such as HD DRAG, the excitation rate of an idling target qubit is, however, not representative of the average gate error for an idling target qubit let alone the gate error of a driven target qubit during simultaneous $X_{\pi/2}$ gates. This is because the excitation rate ignores the ac Stark shift impacting the gate error of an idling target qubit and further the Rabi splitting terms affecting a driven target qubit. To demonstrate this difference, we use RB experiments to estimate the crosstalk-induced gate error of the target qubit in case the target qubit is either idling or driven in~\cref{fig: alternating hd drag }. For measuring the crosstalk-induced gate error on a driven target qubit, we use a simultaneous RB experiment, in which we simultaneously apply a 20-ns off-resonant HD DRAG pulse on the control qubit and a 20-ns cosine DRAG pulse on the target qubit similarly to Sec.~\ref{sec: experimental demonstration on a high-crosstalk pair}. To characterize the crosstalk-induced gate error for an idling target qubit, we perform single-qubit RB sequences on both qubits, but we instead apply the 20-ns HD DRAG pulse on the control qubit before the 20-ns cosine DRAG pulse on the target qubit as shown in Fig.~\ref{fig: alternating hd drag }(a). Using alternating HD DRAG and cosine DRAG pulses, the effective gate duration is increased to \SI{40}{\nano\second}. As a baseline, we also benchmark the gate error for simultaneous 20-ns or 40-ns $X_{\pi/2}$ gates based on cosine DRAG pulses. 

The key results are shown in Fig.~\ref{fig: alternating hd drag }(b). Using alternating HD DRAG and cosine DRAG pulses, the simultaneous $X_{\pi/2}$ error of the target qubit stays approximately constant at $\varepsilon_\mathrm{sim}^\mathrm{T} \approx 5\times 10^{-4}$ across  the  available qubit-qubit detuning range down to a  detuning of \SI{30}{\mega\hertz}, where the calibration of the 20-ns HD DRAG pulse is still viable. As shown in Fig.~\ref{fig: alternating hd drag }(c), the crosstalk-induced excess error $\varepsilon_\mathrm{sim} - \varepsilon_\mathrm{ind}$ is reduced below $1\times 10^{-4}$  across almost the entire detuning range despite the high magnitude of crosstalk $C_{5, 11} = -\SI{13.9}{\decibel}$. As $f_{01}^\mathrm{C}$ approaches either $f_{01}^\mathrm{C}$ or $f_{12}^\mathrm{C}$, alternating HD DRAG and cosine DRAG pulses reduce the crosstalk-induced excess error of the target qubit by over an order of magnitude compared to simultaneous 40-ns cosine DRAG pulses or simultaneous 20-ns HD DRAG and cosine DRAG pulses. 
Compared to simultaneous 20-ns cosine DRAG pulses, the alternating 20-ns pulses reduce the excess crosstalk error $\varepsilon_\mathrm{sim} - \varepsilon_\mathrm{ind}$ by up to two orders of magnitude in line with the simulations of Fig.~\ref{fig: concept of higher-derivative drag}(f).

Ideally, the approach with alternating pulses fully mitigates the cross driving errors modeled in Eq.~\eqref{eq:mt_eps_av random phase} owing to the spectral minima of HD DRAG pulses at the relevant transition frequencies and the absence of the Rabi splitting of the target qubit during the HD DRAG pulse. Thus, the remaining crosstalk-induced error is dominated by the ac Stark shift of the target qubit, which is supported by the good match between the measured crosstalk error $\varepsilon_\mathrm{sim} - \varepsilon_\mathrm{ind}$ and the predicted gate error induced by ac Stark shift in Eq.~\eqref{ac stark error}, see Fig.~\ref{fig: alternating hd drag }(c). For the other pulse shaping techniques, the cross driving errors modeled in Eq.~\eqref{eq:mt_eps_av random phase} dominate the gate error of the target qubit as shown in Fig.~\ref{fig: alternating hd drag }(c), and the ac Stark shift is only a minor correction with an estimated error $\lesssim 10^{-4}$ across the studied range of qubit-qubit detunings. 

The disadvantage of alternating HD DRAG and cosine DRAG pulses is an effective increase of the gate duration by a factor of two. As a result, the gate error of the control qubit is approximately doubled and equal to that of simultaneous 40-ns cosine DRAG pulses as shown in Fig.~\ref{fig: alternating hd drag }(d). As shown in Fig.~\ref{fig: alternating hd drag }(b), the gate error of the target qubit for alternating pulses is also higher compared to simultaneous 20-ns pulses near the middle of the straddling regime, where the crosstalk-induced errors are low and incoherent errors dominate. Thus, alternating HD DRAG and cosine DRAG pulses are expected to provide a benefit on high-crosstalk qubit pairs in frequency-crowded configurations with small qubit-qubit detunings $t_\mathrm{g} |f_{01}^\mathrm{C} - f_\mathrm{nearest}^\mathrm{T}| \sim 1$. In such scenarios, our results demonstrate that alternating HD DRAG and cosine DRAG pulses may outperform twice as long cosine DRAG pulses in terms of crosstalk errors and even compensate for the longer effective gate duration compared to simultaneous HD DRAG and cosine DRAG pulses.

Overall, we have experimentally demonstrated that HD DRAG pulse shaping enables orders-of-magnitude more effective crosstalk error reduction if the target qubit idles instead of being simultaneously driven with the control qubit. This confirms the predictions of our analytical model and numerical simulations. Our experimental results also imply that the gate error and excitation rate of an idling target qubit are not necessarily representative of the gate performance during simultaneous gates when using spectral shaping techniques, such as HD DRAG or those in Refs.~\cite{vesterinen2014mitigating, wang2025suppressing, matsuda2025selective}.

\section{Leakage-aware benchmarking of single-qubit gate errors}
\label{ap: benchmarking of single-qubit gate errors}

In addition to standard single-qubit RB, we utilize leakage RB~\cite{chen2016measuring, wood2018quantification} and purity RB~\cite{wallman2015estimating, feng2016estimating} to estimate the total gate error, the leakage error, the incoherent error and the coherent error of single-qubit gates for some of the experimental demonstrations in Sections~\ref{sec: experimental demonstration on a high-crosstalk pair} and \ref{sec: demonstration of reduced bandwidth on full QPU using pulse shaping techniques}. To robustly estimate the total gate error using leakage RB, we adjust the analysis proposed in Ref.~\cite{wood2018quantification} to avoid fitting a sum of two exponential functions. Furthermore, we account for leakage errors in the analysis of purity RB to precisely measure the incoherent error even in the presence of non-negligible leakage. 
In this section, we first summarize the key steps in our analysis of leakage RB and purity RB. This is followed by a motivation of our analysis based on a rate equation model of a qutrit. 

\subsection{Analysis of leakage RB and purity RB}
\label{ap: analysis of leakage RB and purity RB}

For leakage RB and purity RB, we decompose each single-qubit Clifford gate  into native gates from the set $\left\{I, X_{\pi/2},X_{-\pi/2}, Y_{\pi/2}, Y_{-\pi/2}\right\}$ with an average number of $N_\mathrm{g}=2.167$ $\pi/2$-pulses per Clifford  similarly to Sec.~\ref{sec: experimental relation between crosstalk and 1qb gate error}. In our implementation of leakage RB and purity RB, each Clifford gate sequence is followed by a three-state readout operation, and the state probabilities are corrected with the measured assignment matrix as in, e.g., Refs.~\cite{mckay2017efficient, hyyppa2024reducing}.  This allows us to estimate the leakage per $X_{\pi/2}$ gate by first fitting an exponential function to the $|2\rangle$ state probability as  $P_{|2\rangle}(N_\mathrm{Cl}) = A_\mathrm{f}  + B_\mathrm{f}\lambda_1^{N_\mathrm{Cl}}$, where $N_\mathrm{Cl}$ denotes the number of Clifford gates, and $\{A_\mathrm{f}, B_\mathrm{f}, \lambda_1\}$ are fitting parameters. Then, the leakage per gate is evaluated as $L_\mathrm{g}=L_\mathrm{Cl}/N_\mathrm{g} = -B_\mathrm{f}(1-\lambda_1)/N_\mathrm{g}$. For the leakage estimation, we utilize the amplitude $B_\mathrm{f}$  of the fitted exponential instead of the offset $A_\mathrm{f}$ to obtain a more robust estimate of the steady-state leakage population in the presence of any minor readout drifts. 

To further estimate the total error per gate from leakage RB, we fit an exponential model  $\tilde{P}_{|0\rangle}(N_\mathrm{Cl})  =  A_\mathrm{g}  + B_\mathrm{g}\lambda_2^{N_\mathrm{Cl}}$ to $\tilde{P}_{|0\rangle} = P_{|0\rangle} + P_{|2\rangle}/2$ followed by a computation of the gate error as $\varepsilon_\mathrm{g} = (1-\lambda_2)/(2N_\mathrm{g}) + L_\mathrm{g}/2$.  We avoid fitting a sum of two exponential functions to  $P_{|0\rangle}(N_\mathrm{Cl})$ as in Refs.~\cite{wood2018quantification, werninghaus2021leakage}  since $\tilde{P}_{|0\rangle}(N_\mathrm{Cl})$ decays as a single exponential as motivated by a rate equation model derived in~\cref{ap: motivation of analysis using a rate equation model}. This improves the robustness of estimating the total gate error in the presence of non-negligible leakage errors. Similarly to Refs.~\cite{wood2018quantification, chen2025randomized}, we define the total gate error of an error channel $\Lambda(\rho)$ as the average gate infidelity over initial states $|\psi_\mathrm{C}\rangle$ in the computational subspace, i.e., $\varepsilon = 1 - \int \mathrm{d} \psi_\mathcal{C} \langle \psi_\mathcal{C}|\Lambda(|\psi_\mathcal{C}\rangle \langle \psi_\mathcal{C}|)|\psi_\mathcal{C} \rangle$. Thus, the total gate error is equivalent to a sum of  errors in the computational subspace and leakage to higher levels~\cite{chen2025randomized}.

In purity RB, we append each Clifford gate sequence with a final gate from the set $\{Y_{-\pi/2}, X_{\pi/2}, I\}$ to measure $ \langle \sigma_x\rangle$, $\langle \sigma_y\rangle$ and $\langle \sigma_z\rangle$, respectively, for each Clifford gate sequence. Subsequently, we compute the normalized purity $P_\mathrm{norm} = \langle \sigma_x\rangle^2 +  \langle \sigma_y\rangle^2  +  \langle \sigma_z\rangle^2 $ for each sequence and average over random Clifford sequences of same length to obtain $\bar{P}_\mathrm{norm}$. To estimate the incoherent error within the computational subspace, we fit an exponential model to the average normalized purity $\bar{P}_\mathrm{norm}(N_\mathrm{Cl}) = A' + B'\tilde{u}^{N_\mathrm{Cl}}$ and compute the incoherent error per gate as $\varepsilon_\mathrm{incoh, g} = (1 - \sqrt{\tilde{u} + 2L_\mathrm{Cl}})/(2 N_\mathrm{g})$. To disentangle incoherent errors within the computational subspace from leakage, we correct the fitted $\tilde{u}$ by the measured leakage error per Clifford $L_\mathrm{Cl}$ to estimate the unitarity in the computational subspace as $u= \tilde{u} + 2L_\mathrm{Cl}$, which we motivate below in \cref{ap: motivation of analysis using a rate equation model}. Without this correction, the incoherent error would be overestimated in the presence of leakage since leakage error also contributes to the measured decay rate in a purity RB experiment. Finally, the coherent error within the computational subspace can be estimated as $\varepsilon_\mathrm{coh, g} = \varepsilon_\mathrm{g} - \varepsilon_\mathrm{incoh, g} - L_\mathrm{g}$. In the implementation of purity RB, we estimate all of the error metrics $\varepsilon_\mathrm{g}$, $\varepsilon_\mathrm{incoh, g}$, $\varepsilon_\mathrm{coh, g}$, and $L_\mathrm{g}$ from a single experiment to avoid the impact of temporal drifts on the estimation of $\varepsilon_\mathrm{coh, g}$. To this end, we use the shots with the final gate $I$ to estimate the total gate error $\varepsilon_\mathrm{g}$, and the shots across all final gates to estimate the leakage per gate $L_\mathrm{g}$.

\subsection{Motivation of the analysis using a rate equation model}
\label{ap: motivation of analysis using a rate equation model}

To motivate our leakage-aware analysis, we utilize a rate equation model of a qutrit describing the state probabilities of $|0\rangle$, $|1\rangle$, and $|2\rangle$ after the final inverting Clifford gate. We assume that the qutrit suffers from incoherent  errors within the computational subspace with a rate per Clifford $r/2$. In the absence of leakage errors, this corresponds to a depolarizing channel $\Lambda(\rho) = (1 - r) \rho + r \mathds{1}/2$ in the $\{|0\rangle, |1\rangle\}$ subspace. In addition to errors in the computational subspace, there is also leakage from the computational subspace to the $|2\rangle$ state with a leakage rate per Clifford $\gamma_\uparrow = L_\mathrm{Cl}$ and seepage back from the $|2\rangle$ state to the computational subspace with a seepage rate per Clifford $\gamma_\downarrow$. Since we model the probabilities $P_{|0\rangle}$, $P_{|1\rangle}$, and $P_{|2\rangle}$ after the final inverting Clifford gate, both states $|0\rangle$ and $|1\rangle$ are subject to a leakage rate of $\gamma_\uparrow$, while the seepage from $|2\rangle$ occurs to $|0\rangle$ or $|1\rangle$ with an equal probability of $\gamma_\downarrow/2$. This is because each Clifford gate in the RB sequence effectively randomizes the qubit state within the computational subspace.    Incorporating both errors within the computational subspace and leakage, the total gate error per Clifford is thus given by $\varepsilon_\mathrm{Cl} = r/2 + \gamma_\uparrow$, which is consistent with the convention of Refs.~\cite{wood2018quantification, chen2025randomized}. 
To connect $r$ and $\gamma_\uparrow$ to experimentally measurable RB decay rates, we formulate and solve the rate equation model. 

In the rate equation model, one additional Clifford gate changes the probabilities $P_{|0\rangle}$, $P_{|1\rangle}$, and $P_{|2\rangle}$ as
\begin{align}
    \frac{\mathrm{d} P_{|0\rangle}}{\mathrm{d} m} &= -\left(\frac{r}{2} + \gamma_\uparrow\right)P_{|0\rangle}(m) + \frac{r}{2}P_{|1\rangle}(m) + \frac{\gamma_\downarrow}{2} P_{|2\rangle}(m), \nonumber \\
    \frac{\mathrm{d} P_{|1\rangle}}{\mathrm{d} m} &= \frac{r}{2}P_{|0\rangle}(m) -\left(\frac{r}{2} + \gamma_\uparrow\right)P_{|1\rangle}(m) + \frac{\gamma_\downarrow}{2} P_{|2\rangle}(m), \nonumber \\
    \frac{\mathrm{d} P_{|2\rangle}}{\mathrm{d} m} &= \gamma_\uparrow P_{|0\rangle}(m) + \gamma_\uparrow P_{|1\rangle}(m) - \gamma_\downarrow P_{|2\rangle}(m),  \label{eq: rate equation model of qutrit} 
\end{align}
where we have denoted the number of Clifford gates as $m=N_\mathrm{Cl}$ for brevity, and taken the continuum limit $m \rightarrow 0$ using $\mathrm{d} P_{|i\rangle}/\mathrm{d} m \approx P_{|i\rangle}(m+1) - P_{|i\rangle}(m)$. Here, we have taken the continuum limit to solve the time evolution using standard techniques of differential equations. Alternatively, the discrete-time dynamics could be described as a discrete-time Markov chain, which would lead to an equivalent solution.

Noting that $P_{|0\rangle}(m) + P_{|1\rangle}(m) = 1 - P_{|2\rangle}(m)$, the equation for $P_{|2\rangle}(m)$ can be solved as
\begin{align}
    P_{|2\rangle}(m) &= \frac{\gamma_\uparrow}{\gamma_\Sigma} \left( 1 - e^{-\gamma_\Sigma m }\right) \label{eq: Pf continuum limit} \\
    &\approx  \frac{\gamma_\uparrow}{\gamma_\Sigma} \left( 1 - (1 - \gamma_\Sigma)^m \right), \label{eq: Pf discrete}
\end{align}
where $\gamma_\Sigma = \gamma_\uparrow + \gamma_\downarrow$. As proposed in Ref.~\cite{wood2018quantification}, the leakage error can be estimated  by fitting an exponential model of the form $P_{|2\rangle}(m) = A_\mathrm{f}  + B_\mathrm{f}\lambda_1^{m}$ to the leakage population and computing the leakage per gate as $L_\mathrm{g}=\gamma_\uparrow/N_\mathrm{g} = -B_\mathrm{f}(1-\lambda_1)/N_\mathrm{g}$.

Subsequently, $P_{|0\rangle}$ and $P_{|1\rangle}$ can be solved from the rate equation model in Eq.~\eqref{eq: rate equation model of qutrit} by inserting back the solution of $P_{|2\rangle}(m)$ in Eq.~\eqref{eq: Pf continuum limit} and assuming the solution to take a form $P_{|i\rangle} = A_i + B_ie^{-r_1m} + C_ie^{-\gamma_\Sigma m}$. As a result, we obtain 
\begin{align}
    A_0 &= A_1 = \frac{1}{2}\left(1 - \frac{\gamma_\uparrow}{\gamma_\Sigma} \right), \nonumber \\
    B_0 &= -B_1 = \frac{1}{2}, \nonumber \\
    C_0 &= C_1 =  \frac{\gamma_\uparrow}{2\gamma_\Sigma},  \nonumber \\
    r_1 &=r + \gamma_\uparrow.  \label{eq: P0 and P1 coefs}
\end{align}
Thus, we observe that $P_{|0\rangle}$ behaves as a sum of two exponential functions. In the presence of non-negligible leakage errors, the decay rates of the two exponential functions may be of similar order of magnitude, which may hamper robust fitting. Therefore, we  extract $r$ by considering $\tilde{P}_{|0\rangle} = P_{|0\rangle} + P_{|2\rangle}/2$, which behaves as a single-exponential function 
\begin{align}
    \tilde{P}_{|0\rangle} &= \frac{1}{2} \left(1 - e^{-(r + \gamma_\uparrow)m} \right) \nonumber \\
    &\approx \frac{1}{2} \left(1 - (1 - r - \gamma_\uparrow)^m \right). 
\end{align}
Thus, we fit a single exponential model $\tilde{P}_{|0\rangle}(m) = A_\mathrm{g}  + B_\mathrm{g}\lambda_2^{m}$ to  $\tilde{P}_{|0\rangle}(m)$ and evaluate the error per gate as $\varepsilon_\mathrm{g} = (1-\lambda_2)/(2N_\mathrm{g}) + L_\mathrm{g}/2$. 
We verified in experiments that $\tilde{P}_{|0\rangle}$ accurately approaches $1/2$ as $m$ increases even in the presence of high leakage, as predicted by the rate equation model. 

Finally, we investigate the impact of leakage error on the decay rate measured in purity RB in order to disentangle incoherent errors within the computational subspace from leakage. In purity RB, the incoherent error is inferred from the decay rate of the normalized purity $P_\mathrm{norm} = \langle \sigma_x \rangle^2 + \langle \sigma_y \rangle^2 + \langle \sigma_z \rangle^2$. In the rate equation model without coherent errors, we focus on the evolution of $\langle \sigma_z \rangle$ as $\langle \sigma_x \rangle = 0$ and $\langle \sigma_y \rangle = 0$. Using the solved time evolution of $P_{|0\rangle}(m)$ and $P_{|1\rangle}(m)$ given by Eq.~\eqref{eq: P0 and P1 coefs}, we observe that the average normalized purity $\bar{P}_\mathrm{norm}$ is given by  a single exponential, but the decay rate is increased due to leakage as
\begin{align}
    \bar{P}_\mathrm{norm} &= \langle \sigma_z \rangle^2 = (P_{|0\rangle}(m) - P_{|1\rangle}(m))^2 \nonumber \\
    &=e^{-2(r + \gamma_\uparrow)m} \approx (1 - 2(r +  \gamma_\uparrow))^m. 
\end{align}
In the absence of leakage errors, the normalized purity decays ideally as $\bar{P}_\mathrm{norm} = u^m$, and the decay is characterized by the unitarity $u=1 - 2r$, which is connected to the incoherent error per Clifford as $\varepsilon_\mathrm{Cl, incoh} = (1 - \sqrt{u})/2$~\cite{wallman2015estimating, feng2016estimating}. In the presence of leakage, we fit an exponential model $\bar{P}_\mathrm{norm}(m) = A' + B'\tilde{u}^{m}$ to the normalized purity and evaluate the unitarity within the computational subspace as $u = \tilde{u} +  2\gamma_\uparrow$ and the incoherent error per gate as $\varepsilon_\mathrm{g, incoh} = (1 - \sqrt{\tilde{u} +  2\gamma_\uparrow})/(2N_\mathrm{g})$. 

Using pulse-level simulations of purity RB, we confirm that the proposed approach is able to precisely estimate the incoherent error even in the presence of high leakage, i.e., $\gamma_\uparrow \sim r$, where the standard purity RB analysis would significantly overestimate the incoherent error. In experiments, we further confirm that the incoherent error estimated using the presented analysis stays approximately constant when moving between regimes with $\gamma_\uparrow \sim r$ and $\gamma_\uparrow \ll r$ by sweeping the DRAG coefficient $\beta$ for a given gate duration. However, we observe that the estimated incoherent error may underestimate the actual incoherent error if leakage error dominates, i.e., $\gamma_\uparrow \gg r$.

\section{Supplementary data for CTS pulse shaping on a high-crosstalk pair (Fig.~\ref{fig: experimental demonstration of off-resonant HD DRAG})}
\label{ap: supplementary fig 5 data}

In \cref{fig: experimental demonstration of off-resonant HD DRAG}, we demonstrated that CTS based on off-resonant HD DRAG pulses can reduce the gate error of the target qubit in a high-crosstalk pair by an order of magnitude during simultaneous drive pulses. 
In this section, we provide supplementary experimental results to additionally demonstrate that CTS pulse shaping can enable low leakage errors on both qubits and a similar gate error on the control qubit as conventional cosine DRAG pulses. The results consist of additional data 
for the off-resonant drive detuning $\Delta f_\mathrm{d}^\mathrm{C}$ sweep of \cref{fig: experimental demonstration of off-resonant HD DRAG}(b) shown in Fig.~\ref{fig: supplementary data on single-pair detuning sweep} and an additional sweep of pulse duration shown in Fig.~\ref{fig: supplementary data on single-pair duration sweep} using the same frequency configuration as in \cref{fig: experimental demonstration of off-resonant HD DRAG}(b). 
The measurements of \cref{fig: experimental demonstration of off-resonant HD DRAG}(b) were carried out at a set frequency configuration with $f_{01}^{\mathrm{Q}_{11}}=\SI{4.014}{\giga\hertz}$, $\alpha^{\mathrm{Q}_{11}}/(2\pi)=-\SI{183}{\mega\hertz}$,  $T_1^{\mathrm{Q}_{11}} = \SI{66}{\micro\second}$, $T_{2, \mathrm{e}}^{\mathrm{Q}_{11}} = \SI{20}{\micro\second}$, $f_{01}^{\mathrm{Q}_{5}}=\SI{4.074}{\giga\hertz}$, $\alpha^{\mathrm{Q}_{5}}/(2\pi)=-\SI{181}{\mega\hertz}$,  $T_1^{\mathrm{Q}_{5}} = \SI{58}{\micro\second}$, $T_{2, \mathrm{e}}^{\mathrm{Q}_{5}} = \SI{33}{\micro\second}$, and $C_{5, 11} = -\SI{13.9}{\decibel}$.

\subsection{Parameters and methods for Fig.~\ref{fig: experimental demonstration of off-resonant HD DRAG}(b)}

\begin{figure}[ht!]
    \centering
\includegraphics{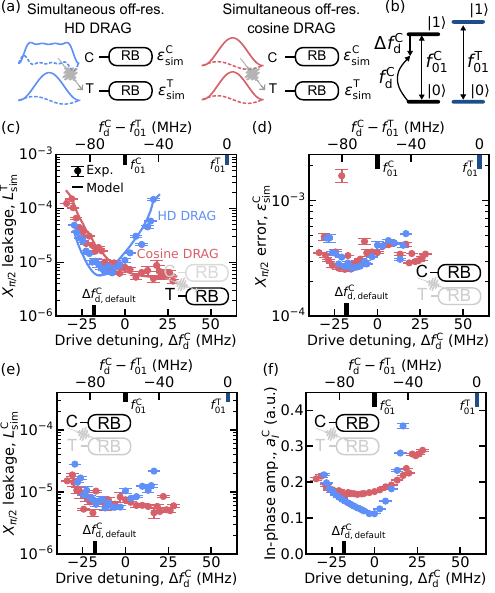}
    \caption{\textbf{Supplementary data for crosstalk transition suppression 
    for the high-crosstalk qubit pair as a function of drive detuning shown in Fig.~\ref{fig: experimental demonstration of off-resonant HD DRAG}(b).} (a) We benchmark the gate error and leakage of the target qubit and the control qubit using simultaneous leakage RB experiments when applying either off-resonant HD DRAG pulses (blue here and later) or off-resonant cosine DRAG pulses (red here and later) on the control qubit. (b) We sweep the off-resonant drive detuning $\Delta f_\mathrm{d}^\mathrm{C} = f_\mathrm{d}^\mathrm{C} - f_{01}^\mathrm{C}$ for a fixed qubit-qubit detuning of $f_{01}^\mathrm{C} - f_{01}^\mathrm{T} = -\SI{60}{\mega\hertz}$. (c)~Measured simultaneous leakage error of 20-ns $X_{\pi/2}$ gates for the target qubit $L_\mathrm{sim}^\mathrm{T}$ (markers) as a function of drive detuning, with an analytical model (solid lines) based on the leakage terms of Eq.~\eqref{eq:mt_eps_av random phase}. The measured leakage error of individual gates has been added to the analytical model. The marker on the lower x axis denotes the magnitude of the default drive detuning $\Delta f_\mathrm{d, default}^\mathrm{C}$ used in the qubit-qubit detuning sweep of \cref{fig: experimental demonstration of off-resonant HD DRAG}(c). (d) Measured simultaneous $X_{\pi/2}$ gate error of the control qubit $\varepsilon_\mathrm{sim}^\mathrm{C}$ as a function drive detuning. (e) Same as (d) but showing the measured simultaneous leakage error of $X_{\pi/2}$ gates for the control qubit $L_\mathrm{sim}^\mathrm{C}$. In panels (c)--(e), the error bars show the 1$\sigma$ uncertainty of mean based on 3 repeated leakage RB experiments. (f) Normalized in-phase amplitude $a_I^\mathrm{C}$ of the control qubit as a function of drive detuning, with the error bars showing the fit uncertainty. The data in all panels is for $\mathrm{Q}_5$--$\mathrm{Q}_{11}$  at the same qubit frequency configuration as \cref{fig: experimental demonstration of off-resonant HD DRAG}(b).
    }
    \label{fig: supplementary data on single-pair detuning sweep}
\end{figure}

\begin{figure}[ht]
    \centering
\includegraphics{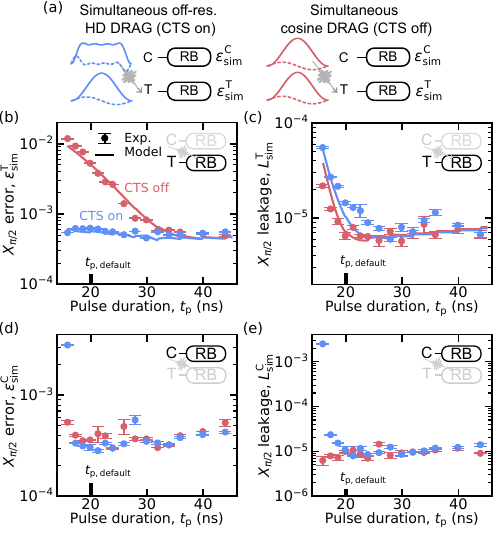}
    \caption{
    \textbf{Impact of crosstalk transition suppression as a function of pulse duration for the high-crosstalk qubit pair.} (a) We compare the gate error and leakage of the target qubit and the control qubit using simultaneous leakage RB experiments for CTS on or off. For CTS on, we apply off-resonant HD DRAG pulses (blue here and later) on the control qubit. For CTS off, we apply resonant cosine DRAG pulses (red here and later) on the control qubit. (b)~Measured simultaneous gate error of $X_{\pi/2}$ gates for the target qubit $\varepsilon_\mathrm{sim}^\mathrm{T}$ (markers) as a function of pulse duration, with an analytical model (solid lines) based on Eq.~\eqref{eq:mt_eps_av random phase}. The measured individual gate error  has been added to the analytical model. (c)~Measured simultaneous leakage error of $X_{\pi/2}$ gates for the target qubit $L_\mathrm{sim}^\mathrm{T}$ (markers) as a function of pulse duration, with an analytical model (solid lines) based on the leakage terms of Eq.~\eqref{eq:mt_eps_av random phase}. The measured leakage error of individual gates has been added to the analytical model.  (d) Measured simultaneous $X_{\pi/2}$ gate error of the control qubit $\varepsilon_\mathrm{sim}^\mathrm{C}$ as a function pulse duration. (e) Same as (d) but showing the measured simultaneous leakage error of $X_{\pi/2}$ gates for the control qubit $L_\mathrm{sim}^\mathrm{C}$. In panels (b)--(d), the error bars show the 1$\sigma$ uncertainty of mean based on 3 repeated leakage RB experiments. 
    The data in all panels is for $\mathrm{Q}_5$--$\mathrm{Q}_{11}$ at the same qubit frequency configuration as \cref{fig: experimental demonstration of off-resonant HD DRAG}(b).
    }
    \label{fig: supplementary data on single-pair duration sweep}
\end{figure}

Here, we provide more details of the methods and parameters used for the experiments in Sec.~\ref{sec: experimental demonstration on a high-crosstalk pair} and for the supplementary experiments in Appendix~\ref{ap: Additional data of the off-resonant drive detuning sweep} and \ref{ap: Pulse duration dependence}.
For each point in the drive detuning and duration sweeps in Figs.~\ref{fig: experimental demonstration of off-resonant HD DRAG}(b), \ref{fig: supplementary data on single-pair detuning sweep}, and \ref{fig: supplementary data on single-pair duration sweep}, we calibrate $X_{\pi/2}$ gates for HD DRAG and cosine DRAG pulses using the DRAG-L technique with the normalized in-phase amplitude $a_I$, the DRAG coefficient $\beta$, and the virtual Z rotation angle $\varphi_z$ as the pulse parameters ~\cite{hyyppa2024reducing}.  As illustrated in \cref{fig: supplementary x90 calibration} of Appendix~\ref{ap: single-qubit gate calibration}, we apply standard error amplification experiments to minimize overrotation errors, phase errors, and leakage. Before the gate calibration experiments, the off-resonant drive detuning $\Delta f_\mathrm{d}^\mathrm{C}$ of the HD DRAG pulse is chosen to shift the drive frequency away from the nearest target qubit transition as conceptually explained in Sec.~\ref{sec: off-resonant control pulses} and detailed in Appendix~\ref{ap: strongly off-resonant control pulses}. We set the suppressed frequencies of the HD DRAG envelope based on the chosen drive frequency $f_\mathrm{d}^\mathrm{C}$ and the measured transition frequencies as $\{f_{\mathrm{s}, j}\} = \{|f_\mathrm{d}^\mathrm{C} - f_{01}^\mathrm{T}|, |f_\mathrm{d}^\mathrm{C} - f_{12}^\mathrm{T}|, |f_\mathrm{d}^\mathrm{C} - f_{12}^\mathrm{C}|\}$ without any further optimization as motivated by Appendix~\ref{ap: selection of suppressed frequencies for HD DRAG}. To benchmark the error $\varepsilon_\mathrm{g}$ and leakage $L_\mathrm{g}$ of the calibrated $X_{\pi/2}$ gates, we use leakage RB~\cite{wood2018quantification} with three-state single-shot readout after each Clifford gate sequence, see Appendix~\ref{ap: benchmarking of single-qubit gate errors} for more details. 

\subsection{Additional data for the off-resonant drive detuning sweep}
\label{ap: Additional data of the off-resonant drive detuning sweep}

In \cref{fig: supplementary data on single-pair detuning sweep}, we present supplementary data for the drive detuning sweep comparing HD DRAG and cosine DRAG pulses with a gate duration of $t_\mathrm{g} = \SI{20}{\nano\second}$. 
For the set qubit-qubit detuning of $f_{01}^\mathrm{C} - f_{01}^\mathrm{T} = -\SI{60}{\mega\hertz}$,  a negative drive detuning $\Delta f_\mathrm{d}^\mathrm{C}$  reduces the energy spectral density of the control qubit pulse around the nearest harmful transition frequency $f_{01}^\mathrm{T}$, which decreases the gate error of the target qubit as discussed in \cref{sec: experimental demonstration on a high-crosstalk pair}. Figure~\ref{fig: supplementary data on single-pair detuning sweep}(c) demonstrates that HD DRAG pulse shaping also enables a low leakage error of the target qubit $L_\mathrm{sim}^\mathrm{T} \sim 10^{-5}$ 
for off-resonant drive detunings of interest $\Delta f_\mathrm{d}^\mathrm{C} \sim -\SI{20}{\mega\hertz}$. 
For larger negative drive detunings, the leakage error of the target qubit increases for both considered pulse shapes since the center drive frequency $f_\mathrm{d}^\mathrm{C}$ moves towards $f_{12}^\mathrm{T}$. 

Importantly, the gate error and leakage of the control qubit stay essentially constant across a large range of drive detunings using HD DRAG pulses as shown in Figs.~\ref{fig: supplementary data on single-pair detuning sweep}(d) and (e). Furthermore, 20-ns HD DRAG pulses enable a comparable gate performance on the control qubit compared to 20-ns cosine DRAG pulses. This implies that crosstalk transition suppression combining HD DRAG pulses and off-resonant driving enables crosstalk error reduction on the target qubit without compromising the gate performance of the control qubit. We suspect that the minor oscillation of the gate error as a function of drive detuning in \cref{fig: supplementary data on single-pair detuning sweep}(d) is caused by uncorrected microwave pulse reflections or distortions. Namely, we observe that the coherent error measured from purity RB varies between $0.5 \times 10^{-4}$ and $2\times 10^{-4}$ as a function of drive detuning, and generally decreases if we increase the zero padding from \SI{0}{\nano\second} to a few tens of ns. 

In \cref{fig: supplementary data on single-pair detuning sweep}(f), we further demonstrate that off-resonant drive pulses require  at most a small increase of drive amplitude compared to  conventional resonant pulses across the relevant range of drive detunings for the given qubit-qubit detuning. For specific parameters, HD DRAG pulses may even reduce the pulse amplitude compared to conventional cosine DRAG pulses, though this depends on the choice of the suppressed frequencies, the qubit-qubit detuning, and the drive detuning. 

\subsection{Pulse duration dependence}
\label{ap: Pulse duration dependence}

Finally, we demonstrate that crosstalk transition suppression using off-resonant HD DRAG pulses helps to significantly mitigate crosstalk errors  as the pulse duration is reduced as shown in \cref{fig: supplementary data on single-pair duration sweep}. Thus, crosstalk transition suppression unlocks faster single-qubit gates, which further reduces incoherent errors on both qubits.
For the pulse duration sweep, we use the same qubit frequency configuration as for the drive detuning sweep, but we sweep the pulse duration $t_\mathrm{p}$. Here, the gate duration $t_\mathrm{g}$ is rounded up as $t_\mathrm{g} = \lceil t_\mathrm{p} / (4~\mathrm{ns}) \rceil \times 4~\mathrm{ns}$ to be compatible with the waveform granularity of the used control electronics. Thus, the gate duration $t_\mathrm{g}$ includes both the pulse duration  $t_\mathrm{p}$ and any zero padding between pulses. For a given pulse duration with CTS on, we adjust the magnitude of drive detuning as $|\Delta f_\mathrm{d}^\mathrm{C}| = \SI{22}{\mega\hertz} \times \SI{20}{\nano\second} / t_\mathrm{p}$ since the range of possible drive detunings scales as $\propto 1/ t_\mathrm{p}$, as shown in \cref{ap: strongly off-resonant control pulses}. 
Figure~\ref{fig: supplementary data on single-pair duration sweep}(b) demonstrates that CTS keeps the gate error of the target qubit below $7 \times 10^{-4}$ down to pulse durations below \SI{20}{\nano\second}, whereas the gate error caused by a conventional cosine DRAG pulse approaches $10^{-2}$. Furthermore, the leakage error of the target qubit also stays low at $L_\mathrm{sim}^\mathrm{T} \sim 10^{-5}$ down to a pulse duration of $t_\mathrm{p} = \SI{20}{\nano\second}$ as demonstrated by Fig.~\ref{fig: supplementary data on single-pair duration sweep}(c).

For the control qubit, CTS on and off lead to a similar gate error apart from the shortest pulse duration of $t_\mathrm{p} = \SI{16}{\nano\second}$ as shown in~\cref{fig: supplementary data on single-pair duration sweep}(d). For CTS on, the leakage error of the control qubit increases for fast gates with $t_\mathrm{p} < \SI{20}{\nano\second}$, see~\cref{fig: supplementary data on single-pair duration sweep}(e). For such  
fast pulses, it is challenging to symmetrically suppress three frequencies within $|\alpha^\mathrm{C}|/(2\pi) \sim \SI{180}{\mega\hertz}$, which leads to fast oscillations in the drive envelope and increased peak amplitudes. 
The FAST DRAG pulse  introduced in Ref.~\cite{hyyppa2024reducing} could help to regularize the pulse shape applied to the control qubit, which has potential to drastically reduce the leakage error of the control qubit at the cost of additional hyperparameters.

\section{Supplementary experimental data for CTS pulse shaping at QPU scale (Fig.~\ref{fig: hd drag on bw minimized config})}
\label{ap: supplementary data on Fig 6}

In this section, we provide supplementary data for the QPU-scale experiments demonstrating CTS pulse shaping based on off-resonant HD DRAG pulses  in \cref{sec: demonstration of reduced bandwidth on full QPU using pulse shaping techniques}. Throughout this section, we benchmark simultaneous single-qubit gate performance on 46 qubits but show the results for the 11 high-crosstalk CTS target qubits in group H and for the 35 other qubits separately. 
This is because CTS pulse shaping attempts to mitigate the crosstalk error on the qubits in group H only, and thus we can isolate the contribution of CTS to the error more clearly. 
Similarly to \cref{sec: demonstration of reduced bandwidth on full QPU using pulse shaping techniques}, we compare CTS pulse shaping on high-crosstalk pairs (CTS on) to conventional control pulses on all qubits (CTS off) as  illustrated in Fig.~\ref{fig: hd drag on bw minimized config}(a). 

\subsection{Stability of gate error with CTS pulse shaping}
\label{ap: stability of error with off-res HD DRAG }

\begin{figure}[ht]
    \centering
\includegraphics{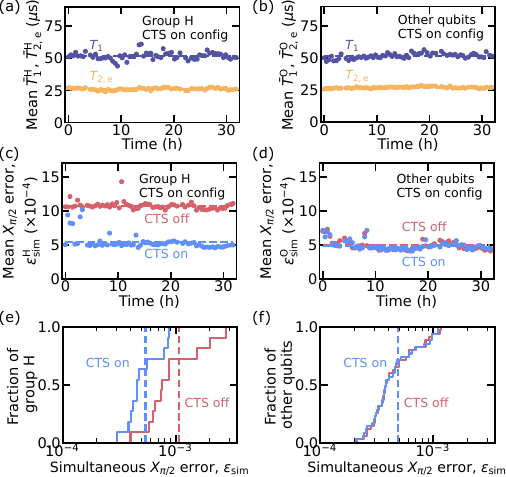}
    \caption{\textbf{Supplementary data for a 30-h stability experiment of gate error and coherence in the CTS-on frequency configuration.} (a) Mean relaxation time $\bar{T}_1^\mathrm{H}$(purple)  and echo coherence time $\bar{T}_{2, \mathrm{e}}^\mathrm{H}$ (yellow) for the 11 high-crosstalk target qubits in group H as a function of time. (b) Same as (a) but for the 35 other qubits. (c) Mean  error of simultaneous 20-ns $X_{\pi/2}$ gates across the 11 qubits in group H for CTS on (blue here and later) or off (red here and later) as a function of time without recalibration. The dashed line shows the mean across the full 30-h measurement. (d) Same as (c) but for the 35 other qubits. (e) Cumulative distribution function of simultaneous $X_{\pi/2}$ error for the 11 qubits in group H averaged across the 30-h measurement for CTS on or off. (f)~Same as (e) but for the 35 other qubits.  
    }
    \label{fig: more data on RB stability 46 qbs}
\end{figure}

To demonstrate systematic and stable crosstalk error reduction using CTS pulse shaping, we monitor the gate error over a time period of 30 hours without recalibration in Fig.~\ref{fig: more data on RB stability 46 qbs}. Additionally we monitor the coherence time in an interleaved fashion with the error. In Figs.~\ref{fig: more data on RB stability 46 qbs}(a) and (b), we show the measured mean relaxation time $\bar{T}_1$ and the mean echo coherence time $\bar{T}_\mathrm{2, e}$ for the 11 qubits in group H and for the 35 other qubits, respectively. For both qubit groups, the mean $\bar{T}_1$ is slightly over 50~$\mu$s and the mean $\bar{T}_\mathrm{2, e}$ exceeds 25~$\mu$s. We further compare the mean gate error of simultaneous 20-ns $X_{\pi/2}$ gates based on repeated standard RB experiments for CTS on and off considering the 11 qubits in group H in \cref{fig: more data on RB stability 46 qbs}(c) and the 35 other qubits in  \cref{fig: more data on RB stability 46 qbs}(d). Cumulative distribution functions of the qubit-wise mean error over the 30-h time period are shown in Figs.~\ref{fig: more data on RB stability 46 qbs}(e) and (f) for the corresponding qubit groups. For the high-crosstalk qubits in group H, CTS pulse shaping reduces the mean simultaneous gate error by a factor of two compared to conventional pulse shaping. Below, we further demonstrate that the gate errors enabled by CTS are close to coherence-limited. 

Importantly, we also do not observe any degradation of the simultaneous gate error of the 35 other qubits when using CTS as shown in \cref{fig: more data on RB stability 46 qbs}(f). 
This is a key feature of the pulse-aware qubit frequency optimization.  The pulse-aware frequency optimization and the local nature of crosstalk on our device ensure that no additional error is incurred even if the energy spectral density of an HD DRAG pulse is increased at frequencies other than the suppressed ones (see~\cref{fig: concept of higher-derivative drag}(b)).

\begin{figure}[t]
    \centering
\includegraphics{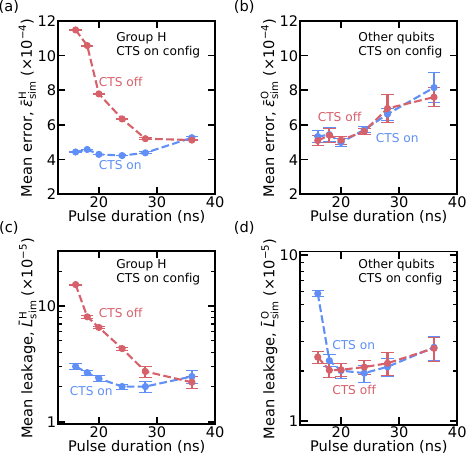}
    \caption{\textbf{Supplementary data for the QPU-scale duration sweep in the CTS-on frequency configuration shown in Fig.~\ref{fig: hd drag on bw minimized config}(d).}  (a) Gate error of simultaneous $X_{\pi/2}$ gates measured from leakage RB and averaged across the 11 high-crosstalk qubits in group H as a function of pulse duration for CTS pulse shaping on (blue here and later) or off (red here and later). (b) Same as (a) but for the 35 other qubits. (c) Leakage error of simultaneous $X_{\pi/2}$ gates measured from leakage RB and averaged across the 11 high-crosstalk qubits in group H as a function of pulse duration  for CTS on  or off.  (d)  Same as (c) but for the 35 other qubits.  In all panels, error bars represent 1$\sigma$ uncertainty of the mean based on 5 repeated leakage RB experiments, and the dashed lines act as a guide for the eye. }
    \label{fig: more data on error vs duration 46 qbs}
\end{figure}

\subsection{Gate error vs duration with and without CTS pulse shaping}
\label{ap: error vs duration using off-res HD DRAG }

We utilize leakage RB experiments to investigate the gate error and leakage of simultaneous $X_{\pi/2}$ gates as a function of pulse duration in the CTS-on configuration shown in Fig.~\ref{fig: hd drag on bw minimized config}(b). Similarly to \cref{ap: supplementary fig 5 data}, we round up the gate duration as $t_\mathrm{g} = \lceil t_\mathrm{p} / (4~\mathrm{ns}) \rceil \times 4~\mathrm{ns}$ and thus, the gate duration includes both the pulse duration and any zero padding between pulses.  In Fig.~\ref{fig: hd drag on bw minimized config}(d), we demonstrated that CTS pulse shaping significantly reduces the gate error on high-crosstalk target qubits in  group H for fast gates with $t_\mathrm{g} \lesssim \SI{30}{\nano\second}$. Here, our goal is to show that CTS pulse shaping also reduces the leakage error in group H, while also preserving a low leakage error on the other qubits. 
For each pulse duration, we re-calibrate the pulse parameters but use the same qubit frequency configuration optimized for 20-ns $X_{\pi/2}$ gates assuming CTS pulse shaping.

In Figs.~\ref{fig: more data on error vs duration 46 qbs}(a) and (b), we demonstrate a significant reduction of both gate error and leakage error on the high-crosstalk target qubits in group H for pulse durations below \SI{30}{\nano\second}. For $t_\mathrm{g}=t_\mathrm{p} = \SI{20}{\nano\second}$, the mean simultaneous leakage error across group H is reduced from $\bar{L}_\mathrm{sim}^\mathrm{H} = 6.5\times 10^{-5}$ without CTS to $\bar{L}_\mathrm{sim}^\mathrm{H} = 2.4\times 10^{-5}$ with CTS. For the 35 other qubits, CTS pulse shaping has practically no impact on the simultaneous gate error that stays unchanged within the error bars across the studied range of pulse durations $t_\mathrm{p} \in [16, 36]$~ns as shown in Fig.~\ref{fig: more data on error vs duration 46 qbs}(c). For the shortest tested pulse duration of $t_\mathrm{p} = \SI{16}{\nano\second}$, the leakage error increases on the control qubits with CTS pulse shaping, thus leading to an elevated mean leakage error for $t_\mathrm{p} = \SI{16}{\nano\second}$ in Fig.~\ref{fig: more data on error vs duration 46 qbs}(d). Therefore, $t_\mathrm{p} = \SI{16}{\nano\second}$ essentially represents a speed limit in our system if CTS is implemented using HD DRAG pulses suppressing three transitions. 

\begin{figure}[t]
    \centering
\includegraphics{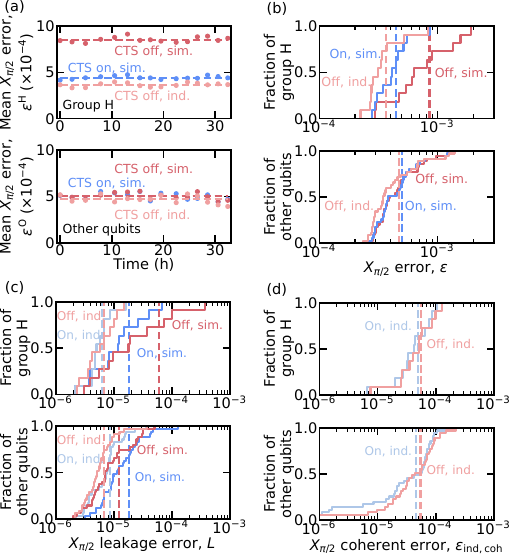}
    \caption{\textbf{Supplementary data for the error budget of simultaneous 20-ns $X_{\pi/2}$ gates in the CTS-on frequency configuration shown in Fig.~\ref{fig: hd drag on bw minimized config}(f).}  (a) Top: Mean $X_{\pi/2}$ error from standard RB experiments averaged across the 11 qubits in group H as a function of time without recalibration. Data is shown for simultaneous gates with CTS on (dark blue here and later), simultaneous gates with CTS off (dark red here and later), and individual gates with CTS off (light red here and later). Bottom: Same but for the 35 other qubits. (b) Top: Cumulative distribution function of $X_{\pi/2}$ error for the 11 qubits in group H averaged across the 30-h stability measurement shown in (a). Bottom: Same  but for the 35 other qubits. (c) Top: Cumulative distribution function of $X_{\pi/2}$ leakage error for the qubits in group H measured from leakage RB for simultaneous and individual (light blue here and later) gates with CTS on, and simultaneous and individual gates with CTS off. Bottom: Same but for the other 35 qubits. (d) Top: Cumulative distribution function of coherent errors of $X_{\pi/2}$ gates for the qubits in group H measured from purity RB for individual gates with CTS on or off. Bottom: Same but for the 35 other qubits. In all panels, dashed lines denote mean values.  }
    \label{fig: cdfs behind error budget}
\end{figure}

\subsection{Error budget for simultaneous single-qubit gates}
\label{ap: error budget}

\begin{figure}[b]
    \centering
\includegraphics{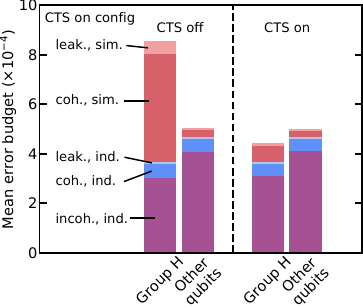}
    \caption{\textbf{Comparison of the $X_{\pi/2}$ error budget for CTS on and off in the CTS-on configuration.} We show the mean error budget separately for the 11 high-crosstalk qubits in group H and for the other 35 qubits. The error budget includes crosstalk-induced leakage (light red), crosstalk-induced errors within computational subspace (dark red), leakage error of individual gates (light blue), coherent error of individual gates (dark blue), and incoherent error (purple).  }
    \label{fig: suppl error budget}
\end{figure}

Finally, we provide supplementary experimental data behind the error budget of simultaneous 20-ns $X_{\pi/2}$ gates shown in \cref{fig: hd drag on bw minimized config}(f). For the error budget, we consider crosstalk-induced leakage errors  $L_\mathrm{ct, coh}$, crosstalk-induced errors within the computational subspace $\varepsilon_\mathrm{ct, coh}$, leakage errors of individual gates $L_\mathrm{ind}$, coherent errors of individual gates within the computational subspace $\varepsilon_\mathrm{ind, coh}$, and incoherent errors within the computational subspace $\varepsilon_\mathrm{ind, incoh}$.  First, we distinguish between crosstalk-induced errors and individual gate errors by performing repeated simultaneous and individual RB experiments for CTS on and off, yielding $\varepsilon_\mathrm{sim}$ and  $\varepsilon_\mathrm{ind}$ for each qubit. We monitor the gate error over a time period of 30 h to average over temporal fluctuations of coherence times as shown in Fig.~\cref{fig: cdfs behind error budget}(a). We measure the individual gate error only without CTS to save time since the individual gate error is approximately equal for both pulse shaping scenarios.  The cumulative distribution functions of gate error in \cref{fig: cdfs behind error budget}(b) demonstrate that CTS pulse shaping brings the simultaneous gate error close to the individual gate error for the qubits in group H, whereas significant microwave crosstalk errors remain using conventional pulse shapes. For the 35 other qubits, the simultaneous gate error is close to the individual gate error for both pulse shaping scenarios.

Subsequently,  we use the leakage-aware purity RB described in Appendix~\ref{ap: benchmarking of single-qubit gate errors}, to measure the leakage error of simultaneous gates $L_\mathrm{sim}$, the leakage error of individual gates  $L_\mathrm{ind}$, and the coherent error of individual gates $\varepsilon_\mathrm{ind, coh}$ as shown in Figs.~\ref{fig: cdfs behind error budget}(c) and (d). 
Finally, we estimate the crosstalk-induced leakage error as $L_\mathrm{ct} = L_\mathrm{sim} - L_\mathrm{ind}$, the crosstalk-induced errors within the computational subspace as $\varepsilon_\mathrm{ct, coh} = \varepsilon_\mathrm{sim} - \varepsilon_\mathrm{ind} - L_\mathrm{ct}$, and the incoherent errors within the computational subspace as $\varepsilon_\mathrm{ind, incoh} = \varepsilon_\mathrm{ind} - \varepsilon_\mathrm{ind, coh}- L_\mathrm{ind}$. We summarize the mean error budget for CTS on and off in \cref{fig: suppl error budget}.

 Based on \cref{fig: cdfs behind error budget}(c), we observe that the mean leakage error of simultaneous gates in group H is an order of magnitude higher than that of individual gates if conventional pulse shapes are used. CTS pulse shaping reduces the mean leakage error of simultaneous gates by a factor of three compared to conventional cosine DRAG pulses. For the 35 other qubits, CTS pulse shaping increases the mean simultaneous leakage error marginally by $0.6 \times 10^{-5}$ compared to conventional pulse shaping. 
 
 Finally, we show that the coherent error of individual gates in \cref{fig: cdfs behind error budget}(d) is approximately equal for CTS on and off.  The mean coherent error of individual gates is approximately at the level of $5 \times 10^{-5}$ for both qubit groups, with the largest errors  exceeding $10^{-4}$. We suspect the remaining coherent errors to be  caused by uncorrected microwave pulse reflections or distortions~\cite{hyyppa2024reducing, guo2024correction}.

\end{document}